\documentclass[a4paper]{article}

\usepackage[english]{babel}
\usepackage[utf8x]{inputenc}
\usepackage[T1]{fontenc}

\usepackage[a4paper,top=3cm,bottom=2cm,left=3cm,right=3cm,marginparwidth=1.75cm]{geometry}

\usepackage[table]{xcolor}
\usepackage{booktabs}
\usepackage{amsmath}
\usepackage{graphicx}
\usepackage[colorinlistoftodos]{todonotes}
\usepackage[colorlinks=true, allcolors=blue]{hyperref}
\usepackage{authblk}
\usepackage[comma,super,sort&compress]{natbib}
\usepackage{nomencl}
\usepackage{booktabs}
\usepackage{tabularx}
\usepackage{textcomp}
\usepackage{amsfonts}
\usepackage{comment} 
\usepackage{outlines} 
\usepackage{subfig}
\usepackage[version=4]{mhchem}
\usepackage{siunitx}
\usepackage{xcolor}
\date{}

\usepackage{lineno}
\usepackage{color}
\usepackage{soul}
\usepackage{cancel}

\title{Graph neural networks for materials science and chemistry}

\author[1,2]{Patrick Reiser}
\author[1]{Marlen Neubert}
\author[1]{Andr\'e Eberhard}
\author[1]{Luca Torresi}
\author[1]{Chen Zhou}
\author[1]{Chen Shao}
\author[1,3]{Houssam Metni}
\author[1,4]{Clint van Hoesel}
\author[1,2]{Henrik Schopmans}
\author[1,5]{Timo Sommer}
\author[1,2]{Pascal Friederich}

\affil[1]{Institute of Theoretical Informatics, Karlsruhe Institute of Technology, Am Fasanengarten 5, 76131 Karlsruhe, Germany}
\affil[2]{Institute of Nanotechnology, Karlsruhe Institute of Technology, Hermann-von-Helmholtz-Platz 1, 76344 Eggenstein-Leopoldshafen, Germany}
\affil[3]{ECPM, Universit\'{e} de Strasbourg, 25 Rue Becquerel, 67087 Strasbourg, France}
\affil[4]{Department of Applied Physics, Eindhoven University of Technology, Groene Loper 19, 5612 AP Eindhoven, The Netherlands}
\affil[5]{Institute for Theory of Condensed Matter, Karlsruhe Institute of Technology, Wolfgang-Gaede-Str. 1, 76131 Karlsruhe, Germany}

\begin{document}
    \maketitle
    \begin{abstract}
    Machine learning plays an increasingly important role in many areas of chemistry and materials science, e.g. to predict materials properties, to accelerate simulations, design new materials, and to predict synthesis routes of new materials.
    Graph neural networks (GNNs) are one of the fastest growing classes of machine learning models.
    They are of particular relevance for chemistry and materials science, as they directly work on a graph or structural representation of molecules and materials and therefore have full access to all relevant information required to characterize materials.
    In this review article, we provide an overview of the basic principles of GNNs, widely used datasets and state-of-the-art architectures, followed by a discussion of a wide range of recent applications of GNNs in chemistry and materials science, and concluding with a road-map for the further development and application of GNNs.
    \end{abstract}

    \section{Introduction}\label{sec:introduction}
    
Data science and machine learning have become an integral part of natural sciences, discussed as the fourth pillar in science, next to experiment, theory, and simulation~\cite{von_Lilienfeld_2020}.
Machine learning methods are increasingly applied in all steps of the materials development cycle, from finding initial candidate materials using property prediction~\cite{PhysRevMaterials.4.093801, vamathevan_applications_2019}, database screening~\cite{Jorissen2005VirtDatabase, ekins_exploiting_2019} or even inverse materials design~\cite{Juhwan2020InvDesign, zunger_inverse_2018}, over the detailed analysis of materials in machine learning accelerated simulations~\cite{PhysRevLett.98.146401,friederich2021machine}, to the prediction of synthesis conditions~\cite{shields2021bayesian,luo2021mof} and automated experimental data analysis~\cite{kalinin2022machine,velasco2021phase} and experimental planning~\cite{HASE2019282}.
Machine learning models applied in chemistry and materials science cover a wide spectrum of methods, ranging from classical machine learning models such as decision tree ensembles to modern deep learning methods such as convolutional neural networks~\cite{LeCun2010IEEE} and sequence models~\cite{schwaller2019molecular} originally developed for challenges in computer vision and natural language processing.

A recent addition to the toolbox of machine learning models for chemistry and materials science are graph neural networks (GNNs), which operate on graph-structured data and have strong ties to the field of geometric deep learning~\cite{Sperduti1997,Gori2005,scarselli-gnn}.
Aside from research on social and citation networks as well as knowledge graphs, chemistry has been one of the main drivers in the development of GNNs~\cite{duvenaud2015convolutional,message-passing-qc}.
Graph neural networks can be interpreted as the generalization of convolutional neural networks to irregular-shaped graph structures.
While other machine learning methods, e.g. convolutional neural networks are at the peak of publication activity, GNNs are still rising exponentially, with hundreds of papers per year since 2019 (see Figure \ref{fig:figure1}).
Their architecture allows them to directly work on natural input representations of molecules and materials, which are chemical graphs of atoms and bonds, or even 3D structures or point clouds of atoms.
Therefore, GNNs have access to a complete representation of materials on the atomic level~\cite{schutt_schnet_2018}, with a lot of flexibility to incorporate physical laws~\cite{Unke2019physnet}, as well as phenomena on larger scales, such as doping and disorder.
Using that information, GNNs can learn internal materials representations that are useful and informative for specific tasks such as the prediction of given materials' properties.
Therefore, GNNs can complement or even replace hand-crafted feature representations which were and are widely used in the context of natural sciences in general.
A similar trend toward representation learning methods has also been observed in other application areas during the last years, where end-to-end trainable models show a systematic advantage over traditional feature-based methods~\cite{SchmidtCrystalAttention2021}.
However, despite promising recent developments toward higher sample efficiency~\cite{batzner20223,klicpera2021gemnet}, this often comes at the cost of higher data requirements, potentially limiting the applicability of existing GNNs to applications where large amounts of data are available.
Overall, GNNs outperformed conventional machine learning models in predicting molecular properties throughout the last years~\cite{schutt_schnet_2018,Klicpera2020DimeNet,schutt2021painn}.
While GNNs are not as widely applied (yet) in materials science as they are in chemistry, there are advantages and the potential to outperform other machine learning methods and thus boost virtual materials design and materials science in general, which will be discussed in this article.

In Section 2, we will introduce the general formalism of GNNs and discuss the way they transform the atomic structure of materials and molecules and use it to predict materials' properties. We will present and compare state-of-the-art architectures and benchmark datasets, as well as summarize initial efforts towards inverse materials design based on GNNs.
Section 3 covers a wide range of current application areas but also open challenges for GNNs in chemistry and materials science.
Section 4 concludes with a perspective on necessary and expected future developments and so far unused potential of GNNs in materials science.

    \section{Graph neural networks in materials science and chemistry}\label{sec:graph-neural-networks-in-materials-science-and-chemistry}

    \subsection{Basic principles}\label{subsec:basic-principles}
    
In the most general sense, graphs are used to describe abstract structures consisting of entities or objects represented as \textit{vertices} (or \textit{nodes}) and their connections, called \textit{edges}.
Formally, a graph is a tuple $G = (V, E)$ of a set of vertices $v \in V$ and a set edges $e_{v,w} = (v, w) \in E$, which defines the connection between vertices. Potential tasks that can be solved using graph neural networks (GNNs) include classification or regression of graph properties on graph level (molecular property prediction), node level (classification of members, i.e. nodes, of a social graph), or edge level (prediction of relations, i.e. edges, between objects in a scene graph). In materials science and chemistry, most tasks involve graph-level predictions, which will be the focus of this paper.

The concept of graphs is used in mathematical chemistry to represent the structure of compounds.
The molecular structure is represented by an undirected graph, where nodes correspond to atoms and edges correspond to chemical bonds.
In fact, chemical graphs were first considered as early as in 1874~\cite{Cayley1874Isomer} and their idea traces back further~\cite{Bonchev1991ChemicalGraph}, which may place them even before the advent of the term \textit{graph} in modern graph theory~\cite{Biggs1986GraphTheory}. The description of molecules as graphs can also be transferred to solid-state materials, even though bonds might not be uniquely defined in crystals, and the exact three-dimensional arrangement of atoms plays a more decisive role.

Since their proposal~\cite{Sperduti1997, Gori2005, scarselli-gnn}, GNNs have become a popular machine 
learning method for processing irregularly shaped data encoded as graphs.
They can be seen as an alternative to approaches, where predefined feature representations of molecules or materials are used as input to conventional machine learning models such as densely connected neural networks, random forest models, or Gaussian process regression models. In the case of GNNs, the full molecular structure or even geometry is used as input and the GNN itself learns informative molecular representations to predict given target properties.
Due to their popularity and wide applicability, a large number of different GNN architectures have been proposed~\cite{Merkwirth2005Chem, kipf-welling-gnn, Gori2005, duvenaud2015convolutional, Defferrard2016, message-passing-qc}.
While the exact architecture type can notably differ, ranging from the initially proposed recursive GNNs~\cite{Gori2005} to spectral neural filters~\cite{Bruna2014, Defferrard2016} and finally to spatial or convolutional GNNs~\cite{kipf-welling-gnn}, most GNNs designed for chemistry and materials
science can be summarized under the framework of Message Passing Graph Neural Networks (MPNN) as suggested by Gilmer et al.~\cite{message-passing-qc}.
In this section, we give an overview of ideas of the message passing framework and discuss how learned graph- or node-level embeddings can be used for materials property prediction.

\begin{figure}[h]
    \includegraphics[width=\linewidth]{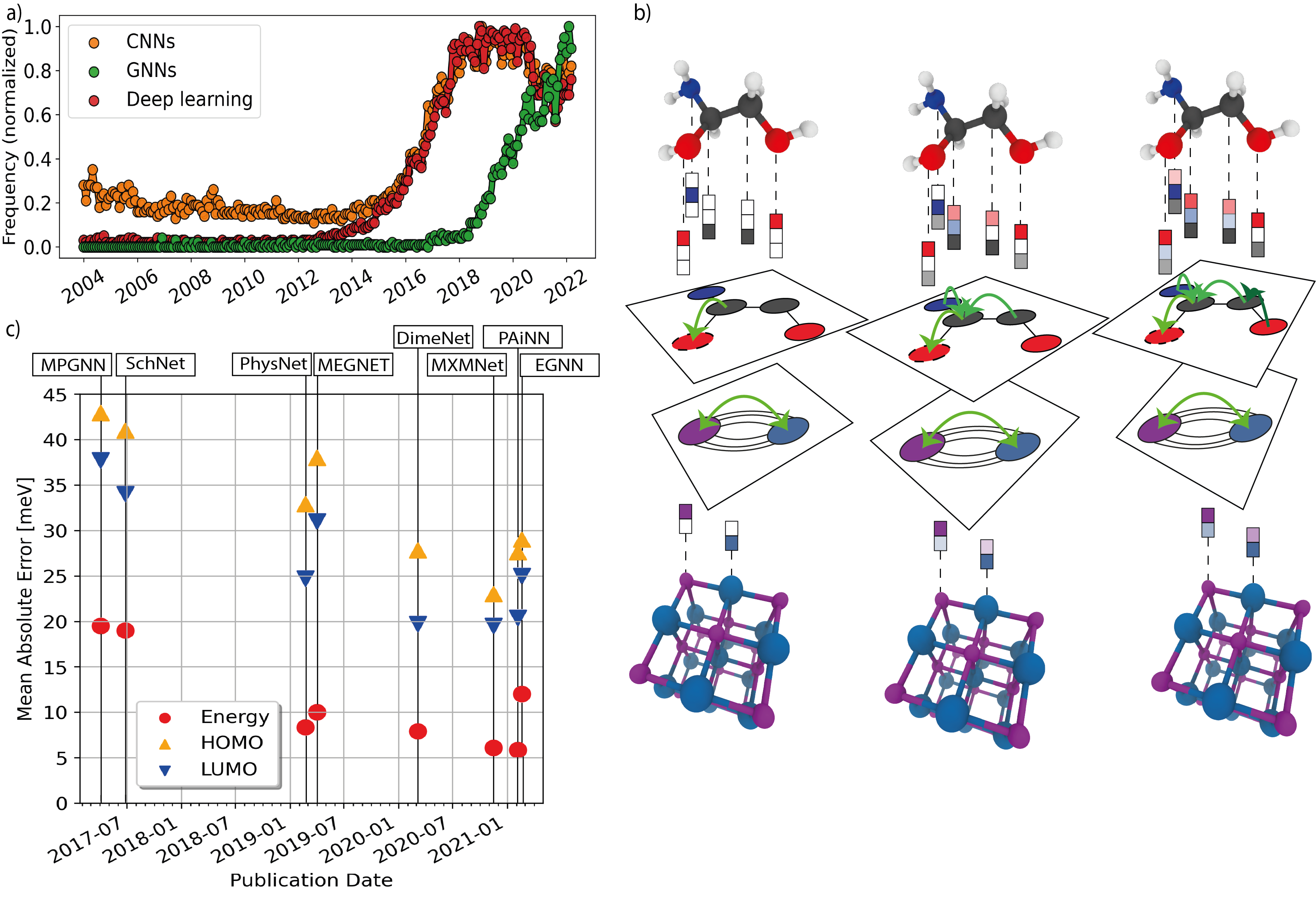}
    \caption{a) Analysis of the keywords "convolutional neural network", "graph neural network", and "deep learning" based on Google Trends (March 2022) b) Schematic depiction of the message passing operation for molecules and crystalline materials. c) QM9 benchmark. Mean absolute error of the prediction of internal, HOMO and LUMO energies for different GNN models since 2017.}
    \label{fig:figure1}
\end{figure}

For MPNNs, associated node or edge information (e.g. atom and bond types) is commonly provided by node attributes $h_{v}^{0}\in \mathbb{R}^{d}$ and edge attributes $h_{e}^{0} \in \mathbb{R}^{c}$.
Details about feature and structure representations are discussed in Section~\ref{subsec:structure-representation}.
Using node and edge features in combination with the graph's structure, GNNs are capable of deriving a node-level embedding of the graph, i.e. learned vectors representing each atom including its individual chemical environment.
This is done in the so-called \textit{message passing} phase, in which node information is propagated in form of messages $m_v$ through edges to neighboring nodes.
The embedding of each node is then updated based on all incoming messages.
The locality of the message passing is sought to be alleviated by repeating the message passing phase $t=1\dots K$ times, in principle allowing information to travel longer distances, i.e.\ within the K-hop neighborhood. In practice, however, information from long-range dependencies can be distorted in node bottlenecks, referred to as \textit{over-squashing}\cite{over-squashing}, or be washed out, leaving indistinguishable representations of neighboring nodes, known as \textit{over-smoothing}\cite{Li_Han_Wu_2018}.
Note that for typical (not fully linear) molecules and crystal unit cells with $n$ atoms, only approximately $\log n$ message passing steps are required to pass information to all other atoms.
The information processing is facilitated by the learnable functions $U_{t}(\cdot)$ for node update and $M_{t}(\cdot)$ for the message generation.
Finally, in the \textit{readout} phase, a graph-level embedding $y$ is obtained by pooling node embeddings of the entire graph via a parametric readout function $R(\cdot)$.
The final representation of the graph is used for training both regression and classification tasks.
In summary, the MPNN scheme reads~\cite{message-passing-qc}:
\begin{align}
    \label{eq:equation_message}
    m^{t+1}_{v} &= \sum_{w\in N(v)}M_{t}(h^{t}_{v}, h^{t}_{w}, e_{vw}) \\ 
    h^{t+1}_{v} &= U_{t}(h^{t}_{v}, m^{t+1}_{v}) \\
    y &= R(\{ h^{K}_{v}| v \in G\}) \;,
\end{align}
were $N(v) = \{u \in V | (v, u) \in E\}$ denotes the set of neighbors of node $v$.
Note that readout and aggregation can be in principle any mathematical operation that is permutation-invariant, e.g. a sum, mean or maximum operation similar to Equation~\ref{eq:equation_message} or learnable such as the \textit{Set2Set} encoder proposed by Vinyals et al.~\cite{set2set}, which was originally used for the readout $R(\cdot)$.
The learnable functions are mostly neural networks and eventually determine the performance characteristics of the GNN, both in prediction accuracy and computational cost.
Figure~\ref{fig:figure1}b shows a schematic of the message passing scheme for the example of a molecular graph.
Message passing can also be understood as a convolution operation running over each node in a graph.
Different extensions and modifications of the message passing schemes are discussed in Section~\ref{subsec:datasets-and-state-of-the-art-architectures} and include edge updates~\cite{Yang2019DMPNN}, skip connections~\cite{Schutt2018schnetpack}, and geometric information~\cite{klicpera2021gemnet,shui2020heterogeneous,finzi2020lieconv}.

A main open research question of GNNs revolves around their limited expressive performance for specific tasks~\cite{bodnar2021weisfeiler} and how GNNs compare with Weisfeiler-Lehman hierarchy for graph isomorphism testing~\cite{NEURIPS2019_bb04af0f, gin}. With regard to this topic, there are many promising extensions to GNNs proposed in literature, such as hypergraph representations~\cite{feng2019hypergraph, JoEdge2021}, universal equivariant models~\cite{dym2021on} or higher-order graph networks~\cite{morris2021weisfeiler}. Furthermore, the challenge of over-smoothing due to commonly used aggregation functions~\cite{Li_Han_Wu_2018}, transfer and multitask learning~\cite{MAT2020_multi}, as well as training (in)stability~\cite{CHEN2022613} are subject of current research.

    \subsection{Structure representation}\label{subsec:structure-representation}
    
Many graph networks directly use the chemical graph as input, representing both molecules~\cite{message-passing-qc} and inorganic compounds~\cite{PhysRevLett.120.145301, SchmidtCrystalAttention2021}, and offering advantages over compositional or fixed-sized vector representations in terms of flexibility and scalability.
Consequently, GNNs can be applied for tasks such as drug design or material screening~\cite{Zhaoping2020AttentiveFP}, which require knowledge about functional groups, scaffolds~\cite{Kruger2020Scaffold} or the full chemical structure and its topology.
In molecular applications, the chemical graph is often extracted from SMILES codes and augmented with features that can be readily obtained from cheminformatics software such as RDkit~\cite{rdkit} or OpenBabel~\cite{oboyle_open_2011}.
Common features for atoms, bonds, and the molecule itself are listed in Table~\ref{tab:mol_features}.
Besides hand-crafted input features, learned embeddings of molecules and materials motivated by word embedding techniques in natural language processing have been explored which can be used for downstream tasks~\cite{JoEdge2021,Chen2020DrugDeepWalk,Perozzi2014DeepWalk,Grover2016Node2vec}. For specific tasks in chemistry, the connectivity of atoms in molecules (i.e. the molecular graph) contains sufficient and complete information to predict given molecular properties which do not depend on the exact geometry.
Geometry or stereochemical information can be taken into account e.g. in form of additional edge features representing the distance between atoms~\cite{chenGraphNetworksUniversal2019}.
In contrast to that, in materials applications, atom connectivity is not well defined in most cases (apart from e.g. covalently linked frameworks) and graphs have to be extracted from crystal structures based on distance heuristics.\\
\begin{table}[h]
    \centering
    \begin{tabular}{c|c|c}
    Graph-level  &  Attributes & Description \\
    \hline
    nodes & atom-type & type of atoms (one-hot) \\
    & chirality & R or S (one-hot or null) \\
    & degree & number of covalent bonds (one-hot) \\
    & radical & number of radical electrons (integer) \\
    & hybridization & sp, sp$^2$, sp$^3 \dots$ (one-hot) \\ 
    & aromaticity & part of an aromatic system (binary)\\
    & charge & formal charge (integer) \\
    edges & bond-type & single, double, $\dots$ (one-hot) \\
    & conjugation & is conjugated (binary) \\
    & ring & bond is part of a ring (binary) \\
    & stereo & None, Any, Z, E (one-hot) \\
    graph & weight & average atomic weight (float) \\
    & bonds & average bonds per atom (float)
    \end{tabular}
    \caption{Table of typical (molecular) graph features used in literature~\cite{Zhaoping2020AttentiveFP, Pocha2020AtomReprs}. They can be further combined with geometric features~\cite{chenGraphNetworksUniversal2019}.
    }
    \label{tab:mol_features}
\end{table}

The sole chemical graph and its connectivity are often not sufficient to accurately predict quantum-mechanical or electronic-structure properties~\cite{ramakrishnan2014qm9} that strongly depend on the exact molecular geometry, even though ground-state or equilibrium geometries can in principle be inferred from the molecular graph alone.
In tasks that intrinsically involve geometric dependencies, e.g.\ predicting the potential energy surface of molecules and materials, it becomes obvious that geometric information is required.
The representation of positional and geometric information to learn quantum properties has been explored among others in the work of Lilienfeld et al.~\cite{von_lilienfeld_exploring_2020} and Behler et al.~\cite{Behler2011NP} and lead to a large variety of descriptors.
Some examples of descriptors are atomic centered symmetry functions (ACSF)~\cite{PhysRevB.90.024101, Behler2016MLPot}, angular Fourier series (AFS)~\cite{PhysRevB.87.184115},
the smooth overlap of atomic orbitals (SOAP)~\cite{PhysRevB.87.184115}, partial radial distribution function (PRDF)~\cite{PhysRevB.89.205118}, many-body tensor (MBTR)~\cite{huo2018unified}, Spectral London Axilrod-Teller-Muto (SLATM)~\cite{huang2020quantum} and the Faber–Christensen–Huang–Lilienfeld (FCHL)~\cite{Christensen2020FCHL} representation.
Many of those descriptors expand geometric information into symmetry or basis functions.
The resulting vector representation is typically used as input for conventional machine learning models such as neural networks or Gaussian Processes.
Geometric information can also be used for node or edge representations in graph neural networks.
Graph networks have been adopting distances~\cite{chenGraphNetworksUniversal2019}, bond~\cite{klicpera2020dimenetplusplus} and even dihedral angles~\cite{klicpera2021gemnet,flamshepherd2021paths}, motivated by the comparison to force fields~\cite{Pukrittayakamee2009FFfitNN}.
Angles or distances are similarly expanded into Gaussian-like~\cite{schutt_schnet_2018}, radial~\cite{Unke2019physnet} and spherical Fourier-Bessel functions~\cite{Klicpera2020DimeNet}.
Although architectures such as the Behler-Parinello (BP) neural network potentials~\cite{PhysRevLett.98.146401} or SchNet~\cite{schutt_schnet_2018} are not strictly graph networks in terms of the chemical graph, and often do not refer to themselves as such, they can be summarized within the term geometric deep learning~\cite{Bronstein2017GeoDeep, atz_geometric_2021}. 

Under the term geometric deep learning, architectures and descriptors are summarized that focus on manifolds, geometric data or structured data~\cite{Cao2020GeoDeppReview, Monti_2017_CVPR}.
This includes the work on 3D point clouds~\cite{Nguyen2013PointCloud}, which aims at learning segmentation and object detection of a large number of 3D points.
In the case of PointNet\texttt{++}~\cite{qi_pointnet_2017} a graph is constructed which reduces the point set from learned descriptors using the points' features.
Commonly, adjacency matrices are defined by using distance cutoffs between points in 3D clouds, while edges carry explicit information about distances between nodes, i.e. points.
Graph pooling or coarsening algorithms~\cite{Lee2019SAGPool, Zhang2019HierPool} that reduce the input representation and condense structure information are also promising for GNNs to tackle larger molecules such as proteins or polymers.

Eventually, the representation of materials for graph networks can be structural or geometric but must follow certain symmetry considerations~\cite{Li20213DmolNet, Montavon2012InvariantRepr}.
For example molecules without external fields have rotational and translation symmetries.
If they are incorporated into the model and its representation, less data is required and overall performance can be improved.
This concept can be extended to equivariant representations~\cite{Satorras2021Equiv, Nigam2022EquivRepr}, which are equivariant under translation, rotation, and permutation operations, and which can enable the prediction of tensorial properties~\cite{schutt2021painn}.

For solid crystals and periodic structures, the periodicity and space group symmetries are additional symmetries to be added to the representation for GNNs. Periodic extensions of the crystal graph~\cite{PhysRevLett.120.145301, PhysRevMaterials.4.093801} of the unit cell have been introduced~\cite{chenGraphNetworksUniversal2019} and their representation builds on the idea of a k-point mesh of Monkhorst-Pack grid points to sample the Brillouin zone~\cite{chengGeoCGNN2021}.

    \subsection{State-of-the-art architectures and benchmarks}\label{subsec:datasets-and-state-of-the-art-architectures}
    
Different architectures have been proposed in the literature to improve the performance of GNNs on typical tasks arising in chemistry and materials sciences.
Table~\ref{tab:benchmark_datasets} shows a list of popular benchmark datasets for different materials classes, i.e.\ molecules or crystals, and respective supervised tasks, i.e.\ regression or classification.
While some datasets contain experimental data, the largest datasets typically use computational methods to generate labels.
Most datasets in this table can be downloaded from data collections such as TUDatasets~\cite{Morris2020TUDatasets} and MoleculeNet~\cite{Zhenqin2017Moleculenet}.
While a large variety of datasets and tasks exist for chemistry, there are only a few large datasets for materials, limited to crystalline structures. Recent datasets were constructed by filtering the Materials Project (MP)~\cite{jainCommentaryMaterialsProject2013} and Open Quantum Materials Database (OQMD)~\cite{kirklin_open_2015} for specific targets such as electronic band-gap or formation energy while removing incomplete data~\cite{chengGeoCGNN2021}.

\begin{table}[h]
    \centering
    \caption{Table of common benchmark datasets for graph learning tasks.
    Note that this list is not complete and merely serves as an overview of different sizes and supervised learning tasks, which is either regression (R) or classification (C). 
    }
    \begin{tabular}{ccccc}
    \toprule
    Molecules  & Size & Tasks &   Type & Description \\
    \midrule
    QM7 \cite{blum2009qm7} &7,165 & 1 &  R & DFT quantum calculations \\
    QM7b \cite{montavon2013qm7b}   & 7,211 & 13 &  R & DFT quantum calculations\\
    QM9 \cite{ramakrishnan2014qm9} &   133,885 & 12 &  R & DFT quantum calculations \\
    PDBBind \cite{wang2005pdbbind} & 23,496 & 1 &  R & protein binding affinity\\
    MD17 \cite{chiemla2017md17-main-paper,chmiela2018md17-second-paper} & $>$ 100,000 & $\ge$1 &  R &  molecular dynamics trajectories  \\
    FreeSolv \cite{mobley_freesolv_2014}  & 643 & 1 &  R & solvation free energy\\
    Lipop \cite{Zhenqin2017Moleculenet}  & 4,200 & 1 &  R & lipophilicity \\
    Tox21 \cite{Zhenqin2017Moleculenet}  & 8,014 & 12 & C & qualitative toxicity measurement\\
    ToxCast \cite{Richard2016ToxCAST} & 8,615 & 617 &  C & qualitative toxicity measurement\\
    BBBP \cite{Martins2012BBBP} & 2,053 &  1 &  C & blood-brain barrier penetration \\
    HIV \cite{Zhenqin2017Moleculenet} & 41,913 & 1 &  C & inhibition to virus HIV\\
    SIDER \cite{Kuhn2015SIDER, Altae2017SIDER} & 1,427 & 27 &  C & adverse drug reaction  \\
    \midrule
    Crystals  & Size & Tasks &  Type & Description \\
    \midrule
    MP \cite{jainCommentaryMaterialsProject2013}  &  $\sim$ 144,595 & $\ge$1 & R,C & Materials Project (MP)\\
    OQMD \cite{kirklin_open_2015} & $\sim$ 1,022,603 & $\ge$1 & R,C & Open Quantum Materials Database  \\
    OC20 \cite{Chanussot2021OCACS} & $\sim$ 133,934,018 & $\ge$1 & R & Open Catalyst Project \\
	\bottomrule
    \end{tabular}
    \label{tab:benchmark_datasets}
\end{table}

In Section~\ref{subsec:basic-principles}, the message passing framework for GNNs has been illustrated.
Here, we will discuss modified and extended GNN models, which are relevant for materials science and chemistry.
However, listing all graph network architectures would be beyond the scope of this review.

Some of the earliest work on neural networks for molecular graphs dates back to the 90s and 2000s, without explicitly referring to the term graph neural network~\cite{PhysRevLett.98.146401, Merkwirth2005Chem}.
In 2017, a graph convolutional network was proposed by Kipf et al.~\cite{kipf-welling-gnn} for semi-supervised learning, which can be interpreted as a first-order approximation of spectral graph convolutions~\cite{Bruna2014, Defferrard2016}.
The addition of more complicated node aggregation functions such as gated recurrent units~\cite{cho2014properties} or long short-term memories~\cite{Hochreiter1997LSTM} has been employed by GraphSAGE for inductive learning~\cite{Hamilton2017GraphSAGE}.
For graph embedding tasks, a state- or super-node~\cite{message-passing-qc}, which is connected to all nodes, extends the message passing framework to help extract global graph information, in addition to the final node aggregation step.
A message passing neural network (MPNN) with edge features capturing bond information was applied to molecular graphs~\cite{message-passing-qc} and crystal graphs~\cite{PhysRevLett.120.145301}.
A variant of the original MPNN involves directed edge embeddings and message passing between edges in D-MPNN~\cite{Yang2019DMPNN}.
Known from models in natural language processing~\cite{vaswani2017attention}, masked self-attention layers which attend over the node's neighborhood have been suggested for graph attention networks~\cite{velickovic2018GAT} and used explicitly for molecules in Attentive Fingerprint models~\cite{Zhaoping2020AttentiveFP}.

Besides graph models which focus on the chemical graph, there is a large group of models explicitly designed for learning quantum properties.
They commonly take atomic numbers and positions as input and train on data derived from (approximate) solutions of the steady-state Schrödinger equation.
A popular benchmark dataset is the QM9 dataset~\cite{ramakrishnan2014qm9} with 13 quantum properties of small molecules with up to nine atoms apart from hydrogen.
The improvement of graph networks on QM9 property prediction over the past few years is highlighted in Figure~\ref{fig:figure1}c.
Among the first graph networks that reached chemical accuracy on QM9 is SchNet~\cite{schutt_schnet_2018}, which makes use of convolutional filters for inter-atomic distances and applies skip connections between node updates.
One improvement to SchNet was to update the positional features along the graph edges as seen in~\cite{jorgensen2018schnetedge}.
The application of GNNs to crystals using geometric information has been explored by MEGNET~\cite{chenGraphNetworksUniversal2019}, which further leverages global properties such as temperature, which is of importance for solid-state crystalline systems.
The potential energy of molecules depends on bond angles and therefore, in DimeNet~\cite{Klicpera2020DimeNet,klicpera2020dimenetplusplus}, edge embedding uses messages passing steps from atomic triplets and bond pairs in order to incorporate angular features.
This formalism has been adopted in other recent GNNs~\cite{Choudhary2021alignn, zhang2020mxmnet} and can be further extended to include dihedral (or torsion) angles~\cite{hsu2021alignn_d,klicpera2021gemnet, ganea2021geomol}.

For explicit angle plus node information in directed edge updates as in DimeNet~\cite{Klicpera2020DimeNet,klicpera2020dimenetplusplus}, the message passing essentially operates on higher order paths~\cite{flamshepherd2021paths} or $k$-pairs of atoms~\cite{morris2021weisfeiler}.
This is unfeasible for fully connected larger graphs because the number of multi-node interactions that need to be computed is dramatically increasing.
To reduce the computational costs, models like MXMNet~\cite{zhang2020mxmnet} make use of multiplex graphs, which selectively consider only specific edges when going to higher order pathways for calculating bond angles~\cite{Choudhary2021alignn}.

Note that the GNNs mentioned previously are invariant to the translation and rotation of the molecules throughout space. 
Recently, equivariant GNNs have been proposed~\cite{schutt2021painn, Satorras2021Equiv, finzi2020lieconv, anderson2019cormorant}, which transform equivariant under symmetry operations of its (positional) input, meaning the GNN's features or its output undergoes the same operations as well.
This enables efficient consideration of angular information between atoms~\cite{schutt2021painn} without higher-order pathways and enables the prediction of tensor features of general rank~\cite{qiao2021unite, anderson2019cormorant}.

Further adapted message passing steps allow for the determination of the molecular orbitals~\cite{Schutt2019schnorb,Qiao.2020,unke2021se3phisnet, PhysRevMaterials.4.093801}.
Molecular orbital interactions can in turn be used for improving the prediction performance~\cite{PhysRevMaterials.4.093801}.
Lastly, the mapping of atoms to non-Euclidean space such as in the proposed hyperbolic GNNs~\cite{liu2019hyperbolic} can lead to gains in representational efficiency. For a more in-depth discussion of graph variants~\cite{wieder_compact_2020} and graph taxonomy~\cite{ZHOU202057} that goes beyond Table~\ref{tab:model_architectures}, we refer to more general articles about GNNs~\cite{SunReview2019, zhang_graph_2019}, e.g Zhou et al.~\cite{ZHOU202057} and Wu et al.~\cite{ComprehensiveSurvey}.

With regard to the QM9 benchmark in Figure~\ref{fig:figure1}c some models have slightly lower performance for the total energy but can be superior in other QM9 properties or achieve similar results with much less computational effort. 
Some other factors that complicate a stringent comparison are differences in train-test splits, cleaning steps, e.g. of ill-converged molecules in QM9, multi-task vs. single task settings, where a separate model for each QM9 target is usually trained~\cite{chenGraphNetworksUniversal2019}, and differences in used loss metrics (a mean absolute error loss was found to yield lower overall test errors~\cite{Unke2019physnet} than the mean squared error loss used in previous models~\cite{schutt_schnet_2018}, although the mean absolute error is typically given as a benchmark reference).
It has to be noted that hyperparameters are generally very important and are often not exhaustively optimized for GNNs which can cause differences in performance apart from the model architecture~\cite{Yuan2021GNNhyper, jiang_could_2021, Banitalebi2021GCNNAllyouNeed}. 

\begin{table}[h]
    \centering
    \caption{Table of GNN models are sorted by categories. It is to note, that some models can also fall into more than one category and that this table can not list all relevant models but only give a grouping of a few popular models mentioned in the text. There is no strict distinction between categories spatial convolution, message passing, and 3D geometric message passing.
    }
    \begin{tabular}{c|c}
    \toprule
    Categories  & GNN architectures \\
    \midrule
    Spectral convolution & LanczosNet~\cite{liao2018lanczosnet}, SpecConv~\cite{SpecConv1, SpecConv2}, CayleyNet~\cite{CayleyNets}, ChebNet~\cite{Defferrard2016}\\
    \midrule
    Spatial convolution and & GCN~\cite{kipf-welling-gnn}, 123-GNN or k-GNN~\cite{morris2021weisfeiler}, R-GCN~\cite{Schlichtkrull2018}, GIN~\cite{gin}\\
    & PatchySan~\cite{patchysan}, C-SGEL~\cite{SpatialMolWang2019}, GraphSAGE~\cite{Hamilton2017GraphSAGE}, OGCNN~\cite{PhysRevMaterials.4.093801}\\
    & CGCNN and iCGCNN~\cite{xieCrystalGraphConvolutional2018, parkDevelopingImprovedCrystal2020}\\
    \arrayrulecolor{black!15}\midrule
    Message passing & MPNN~\cite{message-passing-qc}, D-MPNN~\cite{Yang2019DMPNN}, MPSN~\cite{bodnar2021weisfeiler}, MGN~\cite{Merkwirth2005Chem}\\
    & G-MPNN and MPNN-R~\cite{NEURIPS2020_217eedd1}, PMP~\cite{strathmann2021persistent}\\
    \midrule
    3D geometric message passing & MEGNET~\cite{chenGraphNetworksUniversal2019}, DimeNet~\cite{Klicpera2020DimeNet,klicpera2020dimenetplusplus}, PhysNet~\cite{Unke2019physnet}, MolNet~\cite{Li20213DmolNet, molNetKIM2022}\\
    & PointNet\texttt{++}~\cite{qi_pointnet_2017}, MXMNet~\cite{zhang2020mxmnet}, SchNet~\cite{schutt_schnet_2018, jorgensen2018schnetedge}, ForceNet~\cite{hu2021forcenet},\\
    & GemNet~\cite{klicpera2021gemnet}, Geomol~\cite{ganea2021geomol},  ALIGNN~\cite{Choudhary2021alignn} and ALIGNN-d~\cite{hsu2021alignn_d},\\
    &  GNNFF~\cite{parkAccurateScalableGraph2021}, GeoCGNN~\cite{chengGeoCGNN2021}, SphereNet~\cite{SphereNet2018}, HGCN~\cite{liu2019hyperbolic}\\
    \arrayrulecolor{black}\midrule
    Attention and graph transformer & GAT~\cite{velickovic2018GAT}, GATv2~\cite{GATv2_2021}, MAT~\cite{MAT2020_multi}, AGNN~\cite{2018attentionbased}, AMPNN~\cite{withnall_building_2020}\\
    & CapsGNN~\cite{xinyi2018capsule}, RGAT~\cite{RGAT2019}, AttentiveFP~\cite{Zhaoping2020AttentiveFP}, AGN~\cite{CHEN2022613}\\
    & GACNN~\cite{wang2020bees}, MEGAN~\cite{sacha2021molecule}, 
    SAMPN~\cite{tang_self-attention_2020}, HamNet~\cite{li2021conformationguided}\\
    \midrule
    Equivariant models & PAiNN~\cite{schutt2021painn}, NequIP~\cite{batzner20223}, TFN~\cite{TFEquiv2018}, CGNet~\cite{ClebschGordanNets},\\
    & Cormorant~\cite{anderson2019cormorant}, LieConv~\cite{finzi2020lieconv}, EGNN~\cite{Satorras2021Equiv}, UNiTE~\cite{qiao2021unite}\\
    & SEGNN~\cite{SEGNN2021}, SE(3)T~\cite{NEURIPS2020_15231a7c}, CNN-G~\cite{GCNN, GCNN2, Bekkers2020B-Spline}\\
    \midrule
    Graph pooling & DiffPool~\cite{DiffPool2018}, EdgePool~\cite{EdgePool2019}, gPool~\cite{pmlr-v97-gao19a} \\
    & HGP-SL\cite{Zhang2019HierPool}, SAGPool~\cite{Lee2019SAGPool}, iPool~\cite{Gao2021iPoolI}, EigenPool~\cite{EigenPooling2019}\\
    \midrule
    Generative graph models & CGVAE~\cite{constrained-gvae}, JT-VAE~\cite{jt-vae}, GCPN~\cite{gcpn}, GeoMol~\cite{ganea2021geomol}\\
    & GraphGAN~\cite{wang2017graphgan}, DCGAN~\cite{long2021constrained} \\
	\bottomrule
    \end{tabular}
    \label{tab:model_architectures}
\end{table}

    \subsection{GNNs in generative models and reinforcement learning}\label{subsec:gnns-in-generative-models-and-reinforcement-learning}
    
An important challenge in materials science is inverse materials design, aiming to generate new materials or molecules that possess required properties and fulfill specific criteria\cite{design-generative-models}.
GNN-based generative methods have been suggested to deal with this challenge in the context of chemistry, e.g.\ for drug discovery~\cite{MoFlow2020} and retrosynthesis~\cite{sacha2021molecule}.
In most cases, only the chemical structure of molecules is generated, i.e.\ the connectivity of the molecular graph, without additional information on specific 3D geometry.

Initial graph generative models were designed to generate graphs based on simplified theoretical assumptions, such as the random Erdös-Renyi (ER) model~\cite{gilbert1959random}, and improvements thereof using small-world approaches~\cite{watts_collective_1998} or the Kronecker graph model~\cite{leskovec2010}.
While these traditional approaches attempt to model real-world graphs, they are based on several assumptions about the nature of graphs generated and are thus inflexible for many data-related applications.
Machine learning approaches for graph generation are promising because they can directly learn to generate realistic graphs from the distribution of observed data while accommodating goal-directed tasks such as property optimization.
Examples include variational autoencoders (VAEs)~\cite{KingmaVAE}, generative adversarial networks (GANs)~\cite{NIPS2014_5ca3e9b1}, reinforcement learning~\cite{sutton2018reinforcement}, recurrent neural networks (RNNs)~\cite{You2018} and flow-based generative models~\cite{dinh2014nice,dinh2016density,kingma2018glow}.

Several architectures of VAEs have been developed to work with different types of input data, such as images~\cite{vae-images}, text-based data~\cite{kusner2017grammar,dai2018syntaxdirected}, or graphs~\cite{variational-graph-ae,simonovsky2018graphvae}.
Kipf et al. introduced a variational graph auto-encoder (VGAE) to learn latent representations of undirected graphs and applied it to a link prediction task in citation networks~\cite{variational-graph-ae}.
Liu et al. introduced a Constrained Graph Variational Autoencoder (CGVAE), in which node type information is learned from and added to the latent vectors~\cite{constrained-gvae}.
Starting from a set of these latent node representations, CGVAE iteratively forms valid molecules following hard valency constraints derived from the explicit node types.
Jin et al. introduced Junction Tree VAE (JT-VAE) to work directly on molecular graphs and achieved an improvement over baseline methods in molecular design tasks~\cite{jt-vae}.
The JT-VAE approach encodes and decodes molecules in two steps: First, tree-structured objects called junction trees are generated which represent trees of molecular subgraphs and their arrangements. GNN-based encoders and decoders are used to generate latent embeddings of the junction trees. In parallel, molecular graph embeddings are generated using GNNs, and junction trees are decoded into molecular representations~\cite{message-passing-qc}.
These are then encoded to a latent vector and decoded back to their original representations using graph and tree-based encoders and decoders.
While this scheme works well for molecules, it is hard to adapt for crystalline materials, where the graphs are less tree-like and the definition of scaffolds is not as straightforward.

GANs have shown promising results in a number of fields, such as image~\cite{denton2015deep} or sequence~\cite{yu2017seqgan} generation, and have also been applied to 3D grid representations of materials~\cite{long2021constrained} and graphs\cite{wang2017graphgan}.
De Cao et al. introduced MolGAN~\cite{molgan} as a framework for generating 
molecular graphs using GNNs.
The generator learns to directly output the graph's representation.
While the standard GAN loss forces the generator to generate molecules following
a particular prior distribution, the authors add a reinforcement learning (RL) 
objective to generate molecules with optimized properties.
The generation of invalid molecules is avoided by the assignment of zero reward to them.
While direct prediction of outputs is appealing in methods using VAEs or GANs, they usually 
predict outputs of small, fixed sizes.
Therefore, another branch of deep graph generative models employs sequential decision-making procedures 
to overcome these limitations.

You et al. identify three challenging aspects when graphs are generated directly \cite{You2018}.
First, as these methods need to model the adjacency matrix in some form, the output size
grows as $\mathcal{O}(n^2)$ for graphs with a maximum number of $n$ nodes.
This is especially undesirable for large and sparse graphs as the model dedicates much of
its capacity to learn which nodes are not connected.
Second, graph isomorphism complicates the calculation of reconstruction losses
for VAEs and usually involves expensive graph matching procedures.
Third, VAEs assume the outputs, e.g., the entries of an adjacency matrix, to be i.i.d.,
which is not true in practice.
As a solution, the authors propose GraphRNN, a framework in which graphs are represented as sequences of node and edge additions.
By using recurrent networks, graph constituents are generated conditioned on previous additions, thus taking into account the history of modifications.
Another sequential graph generation scheme was proposed by Li et al.~\cite{Li2018learningdeepgen}.
In this framework, the generation process is decomposed into modular steps, e.g., whether to
add a node or which nodes to connect.
Each module is a fully-connected neural network modeling probabilities of executing particular 
types of modifications.

\begin{figure}[!htbp]
	\centering
	\includegraphics[width=\textwidth]{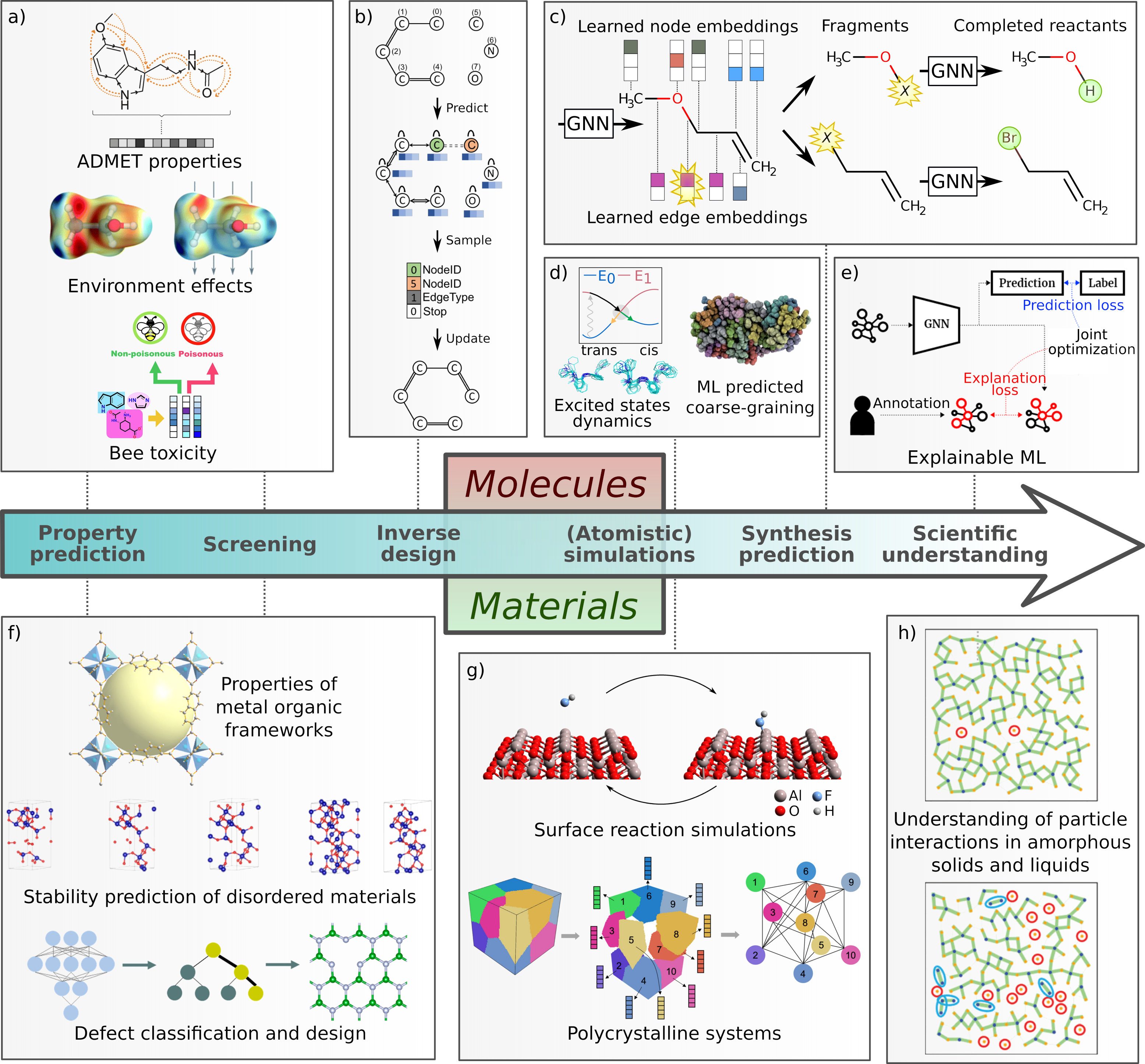}
	\caption{Overview of GNN applications for molecules and materials.
	a) prediction of ADMET properties (adapted with permission from Feiberg et al.\cite{Feinberg2020drug}, Copyright 2020 American Chemical Society), GNNs accounting for environment effects of molecules (reproduced from Ref.~\cite{gastegger2020fieldschnet} with permission from the Royal Society of Chemistry.), GNNs to predict the toxicity of molecules for bees (this illustration was published in Ref.~\cite{wang2020bees}, Copyright Elsevier),
	b) RL-based approach for inverse molecular design based on Graph Convolutional Policy Networks (GCPN) (adapted from Ref.~\cite{gcpn}),
	c) template-free retrosynthesis (adapted from Ref.~\cite{yan2020retroxpert}),
	d) transferable excited states dynamics (reproduced from Ref.~\cite{axelrod2021excited} with permission from Springer Nature), coarse graining (reproduced from Ref.~\cite{li2020coarse} with permission from the Royal Society of Chemistry),
	e) explainable GNNs (adapted from Ref.~\cite{gao2021gnes}),
	f) Crystal GNN to predict methane adsorption volumes in metal organic frameworks (MOFs) (this illustration was published in~\cite{wangCombiningCrystalGraphs2022}, Copyright Elsevier), doped structures (this illustration was published in~\cite{wangDeepTMCDeepLearning2022}, Copyright Elsevier), point defects (adapted with permission from Frey et al.\cite{freyMachineLearningEnabledDesign2020}, Copyright 2020 American Chemical Society)
	g) reactions of \ce{Al2O3} surface in contact with \ce{HF} gas (reproduced from Ref.~\cite{parkAccurateScalableGraph2021} with permission from Springer Nature), GNNs to predict magnetostriction of polycrystalline systems (reproduced from Ref.~\cite{daiGraphNeuralNetworks2021} with permission from Springer Nature),
	h) a GNN classifier to predict if a system is in a liquid or a glassy phase only by the positions of the atoms (reproduced from Ref.~\cite{swansonDeepLearningAutomated2020} with permission from the Royal Society of Chemistry).
	}
	\label{fig:applications}
\end{figure}

You et al. suggested a purely RL-based approach based on Graph Convolutional Policy Networks (GCPN)~\cite{gcpn} (see Figure~\ref{fig:applications}b).
In this setting, the agent explores the chemical space and generates new molecular graphs by modifying the
starting molecules according to the reward function, representing the molecular property to optimize. 
Atance et al. recently introduced a similar RL approach based on Gated GNNs\cite{atance2021},
outperforming other GNN-based approaches in molecular graph generation tasks\cite{Mercado2021}.
Another sequential approach based on conditional graph generative models has been used by Li et al. on drug design tasks\cite{Li2018MultiobjectiveDN}.
The previous two works inspired recent molecular graph generation frameworks such as  \textit{GraphINVENT}\cite{Mercado2021}.

More recently, attention has turned to generative modeling based on normalizing flows (NFs)\cite{papamakarios2021normalizing}, capable of modeling complex target distributions by directly mapping them to simpler ones using revertible bijective transformations.
NFs potentially offer significant improvements in the field of graph generation because they allow exact evaluation and inference of the probability distribution. 
This approach has been applied in a number of molecular tasks\cite{madhawa2019graphnvp,MoFlow2020,bengio2021flow,frey2022fastflows}.

Overall, graph generative models have been extensively applied for molecular materials and stayed up to date with recent developments in the field of graph generation.
However, these remain under-explored for crystalline materials, mainly due to graph representation challenges\cite{wangDeepGenerativeModel2022}.
While finding such a reliable graph representation is still an open question\cite{design-generative-models} and will likely remain case-specific, we believe that using generative models based on GNNs is a promising research direction in inverse design, especially given current breakthroughs such as normalizing flows.

    \section{Applications}\label{sec:applications}

After introducing the basic principles of GNNs as well as selected GNN architectures and benchmark datasets, we will provide a structured overview of GNN applications in chemistry and materials science.
GNNs were successfully applied to a rich variety of different challenges, ranging from property prediction of molecules and materials over accelerated atomistic simulations to predicting reactivity and synthesis routes of molecules and materials.
While other machine learning models such as densely connected neural networks were successfully applied to these tasks as well, state-of-the-art GNNs in many cases currently outperform other models.
However, there exists a range of open challenges, including data requirements and data efficiency, as well as a lack of fully GNN-based generative models for molecular and materials design.

\subsection{Molecular systems}\label{subsec:molecular-systems} 

Some of the first applications of GNNs and probably also one of the main driving forces for the ongoing development of GNN models are challenges in the area of molecular chemistry.
Most prevalent is the task of predicting molecular properties, i.e. a regression or classification task which is challenging to solve with conventional machine learning models, as they typically require predefined molecular representations (e.g.\ molecular feature vectors or molecular fingerprints) which are informative for the label to predict.
GNNs have access to the full chemical graph or even molecular geometry and learn to extract feature representations, which yields an advantage over other ML models. Compared to domain knowledge-informed feature representations combined with conventional ML models, e.g. Gaussian process regression, GNNs often have comparably high data requirements but outperform conventional models when enough data is available.
Once trained, accurate ML models can then be used to accelerate the high-throughput virtual screening of molecules~\cite{Pham2021,John2019HighThroughputPolymer} to find promising candidate molecules for many different applications.
However, property prediction is not the only application of GNNs. They were also successfully applied to provide trainable interatomic potentials to accelerate costly {\textit ab initio} molecular dynamics simulations, as well as to predict the outcome of chemical reactions and synthetic routes.

\subsubsection{Molecular property Prediction}
Among the most relevant molecular properties in the area of drug discovery are the ADMET (absorption, distribution, metabolism, exclusion, and toxicity) properties of potential drug-like molecules (see Figure~\ref{fig:applications}a)~\cite{Feinberg2020drug,Peng2020drug,Montanari2020drug,Yaowen2021drug}. A review on GNNs for drug design can be found in Xiong et al.~\cite{xiong2021graph}. In recent years, one application focus were Covid 19 related challenges, where GNNs were used for e.g. finding new drug candidates~\cite{cheung2020drugcovid} or detecting infections in medical images~\cite{YU2021drugcovid,KUMAR2022drugcovid}. Similar methods are also applicable to other challenges in drug design and medicine.

Furthermore, GNNs were applied to predict electronic and optical properties of molecules.
For many applications such as organic electronics, organic photovoltaics, and organic light-emitting diodes, the energy of the highest occupied molecular orbital (HOMO), the lowest unoccupied MO (LUMO) and the optical gap are of high importance for the device efficiency.
These properties can therefore be found in numerous databases~\cite{ramakrishnan2014qm9,nakata_pubchemqc_2017,montavon2013qm7b,lee2021opt,lu2020optoelec,pronobis2018,atz2021}.
Related properties include (transition) dipole moment~\cite{ramakrishnan2014qm9,nakata_pubchemqc_2020}, ionization potential and electron affinity~\cite{montavon2013qm7b}.
In devices, these properties often depend on the molecular environment, which can be modeled and accounted for with GNNs (see Figure~\ref{fig:applications}a)~\cite{gastegger2020fieldschnet}.
For certain applications such as opto-electronic materials~\cite{lee2021opt,lu2020optoelec}, GNNs have been proposed to complement scalar molecular properties with spectroscopic properties~\cite{hsu2021alignn_d}.

The aforementioned properties are often determined using computationally expensive simulation methods. Cheaper, mostly semi-empirical methods can provide fast but potentially less accurate estimates.
GNNs have been used to represent Kohn-Sham wavefunctions and energy levels in a minimal basis representation~\cite{gastegger2020basis}, as well as for delta-learning from semi-empirical methods to DFT computed molecular properties~\cite{atz2021}. For applications where not enough data is available to train GNNs, the representations learned by GNNs on large generic datasets can also be transferred to supervised tasks with little data~\cite{duvenaud2015convolutional,burkholz2021,li2021fingerprint}, where they are used as input for other ML models such as gradient-boosted models~\cite{deng2021xgraphboost}.

Further application areas of GNNs spread across all application domains of molecular materials, including the prediction of toxicity of molecules to bees (see Figure~\ref{fig:applications}a),\cite{wang2020bees} determining the quality of a material for fuel ignition~\cite{schweidtmann2020fuel} and the classification of different phases of materials, in particular water~\cite{kim2020gcicenet}. It should be noted that another review paper on molecular property prediction utilizing GNNs exists by Wieder et al.~\cite{wieder_compact_2020}.
In many application areas, tools such as GNNExplainer~\cite{ying2019gnnexplainer} are used to validate and analyze GNN predictions, e.g. in the prediction of scents\cite{sanchez2019machine} and for porous organic cages\cite{yuan2022explainable}.

\subsubsection{Dynamics simulations}\label{sec:mol_simulations}

Molecular dynamics simulations are an important tool for understanding dynamic processes and mechanisms on a microscopic level in various areas of chemistry, biology, and materials science.
Besides the prediction of equilibrium properties of molecules and materials, they also offer the possibility to simulate excited states and non-equilibrium dynamics, as well as slow processes and rare events.

In molecular dynamics simulations, total energy and forces are needed in every time step to propagate the system.
Computationally demanding ab initio methods that calculate the energy and forces of a particular atomic configuration of a system at every time step are therefore often too costly.
ML methods can replace ab initio calculations to speed up simulations while ideally retaining their accuracy~\cite{friederich2021machine}.
Therefore, long {\it and} highly accurate MD simulations can be performed based on ML potentials, which have not been possible using classical force fields nor ab initio methods.
GNNs are perfectly suited for this task, as atomic forces depend on the (local) atomic environment and global aggregation is not needed.

The concept of integrating ML models in atomistic simulations was demonstrated multiple times using for example SchNet~\cite{schutt_schnet_2018}, PhysNet~\cite{Unke2019physnet}, DimeNet~\cite{Klicpera2020DimeNet} or DimeNet$++$~\cite{klicpera2020dimenetplusplus}. However, there are several open challenges that need to be overcome in order to move to larger systems, longer time scales, higher data efficiency, better generalization and transferability, and eventually more accurate and realistic applications.
Usually, ML models learn the potential energy surface and calculate forces using derivatives of the energy predictions. This ensures that energy predictions and forces are consistent.
Since only forces are required in MD simulations, architectures are being developed in which these forces are predicted directly - so that the costly derivative calculations are omitted. In the GNN framework (GNNFF~\cite{parkAccurateScalableGraph2021}), a message passing step builds upon an embedding step in which node and edge features include atom type and interatomic distances respectively.
Force magnitudes per atom are then calculated from the sum of the forces of the neighboring atoms.
Evaluation on the ISO17 database reveals higher force accuracies compared to SchNet while being 1.6$\times$ faster.
The approach is also shown to be scalable to larger systems.
Due to the direct prediction of the forces, the energy of the system is however not necessarily preserved, making the model not suitable to predict energy-related properties.

ForceNet~\cite{hu2021forcenet} is based on an encoder-decoder architecture and tries to completely capture 3D geometry through a specific design of the message passing structure. In contrast to models such as SchNet and DimeNet, ForceNet encodes physical information without constraining the model architecture to enforce physical invariances.
Instead, physics-based data augmentation is performed on the data level to achieve rotational invariance.
The evaluation was performed on the OC20 dataset which contains DFT calculated energies and per-atom forces of more than 200 million large atomic structures (20-200 atoms) including non-equilibrium structures from optimization trajectories.
The resulting mean absolute force errors are comparable to DimeNet++, while being faster in training and prediction.

A promising approach to encoding more physical information about a system is the design of equivariant models.
Models that are based on equivariant message passing, e.g.\ PaiNN~\cite{schutt2021painn}, NequIP~\cite{batzner20223}, NewtonNet~\cite{haghighatlari2021newtonnet}, are shown to significantly increase data efficiency and predictive performance compared to models that are based on invariant convolutions.
The Neural Equivariant Interatomic Potential (NequIP~\cite{batzner20223}) predicts both energy and forces utilizing E(3)-equivariant convolutions over geometric tensors.
Evaluated on the MD17 data set its accuracy exceeds those of existing models while needing up to three orders of magnitude less training data.
Due to its data efficiency, it was also used with coupled cluster (CCSD(T)) based training data, showing great potential for applications where a prediction accuracy beyond DFT is needed.
In order to further improve data efficiency, GNNFF\cite{parkAccurateScalableGraph2021} and NewtonNet~\cite{haghighatlari2021newtonnet} introduce more physical priors in the form of latent force vectors as well as operators containing physical information.
This leads to good prediction accuracy with higher computational efficiency at only 1-10\% of the training data compared to other models.

To improve generalization, hybrid models such as SpookyNet~\cite{unke2021spookynet} explicitly include electronic structure information such as total charge or spin state, not included in most ML potentials, by applying self-attention in a transformer architecture.
Empirical augmentations to include non-local and long-range contributions such as electrostatic interaction and dispersion improve transferability and at the same time enable interpretability~\cite{unke2021spookynet}.

\textbf{GNN for large-scale MD simulations.}
Many applications require models that scale to large system sizes.
For example, simulations of biological systems, e.g. to study protein dynamics or drug binding mechanisms involve orders of magnitude more atoms than many other applications, while configuration changes occur on much longer timescales ($10^{-3}-10^3 \mathrm{s}$) than a typical MD timestep ($10^{-15} \mathrm{s}$). One way to address this enormous challenge is the development of models in a QM/MM-inspired approach, where only a small relevant subsystem, e.g.\ a reaction site, needs to be simulated at ab initio accuracy, while the rest of the system can be described using classical force fields.

GNNs also have the potential to support (adaptive) coarse-graining methods.
They were shown to be useful in mapping atoms of a molecule into coarse-grained groups needed for large-scale simulations.
The Deep Supervised Graph Partitioning Model (DSGPM)~\cite{li2020coarse} treats mapping operators as a graph segmentation problem.
It predicted a coarse-grained mapping nearly indistinguishable from human annotations (see Figure~\ref{fig:applications}d).
Furthermore, the ML framework by Wang et al.~\cite{Wang.2019}, which generates a coarse-grained force field, was further improved by Husic et al.~\cite{husic2020coarse} replacing manual input features with a GNN architecture making the models transferable across different molecular systems such as small proteins.

\textbf{Excited states dynamics.}
GNNs were also shown as a very promising tool to tackle the challenging task of simulating excited state dynamics of complex systems~\cite{westermayr_deep_2020}.
Unlike ground-state dynamics, multiple potential energy surfaces as well as their crossings and couplings must be considered, leading to a higher dimensionality and complexity of the problem. Furthermore, even the generation of reliable training data using quantum mechanical calculations is challenging. 

Westermayr et al. developed SchNarc~\cite{Westermayr.2020b} for photodynamics simulations by adapting SchNet for excited states potentials, forces, and couplings, combining it with the MD framework SHARC (surface hopping including arbitrary couplings). While SchNarc is molecule specific and was applied to two compounds, CH$_2$NH$_2^+$ and CSH$_2$, Axelrod et al. developed the first transferable excited state potential (see Figure~\ref{fig:applications}d)\cite{axelrod2021excited}. The diabatic artificial neural network (DANN) is based on PaiNN combined with a physics-informed diabatic model for photodynamic simulations for virtual screening.
The resulting ML potential is transferable among azobenzene derivatives and estimated to be multiple orders of magnitude faster than the underlying quantum mechanical calculation method, even considering computational effort for transfer which required additional training data.

\subsubsection{Reaction Prediction and Retrosynthesis}
While reliable property prediction and simulation methods are crucial for virtual molecular design, synthesis is often one of the main bottlenecks in the overall development process of new molecules. Progress in reaction prediction and retrosynthesis, i.e. the prediction of a reaction outcome and the design of synthetic routes for a desired product, can help to accelerate and also automate~\cite{tabor2018accelerating} the development of new molecules.
However, the two problems are still considered challenging due to the vast chemical space and currently require skills and experience from well-trained chemists.
Therefore, many machine learning algorithms, e.g. seq2seq models and transformers, have been proposed for synthesis prediction and retrosynthesis, aiming at reducing manual effort. In many cases, molecules are embedded as SMILES codes, and reaction predictions as well as retrosynthesis predictions are formulated as natural language processing tasks~\cite{pande2017sm,yang2020sm,lai2020sm,petraglia2020sm}.
Furthermore, fingerprints are widely used as structure encodings and neural networks trained on them are able to predict the most probable transformations for given molecules~\cite{waller2017fp,waller2018fp}.
GNN based graph embeddings have recently attracted growing attention, due to the natural representation of molecular structures as graphs.

Prediction of reactivity with GNNs has been formalized into reaction center identification and graph edit tasks.
Jin et al. developed a GNN-based approach to scoring each pair of connected atoms with bond change likelihood, followed by the selection of bonds with the highest score and transformation to potential products using a Weisfeiler-Lehman Difference Network (WLDN)~\cite{jin2017predicting, jin2019predicting}.
The WLN architecture is also adopted by Struble and coworkers~\cite{struble2020wln} to perform multitask prediction of aromatic C-H functionalization reactions with site selectivity, while Guan et al. predict reaction outcomes with regio-selectivity through combining molecule representations learned by the WLN with on-the-fly calculated quantum mechanical descriptors~\cite{guan2021wln}.
In 2020, Nikitin et al. introduced a strategy to treat reaction prediction as a node classification problem, where the role of each node in the reactant graphs is predicted by a GNN with a self-attention mechanism and pseudo-global nodes~\cite{nikitin2020wln}.

Furthermore, predicting reaction products as the result of a sequence of graph editing actions on the reactant molecules has also been investigated.
One example is the work by Do et al. where graph editing actions are predicted by a reinforcement learning algorithm based on reactant and reagent molecule representations generated by GNNs~\cite{do2018graph}.
In 2019, Bradshaw et al. developed a generative model that uses GNNs to predict a series of electron movements for reactions, through which the products and reaction mechanisms are predicted at the same time~\cite{bradshaw2019generative}.
Apart from product prediction, GNNs are also employed to predict important reaction properties including bond dissociation energies~\cite{wen2021bond,john2020bond}, transition states~\cite{pattanaik2020ts}, and activation energies~\cite{grambow2020ae}.

For the retrosynthesis task, recent studies can be divided into template-based and template-free approaches.
The former matches the graph embedding of the product molecule to a large number of reaction templates, which determine bond changes and thus predict possible reactants, while the latter bypass the templates and directly modifies input graphs to generate synthetic precursors.
Examples of template-based retrosynthesis prediction include the work by Dai and coworkers, who predict the probability distribution of reaction templates to be fitted to the reaction outcomes by a Conditional Graph Logic Network (GLN)~\cite{dai2020retrosynthesis}.
The reactants are generated by the GLN as the result of a probability prediction given the reaction outcome and the selected template.
Another example by Ishida et al. uses GCNs to classify single retrosynthesis steps into reaction templates, where integrated gradients are applied to visualize atom contributions toward the GCN prediction~\cite{ishida2019retro}.
More recently, Chen and coworkers proposed a framework based on MPNNs with a global attention mechanism to predict the reaction center of a target molecule, as well as the corresponding “local” reaction template based on atoms and bonds in the reaction center, which is then applied to generate reactants~\cite{chen2021retro}.

For template-free retrosynthesis, Yan et al. use an edge-enhanced graph attention network to locate the reaction center in product molecules, which are transformed through bond dissociation into molecular fragments called synthons~\cite{yan2020retroxpert}.
The synthon graphs are converted to SMILES and expanded to reactants by a sequence-to-sequence algorithm.
At the same time, Somnath et al. proposed an approach that uses GNNs to predict a series of graph edits that transform a product into synthons that are further completed into reactants with leaving groups predicted by another GNN (see Figure~\ref{fig:applications}d)~\cite{somnath2021learning}.
A similar strategy is adopted by Shi et al., where synthons are expanded to reactants by a variational graph translation~\cite{shi2021graph}.

In 2021, Sacha et al. proposed a model that formulates both retrosynthesis and forward synthesis tasks as a sequence of graph edit actions that are predicted by an encoder-decoder structure constructed by stacking graph convolutional layers~\cite{sacha2021molecule}.

\subsection{Crystalline and solid state systems}\label{subsec:crystalline-systems}

Compared to molecules, crystal structures and solid-state materials have some additional challenges, such as periodic boundary conditions for crystals and multiple kinds of disorder, either in form of perturbations in the crystal structure itself or in the (lack of) long-range ordering of atoms.
We will present different recent applications of GNNs in solid state systems, from predicting global properties of crystal structures with and without disorder, over driving atomistic simulations, to the design of new materials aided by materials synthesis prediction, active learning, and inverse design.
Table~\ref{tab:3.2:datasets} gives an overview of datasets used in the following applications.
We will also discuss approaches in which the trained GNN models have been analyzed to enable further insight into specific scientific questions, which is often equally important as accurate numerical predictions.

\subsubsection{Materials property prediction and materials design}
GNNs can be used to predict a multitude of materials properties, ranging from formation energies~\cite{bartelCriticalExaminationCompound2020,SchmidtCrystalAttention2021,parkDevelopingImprovedCrystal2020,jorgensenMaterialsPropertyPrediction2019,nohUncertaintyQuantifiedHybridMachine2020,pandeyPredictingEnergyStability2021} and synthesizability prediction~\cite{jangStructureBasedSynthesizabilityPrediction2020} over band-gaps~\cite{liGraphNetworkBased2021,omprakashGraphRepresentationalLearning2021,naTuplewiseMaterialRepresentation2020} and other functional properties to mechanical properties~\cite{xieCrystalGraphConvolutional2018, chenGraphNetworksUniversal2019}.
The most straightforward application of property prediction models is the screening of large databases of crystals, where exhaustive screening using conventional simulation techniques (e.g. DFT) is often non-feasible.
Screening using GNNs only requires labels from simulation or experiment for the training set while providing fast predictions on the rest of the database - provided the model generalizes well.

Wang et al. use a crystal GNN to predict methane adsorption volumes in metal-organic frameworks (MOFs)~\cite{wangCombiningCrystalGraphs2022}.
The pooling function leverages domain knowledge by additionally including structural properties (see Figure~\ref{fig:applications}f), e.g. the pore limiting diameter, achieving better performance than previous work~\cite{wangAcceleratingDiscoveryMetalOrganic2020}.
They apply the model to the screening of a hypothetical MOF database by Wilmer et al.~\cite{wilmerLargescaleScreeningHypothetical2012} and find several high-performing candidates~\cite{wangCombiningCrystalGraphs2022}.
Gu et al. use an ensemble of attention crystal GNNs to screen alloy catalysts for \ce{CO2} reduction~\cite{guPracticalDeepLearningRepresentation2020}.
Only bulk-relaxed structures without the adsorbate (e.g. from the Materials Project (MP)~\cite{jainCommentaryMaterialsProject2013} database) are needed as input, removing the costly DFT relaxation from the screening process.
The performance approaches that of base-line models trained on fully relaxed structures~\cite{guPracticalDeepLearningRepresentation2020}.

An interesting application of GNNs was presented by Goodall et al., who use GNNs for representation learning of chemical formulas~\cite{goodallPredictingMaterialsProperties2020}. The chemical formula is represented as a dense weighted graph, in which each node corresponds to a chemical element weighted by the fraction of this element in the chemical formula. This graph representation was used to train a GNN in a supervised way to obtain a mapping from the chemical formula to an embedding.
It was demonstrated that this representation has a better sample efficiency than other structure agnostic approaches.

Schmidt et al. use a crystal graph attention network for stability-screening of non-relaxed hypothetical materials, predicting the convex hull distance~\cite{SchmidtCrystalAttention2021}.
Only graph distances are included, making the model usable for hypothetical materials where the exact coordinates might be unknown.
A vast database is used, combining MP, AFLOW~\cite{curtaroloAFLOWAutomaticFramework2012} and group-internal datapoints.
Transfer learning then allows the screening of 15 million tetragonal perovskites of composition \ce{ABCD2}.
Dai et al. use GNNs to predict magnetostriction of polycrystalline systems (see Figure~\ref{fig:applications}g)~\cite{daiGraphNeuralNetworks2021}.
Instead of using GNNs to model the atoms in the unit cell, individual grains and their interactions to neighboring grains are represented by nodes and edges in the GNN. This shows that GNNs can be used on different scales to predict materials properties.

Typically, screening of materials databases using ML models is only possible if training data is available which covers the target materials distribution, i.e. which adequately allows generalization to the rest of the database.
Active learning offers a promising solution when the available data does not fulfill this criterion.
Based on uncertainty quantification, the training dataset can then be iteratively extended to include previously uncertain data points and thereby efficiently explore the chemical space.
Lu et al. use active learning to search for 2D ferromagnetic materials with a custom crystal graph multilayer descriptor in a small dataset~\cite{luCouplingCrystalGraph2020}.
Further applications of active learning based GNNs in materials science are promising and can be expected in the future.

\subsubsection{Disordered systems and defects}
Disordered materials are a wide and particularly interesting field of study, as the effects of disorder can influence or even dominate material properties.
Disorder ranges from weak disorder (defects, dislocations, grain boundaries) to strong disorder (e.g. glassy systems and inhomogeneous systems such as porous materials) and includes topological/structural disorder, orientational disorder (spins, dipoles), or substitutional disorder (chemical doping, compositional disorder)~\cite{catlow2012defects}.
Due to its inherent multi-scale nature, disorder poses severe challenges not only to materials modeling and simulation, but also to materials synthesis, characterization, and fundamental description.
Graph neural networks can be extended to model various forms of disordered systems and predict their local and global properties. In contrast to quantum mechanical methods such as DFT, crystalline systems with substitutional disorder can be modeled using larger, more representative unit cells or by defining atoms/nodes with mixed occupations. Furthermore, amorphous systems can be modeled explicitly by using representative simulation boxes and aggregating over learned atom representations characteristic of the properties of the amorphous system.
In the following, we will present seminal work in that direction.

One important challenge for machine learning models is how to represent substitutional disorder such as doping and fractional occupancies in the crystal graph. Chen et al. addressed this question with a MEGNet model with trainable 16-dimensional embeddings for each atom which were pre-trained on a dataset with only ordered structures~\cite{chenLearningPropertiesOrdered2021}.
The pre-trained embeddings were used to represent doped crystal sites by doing a weighted average of the elemental embeddings, weighted with (logarithmic or appropriately scaled) occupancies of the elements on the crystal sites.
Additionally, it was also demonstrated how to perform multi-fidelity training with band gaps of different levels of DFT by encoding the DFT level in the global state of the MEGNet.
Similarly, Wang et al. have trained a crystal GNN model on predicting the phase stability of materials (see Figure~\ref{fig:applications}f), classifying materials as metal or insulator and predicting the band gap of semiconductors~\cite{wangDeepTMCDeepLearning2022}.
They train a crystal GNN on undoped crystal structures and use this trained model to predict the properties of doped crystal structures, which they validate with DFT calculations.
They find that the crystal GNN and the DFT usually predict the same trend, despite the crystal GNN not being trained on doped crystal structures~\cite{wangDeepTMCDeepLearning2022}.
Frey et al. use a crystal GNN model to predicted the properties of 2D materials with point defects (see Figure~\ref{fig:applications}f)~\cite{freyMachineLearningEnabledDesign2020}.
However, the crystal GNN is only used to screen the properties of ordered structures to find promising host structures.
The properties of the disordered structures were then partially calculated with DFT to train a random forest model on physics-based features which was used to screen additional structures.
Another study predicted the properties of bcc iron structures with point defects~\cite{cianAtomisticGraphNeural2021}.

A very special application of GNNs is glassy systems.
There have been works that predicted the properties of glasses~\cite{bapstUnveilingPredictivePower2020} and which used inverse design to find glasses with new properties~\cite{wangInverseDesignGlass2021}.
Swanson et al. trained a classifier on predicting if a system is in a liquid or a glassy phase only by the positions of the atoms (see Figure~\ref{fig:applications}h)~\cite{swansonDeepLearningAutomated2020}.
They verified that a GNN was significantly better than a CNN for this task and used self-attention to interpret the reasoning of the trained classifier.
Consequently, the trained GNN was interpreted to find previously unknown relationships. Three simple and previously unknown formulas were developed, which describe the reasoning of the classifier and which could be used to differentiate the two phases.
Initial work on predicting polymer properties using GNNs was based on learning representations of single monomers~\cite{park2022prediction} or of unit cells of crystalline polymers~\cite{zeng2018graph}.

\subsubsection{Aiding simulations}
Analogous to atomistic dynamics simulations of molecular systems (see Section~\ref{sec:mol_simulations}, GNNs can also be used to simulate the dynamic behavior of crystals, i.e. to predict potential energy, forces, and partial charges, in order to drive molecular dynamics simulations.
To predict forces for MD of crystals, most approaches use conventional ML methods~\cite{deringerGeneralpurposeMachinelearningForce2020}, but there are first examples that use GNNs~\cite{parkAccurateScalableGraph2021, wangSymmetryadaptedGraphNeural2021}.
Raza et al. use a GNN with node-level readouts to predict partial charges of MOFs for molecular simulations of gas adsorption~\cite{razaMessagePassingNeural2020}.
For each atom, a probability distribution of charge is learned and optimized globally using the maximum likelihood principle under a charge neutrality constraint.
Park et al. predict forces for MD directly using a GNN with edge-level readouts that are used to predict the magnitude of the force between atoms~\cite{parkAccurateScalableGraph2021} (see also Section~\ref{subsec:molecular-systems}).
This avoids calculating the derivative of the potential energy and speeds up the simulation, yielding good accuracy in chemical reactions occurring at the surface of a test system of \ce{Al2O3} in contact with \ce{HF} gas (see Figure~\ref{fig:applications}g).
Also, good scaling accuracy transferring a model trained on a 1x2x1 supercell to a 1x2x2 supercell (\ce{Li7P3S11}) is achieved.
This approach allows simulations of a larger scale than possible using ab initio methods with similar accuracy and might make it possible to simulate mesoscopic phenomena such as grain boundaries in the future.

\subsubsection{Solid state synthesis prediction}
While predicting synthesizability overall~\cite{jangStructureBasedSynthesizabilityPrediction2020} is important in the design process of new materials, predicting the products of solid state synthesis is a challenging task and interesting application area of GNNs. Malik et al. have developed a pipeline to predict the major outcome of solid-state reactions~\cite{malikPredictingOutcomesMaterial2021}.
In this pipeline, GNNs were used to represent the precursors of a reaction in a set-like fashion as a dense graph while a long short-term memory layer is used to represent the series of processing actions in the reaction.
Current limitations in the availability of systematic (experimental) synthesis data, including reaction conditions~\cite{luo2021mof}, hinder further progress in this area. The use of electronic lab notebooks and repositories specifically designed for chemistry and materials science has a large potential to alleviate that challenge~\cite{tremouilhac2017chemotion,brandt2021kadi4mat}.

\subsubsection{Periodic graph generation}
A major requirement for the graph-based inverse design of crystal structures is the possibility to generate new periodic graphs based on a vector representation. Recently, this problem has been addressed with a new architecture called PGD-VAE~\cite{wangDeepGenerativeModel2022}, a variational autoencoder capable of generating new periodic graph structures. Another work using a VAE focuses on predicting stable crystal structures using GNNs~\cite{xieCrystalDiffusionVariational2021}. New structures are generated in a three-step approach: First, the composition, lattice, and number of atoms of a sampled point in the latent space are decoded using an MLP. Afterward, a random structure with these properties is assembled and a GNN trained on moving atoms to their equilibrium positions is used to generate stable structures. Despite these promising efforts, the design of solids state materials and crystal structures using GNNs is only at the beginning. Multiple challenges need to be solved to achieve wide applicability and transferability of methods to multiple classes of materials, ranging from densely packed crystals over porous materials to amorphous systems need to be solved.

\newcolumntype{b}{X}
\newcolumntype{s}{>{\hsize=.15\hsize}X}
\newcolumntype{m}{>{\hsize=.7\hsize}X}
\begin{table}[h]
	\centering
	\caption{Selected application-specific datasets for solid-state systems.}
    \def\arraystretch{1.4}
	\begin{tabularx}{\textwidth}{ bs }
		\toprule
		Dataset & Size \\
		\midrule
		
	    AFLOW~\cite{curtaroloAFLOWAutomaticFramework2012} - calculated properties of materials & $>$ 3,400,000 \\
	    
		Inorganic Crystal Structure Database (ICSD)\cite{bergerhoffInorganicCrystalStructure1983} - extensive and well curated experimental database & $\approx$ 210,000 \\
		
		Pure carbon and C-H-N-O structures at different pressures\cite{pickardAIRSSDataCarbon2020} & $\approx$ 200,000 \\
		
	    Materials Project~\cite{jainCommentaryMaterialsProject2013} - calculated properties of materials & $\approx$ 145,000 \\
	    
		Hypothetical MOF database~\cite{wilmerLargescaleScreeningHypothetical2012} & 137,953 \\
		
		NREL Materials Database (NRELMatDB)\cite{NRELMaterialsDatabase} - computational materials database focused on renewable energy applications & $\approx$ 60,000  \\
		
	    CO and H surface binding energy dataset~\cite{tranActiveLearningIntermetallics2018} & $\approx$ 40,000 \\
	    
		Inorganic materials synthesis recipes\cite{kononovaTextminedDatasetInorganic2019} & 19,488 \\
		
		Perovskite structures and energies\cite{castelliNewCubicPerovskites2012} & 18,928 \\

		CoRE MOF database~\cite{chungAdvancesUpdatesAnalytics2019} - Experimental MOF database & $>$ 14,000 \\
		
		bcc iron structures with energies and various kinds of defects\cite{dragoniAchievingDFTAccuracy2018} & 12,193 \\

	    MOF methane adsorption volume of CoRE MOFs (from GCMC)~\cite{wangCombiningCrystalGraphs2022} & 10,102 \\
	    
		Elemental boron structures with energies\cite{deringerDataDrivenLearningTotal2018} & 5,038 \\
		
		Computational 2D Materials Database (C2DB)\cite{haastrupComputational2DMaterials2018,rasmussenComputational2DMaterials2015} & $\approx$ 4000 \\

		DDEC MOF point charges~\cite{nazarianComprehensiveSetHighQuality2016} & 2,932 \\
		\bottomrule
	\end{tabularx}
	\label{tab:3.2:datasets}
\end{table}

    \section{Outlook}\label{sec:outlook}
    
GNNs became a very versatile and important tool very quickly. A lot has been achieved already, not only in terms of fundamental method development tailor-made for requirements of materials science and chemistry (see Section~\ref{sec:graph-neural-networks-in-materials-science-and-chemistry}), but also in terms of applications (see Section~\ref{sec:applications}), where GNNs were successfully applied for materials simulation, screening, design, and synthesis.
However, there is a wide range of open questions and challenges which need to be solved in order to leverage the full potential of GNNs in materials science and chemistry. 

Despite the growing amount of research and publications on GNN model development and application, GNN models remain expert tools, i.e. they are comparably hard to implement, adapt, transfer, train, and apply to given datasets and applications.
Libraries such as PyTorch Geometric~\cite{fey2019fast}, DGL~\cite{DGLLib} or the Keras based KGCNN~\cite{reiser2021graph} implement a selection of state-of-the-art GNN layers and models.
However, the use of such libraries in many cases requires expert knowledge which goes beyond the knowledge needed for the successful application of more established machine learning models.
One of the reasons for this is certainly related to the fast development of new GNN variants which are partially hard to compare.
The widespread use and further development of common benchmarks (e.g. the QM9 dataset) as well as open communication of models and training workflows, including open-source code, inspired by common practice in the machine learning community, are essential for a reliable quantitative comparison of future developments.
However, non-optimal hyperparameters and the (prohibitively) high computational cost of hyperparameter optimization is an open challenge.
A further challenge hindering the transfer of state-of-the-art models to applications is the discrepancy in the data distribution between widely used benchmark datasets (e.g. QM9) and actually relevant datasets, which typically contain larger molecules (i.e. larger graphs), which are more diverse, e.g. in terms of the chemical elements used and in the size variance, and which sample the chemical or materials space less densely then QM9.\cite{gasteiger2022graph}

Generally, there is more research and model development activity for molecules and chemistry, compared to materials science. As a consequence, the transfer of existing GNNs to new application areas (e.g. porous or amorphous materials) might be challenging and requires further development of GNN models.
To some extent, this can be attributed to a lack of generic benchmark datasets for crystalline materials and, even more importantly, (partially) disordered structures and amorphous materials.
OC20 is one of the few examples of datasets covering both materials science and chemistry~\cite{Chanussot2021OCACS}.
Nonetheless, there are many promising application areas of GNNs for solid-state materials.
In contrast to quantum mechanical simulation methods such as DFT, GNNs are particularly suited for representing disordered systems and predicting their properties, e.g. systems with compositional disorder such as high-entropy alloys.
However, there is a lack of large datasets of disordered systems, particularly labeled datasets with consistent measured or simulated properties.
Overall, screening of (hypothetical) materials spaces with predictive ML models allows the rapid discovery of new materials with optimized target properties, especially when combined with transfer and active learning to improve data efficiency and reduce computational cost.

The first steps towards GNN-based generative models exist, but there are many open challenges of reliability and transferability.
Despite the high potential of GNN-based algorithms for the inverse design of new molecules and materials, convincing success stories of ML-designed materials are rare.
To make generative models more application relevant, new methods are required that e.g. allow to include constraints in the design process, in the simplest case symmetries of generated molecules and materials, or in more complex scenarios additional (empirical or analytical) objectives such as synthesizability. A large step in that direction is new representations, not only for organic molecules but also for (3D) materials~\cite{krenn2022selfies,brammer2022tucan,yao2021inverse}.

Finally, more research on the explainability and interpretability of GNNs and machine learning in general will help to better understand underlying correlations and eventually causal relations in large and complex datasets, eventually contributing to scientific understanding and progress~\cite{friederich2021scientific,krenn2022scientific,lavin2021simulation}.

    \printnomenclature

\bibliographystyle{naturemag} 

\clearpage
\bibliography{ms}

\begin{thebibliography}{100}
\expandafter\ifx\csname url\endcsname\relax
  \def\url#1{\texttt{#1}}\fi
\expandafter\ifx\csname urlprefix\endcsname\relax\def\urlprefix{URL }\fi
\providecommand{\bibinfo}[2]{#2}
\providecommand{\eprint}[2][]{\url{#2}}

\bibitem{von_Lilienfeld_2020}
\bibinfo{author}{von Lilienfeld, O.~A.}
\newblock \bibinfo{title}{Introducing machine learning: Science and
  technology}.
\newblock \emph{\bibinfo{journal}{Machine Learning: Science and Technology}}
  \textbf{\bibinfo{volume}{1}}, \bibinfo{pages}{010201} (\bibinfo{year}{2020}).
\newblock \urlprefix\url{https://doi.org/10.1088/2632-2153/ab6d5d}.

\bibitem{PhysRevMaterials.4.093801}
\bibinfo{author}{Karamad, M.} \emph{et~al.}
\newblock \bibinfo{title}{Orbital graph convolutional neural network for
  material property prediction}.
\newblock \emph{\bibinfo{journal}{Phys. Rev. Materials}}
  \textbf{\bibinfo{volume}{4}}, \bibinfo{pages}{093801} (\bibinfo{year}{2020}).
\newblock
  \urlprefix\url{https://link.aps.org/doi/10.1103/PhysRevMaterials.4.093801}.

\bibitem{vamathevan_applications_2019}
\bibinfo{author}{Vamathevan, J.} \emph{et~al.}
\newblock \bibinfo{title}{Applications of machine learning in drug discovery
  and development}.
\newblock \emph{\bibinfo{journal}{Nature reviews Drug discovery}}
  \textbf{\bibinfo{volume}{18}}, \bibinfo{pages}{463--477}
  (\bibinfo{year}{2019}).

\bibitem{Jorissen2005VirtDatabase}
\bibinfo{author}{Jorissen, R.~N.} \& \bibinfo{author}{Gilson, M.~K.}
\newblock \bibinfo{title}{Virtual screening of molecular databases using a
  support vector machine}.
\newblock \emph{\bibinfo{journal}{Journal of Chemical Information and
  Modeling}} \textbf{\bibinfo{volume}{45}}, \bibinfo{pages}{549--561}
  (\bibinfo{year}{2005}).
\newblock \urlprefix\url{https://doi.org/10.1021/ci049641u}.
\newblock \bibinfo{note}{PMID: 15921445},
  \eprint{https://doi.org/10.1021/ci049641u}.

\bibitem{ekins_exploiting_2019}
\bibinfo{author}{Ekins, S.} \emph{et~al.}
\newblock \bibinfo{title}{Exploiting machine learning for end-to-end drug
  discovery and development}.
\newblock \emph{\bibinfo{journal}{Nature materials}}
  \textbf{\bibinfo{volume}{18}}, \bibinfo{pages}{435--441}
  (\bibinfo{year}{2019}).

\bibitem{Juhwan2020InvDesign}
\bibinfo{author}{Noh, J.}, \bibinfo{author}{Gu, G.~H.}, \bibinfo{author}{Kim,
  S.} \& \bibinfo{author}{Jung, Y.}
\newblock \bibinfo{title}{Machine-enabled inverse design of inorganic solid
  materials: promises and challenges}.
\newblock \emph{\bibinfo{journal}{Chem. Sci.}} \textbf{\bibinfo{volume}{11}},
  \bibinfo{pages}{4871--4881} (\bibinfo{year}{2020}).
\newblock \urlprefix\url{http://dx.doi.org/10.1039/D0SC00594K}.

\bibitem{zunger_inverse_2018}
\bibinfo{author}{Zunger, A.}
\newblock \bibinfo{title}{Inverse design in search of materials with target
  functionalities}.
\newblock \emph{\bibinfo{journal}{Nature Reviews Chemistry}}
  \textbf{\bibinfo{volume}{2}}, \bibinfo{pages}{1--16} (\bibinfo{year}{2018}).

\bibitem{PhysRevLett.98.146401}
\bibinfo{author}{Behler, J.} \& \bibinfo{author}{Parrinello, M.}
\newblock \bibinfo{title}{Generalized neural-network representation of
  high-dimensional potential-energy surfaces}.
\newblock \emph{\bibinfo{journal}{Phys. Rev. Lett.}}
  \textbf{\bibinfo{volume}{98}}, \bibinfo{pages}{146401}
  (\bibinfo{year}{2007}).
\newblock
  \urlprefix\url{https://link.aps.org/doi/10.1103/PhysRevLett.98.146401}.

\bibitem{friederich2021machine}
\bibinfo{author}{Friederich, P.}, \bibinfo{author}{H{\"a}se, F.},
  \bibinfo{author}{Proppe, J.} \& \bibinfo{author}{Aspuru-Guzik, A.}
\newblock \bibinfo{title}{Machine-learned potentials for next-generation matter
  simulations}.
\newblock \emph{\bibinfo{journal}{Nature Materials}}
  \textbf{\bibinfo{volume}{20}}, \bibinfo{pages}{750--761}
  (\bibinfo{year}{2021}).

\bibitem{shields2021bayesian}
\bibinfo{author}{Shields, B.~J.} \emph{et~al.}
\newblock \bibinfo{title}{Bayesian reaction optimization as a tool for chemical
  synthesis}.
\newblock \emph{\bibinfo{journal}{Nature}} \textbf{\bibinfo{volume}{590}},
  \bibinfo{pages}{89--96} (\bibinfo{year}{2021}).

\bibitem{luo2021mof}
\bibinfo{author}{Luo, Y.} \emph{et~al.}
\newblock \bibinfo{title}{Mof synthesis prediction enabled by automatic data
  mining and machine learning}.
\newblock \emph{\bibinfo{journal}{Angewandte Chemie International Edition}}
  \bibinfo{pages}{e202200242} (\bibinfo{year}{2021}).

\bibitem{kalinin2022machine}
\bibinfo{author}{Kalinin, S.~V.} \emph{et~al.}
\newblock \bibinfo{title}{Machine learning in scanning transmission electron
  microscopy}.
\newblock \emph{\bibinfo{journal}{Nature Reviews Methods Primers}}
  \textbf{\bibinfo{volume}{2}}, \bibinfo{pages}{1--28} (\bibinfo{year}{2022}).

\bibitem{velasco2021phase}
\bibinfo{author}{Velasco, L.} \emph{et~al.}
\newblock \bibinfo{title}{Phase--property diagrams for multicomponent oxide
  systems toward materials libraries}.
\newblock \emph{\bibinfo{journal}{Advanced Materials}}
  \textbf{\bibinfo{volume}{33}}, \bibinfo{pages}{2102301}
  (\bibinfo{year}{2021}).

\bibitem{HASE2019282}
\bibinfo{author}{Häse, F.}, \bibinfo{author}{Roch, L.~M.} \&
  \bibinfo{author}{Aspuru-Guzik, A.}
\newblock \bibinfo{title}{Next-generation experimentation with self-driving
  laboratories}.
\newblock \emph{\bibinfo{journal}{Trends in Chemistry}}
  \textbf{\bibinfo{volume}{1}}, \bibinfo{pages}{282--291}
  (\bibinfo{year}{2019}).
\newblock
  \urlprefix\url{https://www.sciencedirect.com/science/article/pii/S258959741930019X}.

\bibitem{LeCun2010IEEE}
\bibinfo{author}{LeCun, Y.}, \bibinfo{author}{Kavukcuoglu, K.} \&
  \bibinfo{author}{Farabet, C.}
\newblock \bibinfo{title}{Convolutional networks and applications in vision}.
\newblock In \emph{\bibinfo{booktitle}{Proceedings of 2010 IEEE International
  Symposium on Circuits and Systems}}, \bibinfo{pages}{253--256}
  (\bibinfo{year}{2010}).

\bibitem{schwaller2019molecular}
\bibinfo{author}{Schwaller, P.} \emph{et~al.}
\newblock \bibinfo{title}{Molecular transformer: a model for
  uncertainty-calibrated chemical reaction prediction}.
\newblock \emph{\bibinfo{journal}{ACS central science}}
  \textbf{\bibinfo{volume}{5}}, \bibinfo{pages}{1572--1583}
  (\bibinfo{year}{2019}).

\bibitem{Sperduti1997}
\bibinfo{author}{Sperduti, A.} \& \bibinfo{author}{Starita, A.}
\newblock \bibinfo{title}{Supervised neural networks for the classification of
  structures}.
\newblock \emph{\bibinfo{journal}{IEEE Transactions on Neural Networks}}
  \textbf{\bibinfo{volume}{8}}, \bibinfo{pages}{714--735}
  (\bibinfo{year}{1997}).

\bibitem{Gori2005}
\bibinfo{author}{Gori, M.}, \bibinfo{author}{Monfardini, G.} \&
  \bibinfo{author}{Scarselli, F.}
\newblock \bibinfo{title}{A new model for learning in graph domains}.
\newblock In \emph{\bibinfo{booktitle}{Proceedings. 2005 IEEE International
  Joint Conference on Neural Networks, 2005.}}, vol.~\bibinfo{volume}{2},
  \bibinfo{pages}{729--734 vol. 2} (\bibinfo{year}{2005}).

\bibitem{scarselli-gnn}
\bibinfo{author}{Scarselli, F.}, \bibinfo{author}{Gori, M.},
  \bibinfo{author}{Tsoi, A.~C.}, \bibinfo{author}{Hagenbuchner, M.} \&
  \bibinfo{author}{Monfardini, G.}
\newblock \bibinfo{title}{The graph neural network model}.
\newblock \emph{\bibinfo{journal}{IEEE Transactions on Neural Networks}}
  \textbf{\bibinfo{volume}{20}}, \bibinfo{pages}{61--80}
  (\bibinfo{year}{2009}).

\bibitem{duvenaud2015convolutional}
\bibinfo{author}{Duvenaud, D.} \emph{et~al.}
\newblock \bibinfo{title}{Convolutional networks on graphs for learning
  molecular fingerprints}.
\newblock In \bibinfo{editor}{Cortes, C.}, \bibinfo{editor}{Lawrence, N.~D.},
  \bibinfo{editor}{Lee, D.~D.}, \bibinfo{editor}{Sugiyama, M.} \&
  \bibinfo{editor}{Garnett, R.} (eds.) \emph{\bibinfo{booktitle}{Advances in
  Neural Information Processing Systems 28: Annual Conference on Neural
  Information Processing Systems 2015, December 7-12, 2015, Montreal, Quebec,
  Canada}}, \bibinfo{pages}{2224--2232} (\bibinfo{year}{2015}).
\newblock
  \urlprefix\url{https://proceedings.neurips.cc/paper/2015/hash/f9be311e65d81a9ad8150a60844bb94c-Abstract.html}.

\bibitem{message-passing-qc}
\bibinfo{author}{Gilmer, J.}, \bibinfo{author}{Schoenholz, S.~S.},
  \bibinfo{author}{Riley, P.~F.}, \bibinfo{author}{Vinyals, O.} \&
  \bibinfo{author}{Dahl, G.~E.}
\newblock \bibinfo{title}{Neural message passing for quantum chemistry}.
\newblock In \emph{\bibinfo{booktitle}{Proceedings of the 34th International
  Conference on Machine Learning - Volume 70}}, ICML'17,
  \bibinfo{pages}{1263–1272} (\bibinfo{publisher}{JMLR.org},
  \bibinfo{year}{2017}).

\bibitem{schutt_schnet_2018}
\bibinfo{author}{Schütt, K.~T.}, \bibinfo{author}{Sauceda, H.~E.},
  \bibinfo{author}{Kindermans, P.-J.}, \bibinfo{author}{Tkatchenko, A.} \&
  \bibinfo{author}{Müller, K.-R.}
\newblock \bibinfo{title}{{SchNet} – {A} deep learning architecture for
  molecules and materials}.
\newblock \emph{\bibinfo{journal}{The Journal of Chemical Physics}}
  \textbf{\bibinfo{volume}{148}}, \bibinfo{pages}{241722}
  (\bibinfo{year}{2018}).
\newblock \urlprefix\url{https://aip.scitation.org/doi/full/10.1063/1.5019779}.
\newblock \bibinfo{note}{Publisher: American Institute of Physics}.

\bibitem{Unke2019physnet}
\bibinfo{author}{Unke, O.~T.} \& \bibinfo{author}{Meuwly, M.}
\newblock \bibinfo{title}{Physnet: A neural network for predicting energies,
  forces, dipole moments, and partial charges}.
\newblock \emph{\bibinfo{journal}{Journal of Chemical Theory and Computation}}
  \textbf{\bibinfo{volume}{15}}, \bibinfo{pages}{3678–3693}
  (\bibinfo{year}{2019}).
\newblock \urlprefix\url{http://dx.doi.org/10.1021/acs.jctc.9b00181}.

\bibitem{SchmidtCrystalAttention2021}
\bibinfo{author}{Schmidt, J.}, \bibinfo{author}{Pettersson, L.},
  \bibinfo{author}{Verdozzi, C.}, \bibinfo{author}{Botti, S.} \&
  \bibinfo{author}{Marques, M. A.~L.}
\newblock \bibinfo{title}{Crystal graph attention networks for the prediction
  of stable materials}.
\newblock \emph{\bibinfo{journal}{Science Advances}}
  \textbf{\bibinfo{volume}{7}}, \bibinfo{pages}{eabi7948}
  (\bibinfo{year}{2021}).
\newblock
  \urlprefix\url{https://www.science.org/doi/abs/10.1126/sciadv.abi7948}.
\newblock \eprint{https://www.science.org/doi/pdf/10.1126/sciadv.abi7948}.

\bibitem{batzner20223}
\bibinfo{author}{Batzner, S.} \emph{et~al.}
\newblock \bibinfo{title}{E(3)-equivariant graph neural networks for
  data-efficient and accurate interatomic potentials}.
\newblock \emph{\bibinfo{journal}{Nature communications}}
  \textbf{\bibinfo{volume}{13}}, \bibinfo{pages}{1--11} (\bibinfo{year}{2022}).

\bibitem{klicpera2021gemnet}
\bibinfo{author}{Klicpera, J.}, \bibinfo{author}{Becker, F.} \&
  \bibinfo{author}{G{\"u}nnemann, S.}
\newblock \bibinfo{title}{Gemnet: Universal directional graph neural networks
  for molecules}.
\newblock In \bibinfo{editor}{Beygelzimer, A.}, \bibinfo{editor}{Dauphin, Y.},
  \bibinfo{editor}{Liang, P.} \& \bibinfo{editor}{Vaughan, J.~W.} (eds.)
  \emph{\bibinfo{booktitle}{Advances in Neural Information Processing Systems}}
  (\bibinfo{year}{2021}).
\newblock \urlprefix\url{https://openreview.net/forum?id=HS_sOaxS9K-}.

\bibitem{Klicpera2020DimeNet}
\bibinfo{author}{Klicpera, J.}, \bibinfo{author}{Gro{\ss}, J.} \&
  \bibinfo{author}{G{\"{u}}nnemann, S.}
\newblock \bibinfo{title}{Directional message passing for molecular graphs}.
\newblock In \emph{\bibinfo{booktitle}{8th International Conference on Learning
  Representations, {ICLR} 2020, Addis Ababa, Ethiopia, April 26-30, 2020}}
  (\bibinfo{publisher}{OpenReview.net}, \bibinfo{year}{2020}).
\newblock \urlprefix\url{https://openreview.net/forum?id=B1eWbxStPH}.

\bibitem{schutt2021painn}
\bibinfo{author}{Sch{\"u}tt, K.}, \bibinfo{author}{Unke, O.} \&
  \bibinfo{author}{Gastegger, M.}
\newblock \bibinfo{title}{Equivariant message passing for the prediction of
  tensorial properties and molecular spectra}.
\newblock In \emph{\bibinfo{booktitle}{International Conference on Machine
  Learning}}, \bibinfo{pages}{9377--9388} (\bibinfo{organization}{PMLR},
  \bibinfo{year}{2021}).

\bibitem{Cayley1874Isomer}
\bibinfo{author}{F.R.S., P.~C.}
\newblock \bibinfo{title}{Lvii. on the mathematical theory of isomers}.
\newblock \emph{\bibinfo{journal}{The London, Edinburgh, and Dublin
  Philosophical Magazine and Journal of Science}}
  \textbf{\bibinfo{volume}{47}}, \bibinfo{pages}{444--447}
  (\bibinfo{year}{1874}).
\newblock \urlprefix\url{https://doi.org/10.1080/14786447408641058}.
\newblock \eprint{https://doi.org/10.1080/14786447408641058}.

\bibitem{Bonchev1991ChemicalGraph}
\bibinfo{author}{Bonchev, D.} \& \bibinfo{author}{Rouvray, D.~H.}
\newblock \emph{\bibinfo{title}{Chemical Graph Theory: Introduction and
  Fundamentals}} (\bibinfo{publisher}{Routledge}, \bibinfo{address}{London},
  \bibinfo{year}{1991}).
\newblock \urlprefix\url{https://doi.org/10.1201/9781315139104}.

\bibitem{Biggs1986GraphTheory}
\bibinfo{author}{Biggs, N.}, \bibinfo{author}{Lloyd, E.~K.} \&
  \bibinfo{author}{Wilson, R.~J.}
\newblock \emph{\bibinfo{title}{Graph Theory, 1736-1936}}
  (\bibinfo{publisher}{Clarendon Press}, \bibinfo{address}{USA},
  \bibinfo{year}{1986}).

\bibitem{Merkwirth2005Chem}
\bibinfo{author}{Merkwirth, C.} \& \bibinfo{author}{Lengauer, T.}
\newblock \bibinfo{title}{Automatic generation of complementary descriptors
  with molecular graph networks}.
\newblock \emph{\bibinfo{journal}{Journal of Chemical Information and
  Modeling}} \textbf{\bibinfo{volume}{45}}, \bibinfo{pages}{1159--1168}
  (\bibinfo{year}{2005}).
\newblock \urlprefix\url{https://doi.org/10.1021/ci049613b}.
\newblock \bibinfo{note}{PMID: 16180893},
  \eprint{https://doi.org/10.1021/ci049613b}.

\bibitem{kipf-welling-gnn}
\bibinfo{author}{Kipf, T.~N.} \& \bibinfo{author}{Welling, M.}
\newblock \bibinfo{title}{Semi-supervised classification with graph
  convolutional networks}.
\newblock In \emph{\bibinfo{booktitle}{5th International Conference on Learning
  Representations, {ICLR} 2017, Toulon, France, April 24-26, 2017, Conference
  Track Proceedings}} (\bibinfo{publisher}{OpenReview.net},
  \bibinfo{year}{2017}).
\newblock \urlprefix\url{https://openreview.net/forum?id=SJU4ayYgl}.

\bibitem{Defferrard2016}
\bibinfo{author}{Defferrard, M.}, \bibinfo{author}{Bresson, X.} \&
  \bibinfo{author}{Vandergheynst, P.}
\newblock \bibinfo{title}{Convolutional neural networks on graphs with fast
  localized spectral filtering}.
\newblock In \bibinfo{editor}{Lee, D.}, \bibinfo{editor}{Sugiyama, M.},
  \bibinfo{editor}{Luxburg, U.}, \bibinfo{editor}{Guyon, I.} \&
  \bibinfo{editor}{Garnett, R.} (eds.) \emph{\bibinfo{booktitle}{Advances in
  Neural Information Processing Systems}}, vol.~\bibinfo{volume}{29}
  (\bibinfo{publisher}{Curran Associates, Inc.}, \bibinfo{year}{2016}).
\newblock
  \urlprefix\url{https://proceedings.neurips.cc/paper/2016/file/04df4d434d481c5bb723be1b6df1ee65-Paper.pdf}.

\bibitem{Bruna2014}
\bibinfo{author}{Estrach, J.~B.}, \bibinfo{author}{Zaremba, W.},
  \bibinfo{author}{Szlam, A.} \& \bibinfo{author}{LeCun, Y.}
\newblock \bibinfo{title}{Spectral networks and deep locally connected networks
  on graphs}.
\newblock In \emph{\bibinfo{booktitle}{2nd international conference on learning
  representations, ICLR}}, vol. \bibinfo{volume}{2014} (\bibinfo{year}{2014}).

\bibitem{over-squashing}
\bibinfo{author}{Alon, U.} \& \bibinfo{author}{Yahav, E.}
\newblock \bibinfo{title}{On the bottleneck of graph neural networks and its
  practical implications}.
\newblock In \emph{\bibinfo{booktitle}{9th International Conference on Learning
  Representations, {ICLR} 2021, Virtual Event, Austria, May 3-7, 2021}}
  (\bibinfo{publisher}{OpenReview.net}, \bibinfo{year}{2021}).
\newblock \urlprefix\url{https://openreview.net/forum?id=i80OPhOCVH2}.

\bibitem{Li_Han_Wu_2018}
\bibinfo{author}{Li, Q.}, \bibinfo{author}{Han, Z.} \& \bibinfo{author}{Wu, X.}
\newblock \bibinfo{title}{Deeper insights into graph convolutional networks for
  semi-supervised learning}.
\newblock In \bibinfo{editor}{McIlraith, S.~A.} \& \bibinfo{editor}{Weinberger,
  K.~Q.} (eds.) \emph{\bibinfo{booktitle}{Proceedings of the Thirty-Second
  {AAAI} Conference on Artificial Intelligence, (AAAI-18), the 30th innovative
  Applications of Artificial Intelligence (IAAI-18), and the 8th {AAAI}
  Symposium on Educational Advances in Artificial Intelligence (EAAI-18), New
  Orleans, Louisiana, USA, February 2-7, 2018}}, \bibinfo{pages}{3538--3545}
  (\bibinfo{publisher}{{AAAI} Press}, \bibinfo{year}{2018}).
\newblock
  \urlprefix\url{https://www.aaai.org/ocs/index.php/AAAI/AAAI18/paper/view/16098}.

\bibitem{set2set}
\bibinfo{author}{Vinyals, O.}, \bibinfo{author}{Bengio, S.} \&
  \bibinfo{author}{Kudlur, M.}
\newblock \bibinfo{title}{Order matters: Sequence to sequence for sets}.
\newblock In \bibinfo{editor}{Bengio, Y.} \& \bibinfo{editor}{LeCun, Y.} (eds.)
  \emph{\bibinfo{booktitle}{4th International Conference on Learning
  Representations, {ICLR} 2016, San Juan, Puerto Rico, May 2-4, 2016,
  Conference Track Proceedings}} (\bibinfo{year}{2016}).
\newblock \urlprefix\url{http://arxiv.org/abs/1511.06391}.

\bibitem{Yang2019DMPNN}
\bibinfo{author}{Yang, K.} \emph{et~al.}
\newblock \bibinfo{title}{Analyzing learned molecular representations for
  property prediction}.
\newblock \emph{\bibinfo{journal}{Journal of Chemical Information and
  Modeling}} \textbf{\bibinfo{volume}{59}}, \bibinfo{pages}{3370--3388}
  (\bibinfo{year}{2019}).
\newblock \urlprefix\url{https://doi.org/10.1021/acs.jcim.9b00237}.
\newblock \bibinfo{note}{PMID: 31361484},
  \eprint{https://doi.org/10.1021/acs.jcim.9b00237}.

\bibitem{Schutt2018schnetpack}
\bibinfo{author}{Sch\"{u}tt, K.~T.} \emph{et~al.}
\newblock \bibinfo{title}{{SchNetPack}: A deep learning toolbox for atomistic
  systems}.
\newblock \emph{\bibinfo{journal}{Journal of Chemical Theory and Computation}}
  \textbf{\bibinfo{volume}{15}}, \bibinfo{pages}{448--455}
  (\bibinfo{year}{2018}).
\newblock \urlprefix\url{https://doi.org/10.1021/acs.jctc.8b00908}.

\bibitem{shui2020heterogeneous}
\bibinfo{author}{Shui, Z.} \& \bibinfo{author}{Karypis, G.}
\newblock \bibinfo{title}{Heterogeneous molecular graph neural networks for
  predicting molecule properties}.
\newblock In \bibinfo{editor}{Plant, C.}, \bibinfo{editor}{Wang, H.},
  \bibinfo{editor}{Cuzzocrea, A.}, \bibinfo{editor}{Zaniolo, C.} \&
  \bibinfo{editor}{Wu, X.} (eds.) \emph{\bibinfo{booktitle}{20th {IEEE}
  International Conference on Data Mining, {ICDM} 2020, Sorrento, Italy,
  November 17-20, 2020}}, \bibinfo{pages}{492--500}
  (\bibinfo{publisher}{{IEEE}}, \bibinfo{year}{2020}).
\newblock \urlprefix\url{https://doi.org/10.1109/ICDM50108.2020.00058}.

\bibitem{finzi2020lieconv}
\bibinfo{author}{Finzi, M.}, \bibinfo{author}{Stanton, S.},
  \bibinfo{author}{Izmailov, P.} \& \bibinfo{author}{Wilson, A.~G.}
\newblock \bibinfo{title}{Generalizing convolutional neural networks for
  equivariance to lie groups on arbitrary continuous data}.
\newblock In \bibinfo{editor}{III, H.~D.} \& \bibinfo{editor}{Singh, A.} (eds.)
  \emph{\bibinfo{booktitle}{Proceedings of the 37th International Conference on
  Machine Learning}}, vol. \bibinfo{volume}{119} of
  \emph{\bibinfo{series}{Proceedings of Machine Learning Research}},
  \bibinfo{pages}{3165--3176} (\bibinfo{publisher}{PMLR},
  \bibinfo{year}{2020}).
\newblock \urlprefix\url{https://proceedings.mlr.press/v119/finzi20a.html}.

\bibitem{bodnar2021weisfeiler}
\bibinfo{author}{Bodnar, C.} \emph{et~al.}
\newblock \bibinfo{title}{Weisfeiler and lehman go topological: Message passing
  simplicial networks}.
\newblock In \emph{\bibinfo{booktitle}{ICLR 2021 Workshop on Geometrical and
  Topological Representation Learning}} (\bibinfo{year}{2021}).
\newblock \urlprefix\url{https://openreview.net/forum?id=RZgbB-O3w6Z}.

\bibitem{NEURIPS2019_bb04af0f}
\bibinfo{author}{Maron, H.}, \bibinfo{author}{Ben-Hamu, H.},
  \bibinfo{author}{Serviansky, H.} \& \bibinfo{author}{Lipman, Y.}
\newblock \bibinfo{title}{Provably powerful graph networks}.
\newblock In \bibinfo{editor}{Wallach, H.} \emph{et~al.} (eds.)
  \emph{\bibinfo{booktitle}{Advances in Neural Information Processing
  Systems}}, vol.~\bibinfo{volume}{32} (\bibinfo{publisher}{Curran Associates,
  Inc.}, \bibinfo{year}{2019}).
\newblock
  \urlprefix\url{https://proceedings.neurips.cc/paper/2019/file/bb04af0f7ecaee4aae62035497da1387-Paper.pdf}.

\bibitem{gin}
\bibinfo{author}{Xu, K.}, \bibinfo{author}{Hu, W.}, \bibinfo{author}{Leskovec,
  J.} \& \bibinfo{author}{Jegelka, S.}
\newblock \bibinfo{title}{How powerful are graph neural networks?}
\newblock In \emph{\bibinfo{booktitle}{7th International Conference on Learning
  Representations, {ICLR} 2019, New Orleans, LA, USA, May 6-9, 2019}}
  (\bibinfo{publisher}{OpenReview.net}, \bibinfo{year}{2019}).
\newblock \urlprefix\url{https://openreview.net/forum?id=ryGs6iA5Km}.

\bibitem{feng2019hypergraph}
\bibinfo{author}{Feng, Y.}, \bibinfo{author}{You, H.}, \bibinfo{author}{Zhang,
  Z.}, \bibinfo{author}{Ji, R.} \& \bibinfo{author}{Gao, Y.}
\newblock \bibinfo{title}{Hypergraph neural networks}.
\newblock \emph{\bibinfo{journal}{Proceedings of the AAAI Conference on
  Artificial Intelligence}} \textbf{\bibinfo{volume}{33}},
  \bibinfo{pages}{3558--3565} (\bibinfo{year}{2019}).
\newblock
  \urlprefix\url{https://ojs.aaai.org/index.php/AAAI/article/view/4235}.

\bibitem{JoEdge2021}
\bibinfo{author}{Jo, J.} \emph{et~al.}
\newblock \bibinfo{title}{Edge representation learning with hypergraphs}.
\newblock \emph{\bibinfo{journal}{Advances in Neural Information Processing
  Systems}} \textbf{\bibinfo{volume}{34}}, \bibinfo{pages}{7534--7546}
  (\bibinfo{year}{2021}).

\bibitem{dym2021on}
\bibinfo{author}{Dym, N.} \& \bibinfo{author}{Maron, H.}
\newblock \bibinfo{title}{On the universality of rotation equivariant point
  cloud networks}.
\newblock In \emph{\bibinfo{booktitle}{International Conference on Learning
  Representations}} (\bibinfo{year}{2021}).
\newblock \urlprefix\url{https://openreview.net/forum?id=6NFBvWlRXaG}.

\bibitem{morris2021weisfeiler}
\bibinfo{author}{Morris, C.} \emph{et~al.}
\newblock \bibinfo{title}{Weisfeiler and leman go neural: Higher-order graph
  neural networks}.
\newblock \emph{\bibinfo{journal}{Proceedings of the AAAI Conference on
  Artificial Intelligence}} \textbf{\bibinfo{volume}{33}},
  \bibinfo{pages}{4602--4609} (\bibinfo{year}{2019}).
\newblock
  \urlprefix\url{https://ojs.aaai.org/index.php/AAAI/article/view/4384}.

\bibitem{MAT2020_multi}
\bibinfo{author}{Maziarka, L.} \emph{et~al.}
\newblock \bibinfo{title}{Molecule attention transformer}.
\newblock \emph{\bibinfo{journal}{CoRR}}
  \textbf{\bibinfo{volume}{abs/2002.08264}} (\bibinfo{year}{2020}).
\newblock \urlprefix\url{https://arxiv.org/abs/2002.08264}.
\newblock \eprint{2002.08264}.

\bibitem{CHEN2022613}
\bibinfo{author}{Chen, Y.}, \bibinfo{author}{Tang, X.}, \bibinfo{author}{Qi,
  X.}, \bibinfo{author}{Li, C.-G.} \& \bibinfo{author}{Xiao, R.}
\newblock \bibinfo{title}{Learning graph normalization for graph neural
  networks}.
\newblock \emph{\bibinfo{journal}{Neurocomputing}}
  \textbf{\bibinfo{volume}{493}}, \bibinfo{pages}{613--625}
  (\bibinfo{year}{2022}).
\newblock
  \urlprefix\url{https://www.sciencedirect.com/science/article/pii/S0925231222000030}.

\bibitem{PhysRevLett.120.145301}
\bibinfo{author}{Xie, T.} \& \bibinfo{author}{Grossman, J.~C.}
\newblock \bibinfo{title}{Crystal graph convolutional neural networks for an
  accurate and interpretable prediction of material properties}.
\newblock \emph{\bibinfo{journal}{Phys. Rev. Lett.}}
  \textbf{\bibinfo{volume}{120}}, \bibinfo{pages}{145301}
  (\bibinfo{year}{2018}).
\newblock
  \urlprefix\url{https://link.aps.org/doi/10.1103/PhysRevLett.120.145301}.

\bibitem{Zhaoping2020AttentiveFP}
\bibinfo{author}{Xiong, Z.} \emph{et~al.}
\newblock \bibinfo{title}{Pushing the boundaries of molecular representation
  for drug discovery with the graph attention mechanism}.
\newblock \emph{\bibinfo{journal}{Journal of Medicinal Chemistry}}
  \textbf{\bibinfo{volume}{63}}, \bibinfo{pages}{8749--8760}
  (\bibinfo{year}{2020}).
\newblock \urlprefix\url{https://doi.org/10.1021/acs.jmedchem.9b00959}.
\newblock \bibinfo{note}{PMID: 31408336},
  \eprint{https://doi.org/10.1021/acs.jmedchem.9b00959}.

\bibitem{Kruger2020Scaffold}
\bibinfo{author}{Kruger, F.}, \bibinfo{author}{Stiefl, N.} \&
  \bibinfo{author}{Landrum, G.~A.}
\newblock \bibinfo{title}{rdscaffoldnetwork: The scaffold network
  implementation in rdkit}.
\newblock \emph{\bibinfo{journal}{Journal of Chemical Information and
  Modeling}} \textbf{\bibinfo{volume}{60}}, \bibinfo{pages}{3331--3335}
  (\bibinfo{year}{2020}).
\newblock \urlprefix\url{https://doi.org/10.1021/acs.jcim.0c00296}.
\newblock \bibinfo{note}{PMID: 32584031},
  \eprint{https://doi.org/10.1021/acs.jcim.0c00296}.

\bibitem{rdkit}
\bibinfo{title}{{RDK}it: Open-source cheminformatics}.
\newblock \bibinfo{howpublished}{\url{http://www.rdkit.org}}.
\newblock \bibinfo{note}{[Online; accessed 11-April-2013]}.

\bibitem{oboyle_open_2011}
\bibinfo{author}{O'Boyle, N.~M.} \emph{et~al.}
\newblock \bibinfo{title}{Open {Babel}: {An} open chemical toolbox}.
\newblock \emph{\bibinfo{journal}{Journal of Cheminformatics}}
  \textbf{\bibinfo{volume}{3}}, \bibinfo{pages}{33} (\bibinfo{year}{2011}).
\newblock \urlprefix\url{https://doi.org/10.1186/1758-2946-3-33}.

\bibitem{Chen2020DrugDeepWalk}
\bibinfo{author}{Chen, Z.-H.} \emph{et~al.}
\newblock \bibinfo{title}{Prediction of drug–target interactions from
  multi-molecular network based on deep walk embedding model}.
\newblock \emph{\bibinfo{journal}{Frontiers in Bioengineering and
  Biotechnology}} \textbf{\bibinfo{volume}{8}} (\bibinfo{year}{2020}).
\newblock
  \urlprefix\url{https://www.frontiersin.org/article/10.3389/fbioe.2020.00338}.

\bibitem{Perozzi2014DeepWalk}
\bibinfo{author}{Perozzi, B.}, \bibinfo{author}{Al-Rfou, R.} \&
  \bibinfo{author}{Skiena, S.}
\newblock \bibinfo{title}{Deepwalk: Online learning of social representations}.
\newblock In \emph{\bibinfo{booktitle}{Proceedings of the 20th ACM SIGKDD
  International Conference on Knowledge Discovery and Data Mining}}, KDD '14,
  \bibinfo{pages}{701–710} (\bibinfo{publisher}{Association for Computing
  Machinery}, \bibinfo{address}{New York, NY, USA}, \bibinfo{year}{2014}).
\newblock \urlprefix\url{https://doi.org/10.1145/2623330.2623732}.

\bibitem{Grover2016Node2vec}
\bibinfo{author}{Grover, A.} \& \bibinfo{author}{Leskovec, J.}
\newblock \bibinfo{title}{Node2vec: Scalable feature learning for networks}.
\newblock In \emph{\bibinfo{booktitle}{Proceedings of the 22nd ACM SIGKDD
  International Conference on Knowledge Discovery and Data Mining}}, KDD '16,
  \bibinfo{pages}{855–864} (\bibinfo{publisher}{Association for Computing
  Machinery}, \bibinfo{address}{New York, NY, USA}, \bibinfo{year}{2016}).
\newblock \urlprefix\url{https://doi.org/10.1145/2939672.2939754}.

\bibitem{chenGraphNetworksUniversal2019}
\bibinfo{author}{Chen, C.}, \bibinfo{author}{Ye, W.}, \bibinfo{author}{Zuo,
  Y.}, \bibinfo{author}{Zheng, C.} \& \bibinfo{author}{Ong, S.~P.}
\newblock \bibinfo{title}{Graph networks as a universal machine learning
  framework for molecules and crystals}.
\newblock \emph{\bibinfo{journal}{Chemistry of Materials}}
  \textbf{\bibinfo{volume}{31}}, \bibinfo{pages}{3564--3572}
  (\bibinfo{year}{2019}).

\bibitem{Pocha2020AtomReprs}
\bibinfo{author}{Pocha, A.}, \bibinfo{author}{Danel, T.},
  \bibinfo{author}{Podlewska, S.}, \bibinfo{author}{Tabor, J.} \&
  \bibinfo{author}{Maziarka, L.}
\newblock \bibinfo{title}{Comparison of atom representations in graph neural
  networks for molecular property prediction}.
\newblock In \emph{\bibinfo{booktitle}{International Joint Conference on Neural
  Networks, {IJCNN} 2021, Shenzhen, China, July 18-22, 2021}},
  \bibinfo{pages}{1--8} (\bibinfo{publisher}{{IEEE}}, \bibinfo{year}{2021}).
\newblock \urlprefix\url{https://doi.org/10.1109/IJCNN52387.2021.9533698}.

\bibitem{ramakrishnan2014qm9}
\bibinfo{author}{Ramakrishnan, R.}, \bibinfo{author}{Dral, P.~O.},
  \bibinfo{author}{Rupp, M.} \& \bibinfo{author}{von Lilienfeld, O.~A.}
\newblock \bibinfo{title}{Quantum chemistry structures and properties of 134
  kilo molecules}.
\newblock \emph{\bibinfo{journal}{Scientific Data}}
  \textbf{\bibinfo{volume}{1}}, \bibinfo{pages}{140022} (\bibinfo{year}{2014}).
\newblock \urlprefix\url{https://doi.org/10.1038/sdata.2014.22}.

\bibitem{von_lilienfeld_exploring_2020}
\bibinfo{author}{von Lilienfeld, O.~A.}, \bibinfo{author}{Müller, K.-R.} \&
  \bibinfo{author}{Tkatchenko, A.}
\newblock \bibinfo{title}{Exploring chemical compound space with quantum-based
  machine learning}.
\newblock \emph{\bibinfo{journal}{Nature Reviews Chemistry}}
  \textbf{\bibinfo{volume}{4}}, \bibinfo{pages}{347--358}
  (\bibinfo{year}{2020}).
\newblock \urlprefix\url{https://doi.org/10.1038/s41570-020-0189-9}.

\bibitem{Behler2011NP}
\bibinfo{author}{Behler, J.}
\newblock \bibinfo{title}{Atom-centered symmetry functions for constructing
  high-dimensional neural network potentials}.
\newblock \emph{\bibinfo{journal}{The Journal of Chemical Physics}}
  \textbf{\bibinfo{volume}{134}}, \bibinfo{pages}{074106}
  (\bibinfo{year}{2011}).
\newblock \urlprefix\url{https://doi.org/10.1063/1.3553717}.
\newblock \eprint{https://doi.org/10.1063/1.3553717}.

\bibitem{PhysRevB.90.024101}
\bibinfo{author}{Seko, A.}, \bibinfo{author}{Takahashi, A.} \&
  \bibinfo{author}{Tanaka, I.}
\newblock \bibinfo{title}{Sparse representation for a potential energy
  surface}.
\newblock \emph{\bibinfo{journal}{Phys. Rev. B}} \textbf{\bibinfo{volume}{90}},
  \bibinfo{pages}{024101} (\bibinfo{year}{2014}).
\newblock \urlprefix\url{https://link.aps.org/doi/10.1103/PhysRevB.90.024101}.

\bibitem{Behler2016MLPot}
\bibinfo{author}{Behler, J.}
\newblock \bibinfo{title}{Perspective: Machine learning potentials for
  atomistic simulations}.
\newblock \emph{\bibinfo{journal}{The Journal of Chemical Physics}}
  \textbf{\bibinfo{volume}{145}}, \bibinfo{pages}{170901}
  (\bibinfo{year}{2016}).
\newblock \urlprefix\url{https://doi.org/10.1063/1.4966192}.
\newblock \eprint{https://doi.org/10.1063/1.4966192}.

\bibitem{PhysRevB.87.184115}
\bibinfo{author}{Bart\'ok, A.~P.}, \bibinfo{author}{Kondor, R.} \&
  \bibinfo{author}{Cs\'anyi, G.}
\newblock \bibinfo{title}{On representing chemical environments}.
\newblock \emph{\bibinfo{journal}{Phys. Rev. B}} \textbf{\bibinfo{volume}{87}},
  \bibinfo{pages}{184115} (\bibinfo{year}{2013}).
\newblock \urlprefix\url{https://link.aps.org/doi/10.1103/PhysRevB.87.184115}.

\bibitem{PhysRevB.89.205118}
\bibinfo{author}{Sch\"utt, K.~T.} \emph{et~al.}
\newblock \bibinfo{title}{How to represent crystal structures for machine
  learning: Towards fast prediction of electronic properties}.
\newblock \emph{\bibinfo{journal}{Phys. Rev. B}} \textbf{\bibinfo{volume}{89}},
  \bibinfo{pages}{205118} (\bibinfo{year}{2014}).
\newblock \urlprefix\url{https://link.aps.org/doi/10.1103/PhysRevB.89.205118}.

\bibitem{huo2018unified}
\bibinfo{author}{Huo, H.} \& \bibinfo{author}{Rupp, M.}
\newblock \bibinfo{title}{Unified representation of molecules and crystals for
  machine learning} (\bibinfo{year}{2018}).
\newblock \eprint{1704.06439}.

\bibitem{huang2020quantum}
\bibinfo{author}{Huang, B.} \& \bibinfo{author}{von Lilienfeld, O.~A.}
\newblock \bibinfo{title}{Quantum machine learning using atom-in-molecule-based
  fragments selected on the fly}.
\newblock \emph{\bibinfo{journal}{Nature Chemistry}}
  \textbf{\bibinfo{volume}{12}}, \bibinfo{pages}{945--951}
  (\bibinfo{year}{2020}).

\bibitem{Christensen2020FCHL}
\bibinfo{author}{Christensen, A.~S.}, \bibinfo{author}{Bratholm, L.~A.},
  \bibinfo{author}{Faber, F.~A.} \& \bibinfo{author}{Anatole~von Lilienfeld,
  O.}
\newblock \bibinfo{title}{Fchl revisited: Faster and more accurate quantum
  machine learning}.
\newblock \emph{\bibinfo{journal}{The Journal of Chemical Physics}}
  \textbf{\bibinfo{volume}{152}}, \bibinfo{pages}{044107}
  (\bibinfo{year}{2020}).
\newblock \urlprefix\url{https://doi.org/10.1063/1.5126701}.
\newblock \eprint{https://doi.org/10.1063/1.5126701}.

\bibitem{klicpera2020dimenetplusplus}
\bibinfo{author}{Klicpera, J.}, \bibinfo{author}{Giri, S.},
  \bibinfo{author}{Margraf, J.~T.} \& \bibinfo{author}{Günnemann, S.}
\newblock \bibinfo{title}{Fast and uncertainty-aware directional message
  passing for non-equilibrium molecules} (\bibinfo{year}{2020}).
\newblock \eprint{2011.14115}.

\bibitem{flamshepherd2021paths}
\bibinfo{author}{Flam-Shepherd, D.}, \bibinfo{author}{Wu, T.~C.},
  \bibinfo{author}{Friederich, P.} \& \bibinfo{author}{Aspuru-Guzik, A.}
\newblock \bibinfo{title}{Neural message passing on high order paths}.
\newblock \emph{\bibinfo{journal}{Machine Learning: Science and Technology}}
  \textbf{\bibinfo{volume}{2}}, \bibinfo{pages}{045009} (\bibinfo{year}{2021}).
\newblock \urlprefix\url{https://doi.org/10.1088/2632-2153/abf5b8}.

\bibitem{Pukrittayakamee2009FFfitNN}
\bibinfo{author}{Pukrittayakamee, A.} \emph{et~al.}
\newblock \bibinfo{title}{Simultaneous fitting of a potential-energy surface
  and its corresponding force fields using feedforward neural networks}.
\newblock \emph{\bibinfo{journal}{The Journal of Chemical Physics}}
  \textbf{\bibinfo{volume}{130}}, \bibinfo{pages}{134101}
  (\bibinfo{year}{2009}).
\newblock \urlprefix\url{https://doi.org/10.1063/1.3095491}.
\newblock \eprint{https://doi.org/10.1063/1.3095491}.

\bibitem{Bronstein2017GeoDeep}
\bibinfo{author}{Bronstein, M.~M.}, \bibinfo{author}{Bruna, J.},
  \bibinfo{author}{LeCun, Y.}, \bibinfo{author}{Szlam, A.} \&
  \bibinfo{author}{Vandergheynst, P.}
\newblock \bibinfo{title}{Geometric deep learning: Going beyond euclidean
  data}.
\newblock \emph{\bibinfo{journal}{IEEE Signal Processing Magazine}}
  \textbf{\bibinfo{volume}{34}}, \bibinfo{pages}{18--42}
  (\bibinfo{year}{2017}).

\bibitem{atz_geometric_2021}
\bibinfo{author}{Atz, K.}, \bibinfo{author}{Grisoni, F.} \&
  \bibinfo{author}{Schneider, G.}
\newblock \bibinfo{title}{Geometric deep learning on molecular
  representations}.
\newblock \emph{\bibinfo{journal}{Nature Machine Intelligence}}
  \textbf{\bibinfo{volume}{3}}, \bibinfo{pages}{1023--1032}
  (\bibinfo{year}{2021}).
\newblock \urlprefix\url{https://doi.org/10.1038/s42256-021-00418-8}.

\bibitem{Cao2020GeoDeppReview}
\bibinfo{author}{Cao, W.}, \bibinfo{author}{Yan, Z.}, \bibinfo{author}{He, Z.}
  \& \bibinfo{author}{He, Z.}
\newblock \bibinfo{title}{A comprehensive survey on geometric deep learning}.
\newblock \emph{\bibinfo{journal}{IEEE Access}} \textbf{\bibinfo{volume}{8}},
  \bibinfo{pages}{35929--35949} (\bibinfo{year}{2020}).

\bibitem{Monti_2017_CVPR}
\bibinfo{author}{Monti, F.} \emph{et~al.}
\newblock \bibinfo{title}{Geometric deep learning on graphs and manifolds using
  mixture model cnns}.
\newblock In \emph{\bibinfo{booktitle}{Proceedings of the IEEE Conference on
  Computer Vision and Pattern Recognition (CVPR)}} (\bibinfo{year}{2017}).

\bibitem{Nguyen2013PointCloud}
\bibinfo{author}{Nguyen, A.} \& \bibinfo{author}{Le, B.}
\newblock \bibinfo{title}{3d point cloud segmentation: A survey}.
\newblock In \emph{\bibinfo{booktitle}{2013 6th IEEE Conference on Robotics,
  Automation and Mechatronics (RAM)}}, \bibinfo{pages}{225--230}
  (\bibinfo{year}{2013}).

\bibitem{qi_pointnet_2017}
\bibinfo{author}{Qi, C.~R.}, \bibinfo{author}{Yi, L.}, \bibinfo{author}{Su, H.}
  \& \bibinfo{author}{Guibas, L.~J.}
\newblock \bibinfo{title}{{PointNet}++: {Deep} {Hierarchical} {Feature}
  {Learning} on {Point} {Sets} in a {Metric} {Space}}.
\newblock In \bibinfo{editor}{Guyon, I.} \emph{et~al.} (eds.)
  \emph{\bibinfo{booktitle}{Advances in {Neural} {Information} {Processing}
  {Systems}}}, vol.~\bibinfo{volume}{30} (\bibinfo{publisher}{Curran
  Associates, Inc.}, \bibinfo{year}{2017}).
\newblock
  \urlprefix\url{https://proceedings.neurips.cc/paper/2017/file/d8bf84be3800d12f74d8b05e9b89836f-Paper.pdf}.

\bibitem{Lee2019SAGPool}
\bibinfo{author}{Lee, J.}, \bibinfo{author}{Lee, I.} \& \bibinfo{author}{Kang,
  J.}
\newblock \bibinfo{title}{Self-attention graph pooling}.
\newblock In \bibinfo{editor}{Chaudhuri, K.} \& \bibinfo{editor}{Salakhutdinov,
  R.} (eds.) \emph{\bibinfo{booktitle}{Proceedings of the 36th International
  Conference on Machine Learning}}, vol.~\bibinfo{volume}{97} of
  \emph{\bibinfo{series}{Proceedings of Machine Learning Research}},
  \bibinfo{pages}{3734--3743} (\bibinfo{publisher}{PMLR},
  \bibinfo{year}{2019}).
\newblock \urlprefix\url{https://proceedings.mlr.press/v97/lee19c.html}.

\bibitem{Zhang2019HierPool}
\bibinfo{author}{Zhang, Z.} \emph{et~al.}
\newblock \bibinfo{title}{Hierarchical graph pooling with structure learning}.
\newblock \emph{\bibinfo{journal}{CoRR}}
  \textbf{\bibinfo{volume}{abs/1911.05954}} (\bibinfo{year}{2019}).
\newblock \urlprefix\url{http://arxiv.org/abs/1911.05954}.
\newblock \eprint{1911.05954}.

\bibitem{Li20213DmolNet}
\bibinfo{author}{Li, C.} \emph{et~al.}
\newblock \bibinfo{title}{3dmol-net: Learn 3d molecular representation using
  adaptive graph convolutional network based on rotation invariance}.
\newblock \emph{\bibinfo{journal}{IEEE Journal of Biomedical and Health
  Informatics}} \bibinfo{pages}{1--1} (\bibinfo{year}{2021}).

\bibitem{Montavon2012InvariantRepr}
\bibinfo{author}{Montavon, G.} \emph{et~al.}
\newblock \bibinfo{title}{Learning invariant representations of molecules for
  atomization energy prediction}.
\newblock In \bibinfo{editor}{Pereira, F.}, \bibinfo{editor}{Burges, C. J.~C.},
  \bibinfo{editor}{Bottou, L.} \& \bibinfo{editor}{Weinberger, K.~Q.} (eds.)
  \emph{\bibinfo{booktitle}{Advances in Neural Information Processing
  Systems}}, vol.~\bibinfo{volume}{25} (\bibinfo{publisher}{Curran Associates,
  Inc.}, \bibinfo{year}{2012}).
\newblock
  \urlprefix\url{https://proceedings.neurips.cc/paper/2012/file/115f89503138416a242f40fb7d7f338e-Paper.pdf}.

\bibitem{Satorras2021Equiv}
\bibinfo{author}{Satorras, V.~G.}, \bibinfo{author}{Hoogeboom, E.} \&
  \bibinfo{author}{Welling, M.}
\newblock \bibinfo{title}{E(n) equivariant graph neural networks}.
\newblock In \bibinfo{editor}{Meila, M.} \& \bibinfo{editor}{Zhang, T.} (eds.)
  \emph{\bibinfo{booktitle}{Proceedings of the 38th International Conference on
  Machine Learning}}, vol. \bibinfo{volume}{139} of
  \emph{\bibinfo{series}{Proceedings of Machine Learning Research}},
  \bibinfo{pages}{9323--9332} (\bibinfo{publisher}{PMLR},
  \bibinfo{year}{2021}).
\newblock \urlprefix\url{https://proceedings.mlr.press/v139/satorras21a.html}.

\bibitem{Nigam2022EquivRepr}
\bibinfo{author}{Nigam, J.}, \bibinfo{author}{Willatt, M.~J.} \&
  \bibinfo{author}{Ceriotti, M.}
\newblock \bibinfo{title}{Equivariant representations for molecular
  hamiltonians and n-center atomic-scale properties}.
\newblock \emph{\bibinfo{journal}{The Journal of Chemical Physics}}
  \textbf{\bibinfo{volume}{156}}, \bibinfo{pages}{014115}
  (\bibinfo{year}{2022}).
\newblock \urlprefix\url{https://doi.org/10.1063/5.0072784}.
\newblock \eprint{https://doi.org/10.1063/5.0072784}.

\bibitem{chengGeoCGNN2021}
\bibinfo{author}{Cheng, J.}, \bibinfo{author}{Zhang, C.} \&
  \bibinfo{author}{Dong, L.}
\newblock \bibinfo{title}{A geometric-information-enhanced crystal graph
  network for predicting properties of materials}.
\newblock \emph{\bibinfo{journal}{Communications Materials}}
  \textbf{\bibinfo{volume}{2}}, \bibinfo{pages}{92} (\bibinfo{year}{2021}).
\newblock \urlprefix\url{https://doi.org/10.1038/s43246-021-00194-3}.

\bibitem{Morris2020TUDatasets}
\bibinfo{author}{Morris, C.} \emph{et~al.}
\newblock \bibinfo{title}{Tudataset: {A} collection of benchmark datasets for
  learning with graphs}.
\newblock \emph{\bibinfo{journal}{CoRR}}
  \textbf{\bibinfo{volume}{abs/2007.08663}} (\bibinfo{year}{2020}).
\newblock \urlprefix\url{https://arxiv.org/abs/2007.08663}.
\newblock \eprint{2007.08663}.

\bibitem{Zhenqin2017Moleculenet}
\bibinfo{author}{Wu, Z.} \emph{et~al.}
\newblock \bibinfo{title}{Moleculenet: a benchmark for molecular machine
  learning}.
\newblock \emph{\bibinfo{journal}{Chemical science}}
  \textbf{\bibinfo{volume}{9}}, \bibinfo{pages}{513--530}
  (\bibinfo{year}{2018}).

\bibitem{jainCommentaryMaterialsProject2013}
\bibinfo{author}{Jain, A.} \emph{et~al.}
\newblock \bibinfo{title}{Commentary: {{The Materials Project}}: {{A}}
  materials genome approach to accelerating materials innovation}.
\newblock \emph{\bibinfo{journal}{APL Materials}} \textbf{\bibinfo{volume}{1}},
  \bibinfo{pages}{011002} (\bibinfo{year}{2013}).

\bibitem{kirklin_open_2015}
\bibinfo{author}{Kirklin, S.} \emph{et~al.}
\newblock \bibinfo{title}{The {Open} {Quantum} {Materials} {Database} ({OQMD}):
  assessing the accuracy of {DFT} formation energies}.
\newblock \emph{\bibinfo{journal}{npj Computational Materials}}
  \textbf{\bibinfo{volume}{1}}, \bibinfo{pages}{15010} (\bibinfo{year}{2015}).
\newblock \urlprefix\url{https://doi.org/10.1038/npjcompumats.2015.10}.

\bibitem{blum2009qm7}
\bibinfo{author}{Blum, L.~C.} \& \bibinfo{author}{Reymond, J.-L.}
\newblock \bibinfo{title}{970 million druglike small molecules for virtual
  screening in the chemical universe database {GDB-13}}.
\newblock \emph{\bibinfo{journal}{J. Am. Chem. Soc.}}
  \textbf{\bibinfo{volume}{131}}, \bibinfo{pages}{8732} (\bibinfo{year}{2009}).

\bibitem{montavon2013qm7b}
\bibinfo{author}{Montavon, G.} \emph{et~al.}
\newblock \bibinfo{title}{Machine learning of molecular electronic properties
  in chemical compound space}.
\newblock \emph{\bibinfo{journal}{New Journal of Physics}}
  \textbf{\bibinfo{volume}{15}}, \bibinfo{pages}{095003}
  (\bibinfo{year}{2013}).
\newblock \urlprefix\url{http://stacks.iop.org/1367-2630/15/i=9/a=095003}.

\bibitem{wang2005pdbbind}
\bibinfo{author}{Wang, R.}, \bibinfo{author}{Fang, X.}, \bibinfo{author}{Lu,
  Y.}, \bibinfo{author}{Yang, C.-Y.} \& \bibinfo{author}{Wang, S.}
\newblock \bibinfo{title}{The {PDBbind} database:{\hspace{0.167em}}
  methodologies and updates}.
\newblock \emph{\bibinfo{journal}{Journal of Medicinal Chemistry}}
  \textbf{\bibinfo{volume}{48}}, \bibinfo{pages}{4111--4119}
  (\bibinfo{year}{2005}).
\newblock \urlprefix\url{https://doi.org/10.1021/jm048957q}.

\bibitem{chiemla2017md17-main-paper}
\bibinfo{author}{Chmiela, S.} \emph{et~al.}
\newblock \bibinfo{title}{Machine learning of accurate energy-conserving
  molecular force fields}.
\newblock \emph{\bibinfo{journal}{Science Advances}}
  \textbf{\bibinfo{volume}{3}}, \bibinfo{pages}{e1603015}
  (\bibinfo{year}{2017}).
\newblock
  \urlprefix\url{https://www.science.org/doi/abs/10.1126/sciadv.1603015}.
\newblock \eprint{https://www.science.org/doi/pdf/10.1126/sciadv.1603015}.

\bibitem{chmiela2018md17-second-paper}
\bibinfo{author}{Chmiela, S.}, \bibinfo{author}{Sauceda, H.~E.},
  \bibinfo{author}{M{\"u}ller, K.-R.} \& \bibinfo{author}{Tkatchenko, A.}
\newblock \bibinfo{title}{Towards exact molecular dynamics simulations with
  machine-learned force fields}.
\newblock \emph{\bibinfo{journal}{Nature Communications}}
  \textbf{\bibinfo{volume}{9}}, \bibinfo{pages}{3887} (\bibinfo{year}{2018}).
\newblock \urlprefix\url{https://doi.org/10.1038/s41467-018-06169-2}.

\bibitem{mobley_freesolv_2014}
\bibinfo{author}{Mobley, D.~L.} \& \bibinfo{author}{Guthrie, J.~P.}
\newblock \bibinfo{title}{{FreeSolv}: a database of experimental and calculated
  hydration free energies, with input files}.
\newblock \emph{\bibinfo{journal}{Journal of Computer-Aided Molecular Design}}
  \textbf{\bibinfo{volume}{28}}, \bibinfo{pages}{711--720}
  (\bibinfo{year}{2014}).
\newblock \urlprefix\url{https://doi.org/10.1007/s10822-014-9747-x}.

\bibitem{Richard2016ToxCAST}
\bibinfo{author}{Richard, A.~M.} \emph{et~al.}
\newblock \bibinfo{title}{Toxcast chemical landscape: Paving the road to 21st
  century toxicology}.
\newblock \emph{\bibinfo{journal}{Chemical Research in Toxicology}}
  \textbf{\bibinfo{volume}{29}}, \bibinfo{pages}{1225--1251}
  (\bibinfo{year}{2016}).
\newblock \urlprefix\url{https://doi.org/10.1021/acs.chemrestox.6b00135}.
\newblock \bibinfo{note}{PMID: 27367298},
  \eprint{https://doi.org/10.1021/acs.chemrestox.6b00135}.

\bibitem{Martins2012BBBP}
\bibinfo{author}{Martins, I.~F.}, \bibinfo{author}{Teixeira, A.~L.},
  \bibinfo{author}{Pinheiro, L.} \& \bibinfo{author}{Falcao, A.~O.}
\newblock \bibinfo{title}{A bayesian approach to in silico blood-brain barrier
  penetration modeling}.
\newblock \emph{\bibinfo{journal}{Journal of Chemical Information and
  Modeling}} \textbf{\bibinfo{volume}{52}}, \bibinfo{pages}{1686--1697}
  (\bibinfo{year}{2012}).
\newblock \urlprefix\url{https://doi.org/10.1021/ci300124c}.
\newblock \bibinfo{note}{PMID: 22612593},
  \eprint{https://doi.org/10.1021/ci300124c}.

\bibitem{Kuhn2015SIDER}
\bibinfo{author}{Kuhn, M.}, \bibinfo{author}{Letunic, I.},
  \bibinfo{author}{Jensen, L.~J.} \& \bibinfo{author}{Bork, P.}
\newblock \bibinfo{title}{{The SIDER database of drugs and side effects}}.
\newblock \emph{\bibinfo{journal}{Nucleic Acids Research}}
  \textbf{\bibinfo{volume}{44}}, \bibinfo{pages}{D1075--D1079}
  (\bibinfo{year}{2015}).
\newblock \urlprefix\url{https://doi.org/10.1093/nar/gkv1075}.
\newblock
  \eprint{https://academic.oup.com/nar/article-pdf/44/D1/D1075/16661270/gkv1075.pdf}.

\bibitem{Altae2017SIDER}
\bibinfo{author}{Altae-Tran, H.}, \bibinfo{author}{Ramsundar, B.},
  \bibinfo{author}{Pappu, A.~S.} \& \bibinfo{author}{Pande, V.}
\newblock \bibinfo{title}{Low data drug discovery with one-shot learning}.
\newblock \emph{\bibinfo{journal}{ACS Central Science}}
  \textbf{\bibinfo{volume}{3}}, \bibinfo{pages}{283--293}
  (\bibinfo{year}{2017}).
\newblock \urlprefix\url{https://doi.org/10.1021/acscentsci.6b00367}.
\newblock \bibinfo{note}{PMID: 28470045},
  \eprint{https://doi.org/10.1021/acscentsci.6b00367}.

\bibitem{Chanussot2021OCACS}
\bibinfo{author}{Chanussot, L.} \emph{et~al.}
\newblock \bibinfo{title}{Open catalyst 2020 (oc20) dataset and community
  challenges}.
\newblock \emph{\bibinfo{journal}{ACS Catalysis}}
  \textbf{\bibinfo{volume}{11}}, \bibinfo{pages}{6059--6072}
  (\bibinfo{year}{2021}).
\newblock \urlprefix\url{https://doi.org/10.1021/acscatal.0c04525}.
\newblock \eprint{https://doi.org/10.1021/acscatal.0c04525}.

\bibitem{cho2014properties}
\bibinfo{author}{Cho, K.}, \bibinfo{author}{van Merrienboer, B.},
  \bibinfo{author}{Bahdanau, D.} \& \bibinfo{author}{Bengio, Y.}
\newblock \bibinfo{title}{On the properties of neural machine translation:
  Encoder-decoder approaches}.
\newblock In \bibinfo{editor}{Wu, D.}, \bibinfo{editor}{Carpuat, M.},
  \bibinfo{editor}{Carreras, X.} \& \bibinfo{editor}{Vecchi, E.~M.} (eds.)
  \emph{\bibinfo{booktitle}{Proceedings of SSST@EMNLP 2014, Eighth Workshop on
  Syntax, Semantics and Structure in Statistical Translation, Doha, Qatar, 25
  October 2014}}, \bibinfo{pages}{103--111} (\bibinfo{publisher}{Association
  for Computational Linguistics}, \bibinfo{year}{2014}).
\newblock \urlprefix\url{https://aclanthology.org/W14-4012/}.

\bibitem{Hochreiter1997LSTM}
\bibinfo{author}{Hochreiter, S.} \& \bibinfo{author}{Schmidhuber, J.}
\newblock \bibinfo{title}{Long short-term memory}.
\newblock \emph{\bibinfo{journal}{Neural Computation}}
  \textbf{\bibinfo{volume}{9}}, \bibinfo{pages}{1735--1780}
  (\bibinfo{year}{1997}).

\bibitem{Hamilton2017GraphSAGE}
\bibinfo{author}{Hamilton, W.}, \bibinfo{author}{Ying, Z.} \&
  \bibinfo{author}{Leskovec, J.}
\newblock \bibinfo{title}{Inductive representation learning on large graphs}.
\newblock \emph{\bibinfo{journal}{Advances in neural information processing
  systems}} \textbf{\bibinfo{volume}{30}} (\bibinfo{year}{2017}).

\bibitem{vaswani2017attention}
\bibinfo{author}{Vaswani, A.} \emph{et~al.}
\newblock \bibinfo{title}{Attention is all you need}.
\newblock \emph{\bibinfo{journal}{Advances in neural information processing
  systems}} \textbf{\bibinfo{volume}{30}} (\bibinfo{year}{2017}).

\bibitem{velickovic2018GAT}
\bibinfo{author}{Velickovic, P.} \emph{et~al.}
\newblock \bibinfo{title}{Graph attention networks}.
\newblock In \emph{\bibinfo{booktitle}{6th International Conference on Learning
  Representations, {ICLR} 2018, Vancouver, BC, Canada, April 30 - May 3, 2018,
  Conference Track Proceedings}} (\bibinfo{publisher}{OpenReview.net},
  \bibinfo{year}{2018}).
\newblock \urlprefix\url{https://openreview.net/forum?id=rJXMpikCZ}.

\bibitem{jorgensen2018schnetedge}
\bibinfo{author}{Jørgensen, P.~B.}, \bibinfo{author}{Jacobsen, K.~W.} \&
  \bibinfo{author}{Schmidt, M.~N.}
\newblock \bibinfo{title}{Neural message passing with edge updates for
  predicting properties of molecules and materials} (\bibinfo{year}{2018}).
\newblock \eprint{1806.03146}.

\bibitem{Choudhary2021alignn}
\bibinfo{author}{Choudhary, K.} \& \bibinfo{author}{DeCost, B.}
\newblock \bibinfo{title}{Atomistic line graph neural network for improved
  materials property predictions}.
\newblock \emph{\bibinfo{journal}{npj Computational Materials}}
  \textbf{\bibinfo{volume}{7}} (\bibinfo{year}{2021}).
\newblock \urlprefix\url{https://doi.org/10.1038/s41524-021-00650-1}.

\bibitem{zhang2020mxmnet}
\bibinfo{author}{Zhang, S.}, \bibinfo{author}{Liu, Y.} \& \bibinfo{author}{Xie,
  L.}
\newblock \bibinfo{title}{Molecular mechanics-driven graph neural network with
  multiplex graph for molecular structures} (\bibinfo{year}{2020}).
\newblock \eprint{2011.07457}.

\bibitem{hsu2021alignn_d}
\bibinfo{author}{Hsu, T.} \emph{et~al.}
\newblock \bibinfo{title}{Efficient, interpretable atomistic graph neural
  network representation for angle-dependent properties and its application to
  optical spectroscopy prediction} (\bibinfo{year}{2021}).
\newblock \eprint{2109.11576}.

\bibitem{ganea2021geomol}
\bibinfo{author}{Ganea, O.} \emph{et~al.}
\newblock \bibinfo{title}{Geomol: Torsional geometric generation of molecular
  3d conformer ensembles}.
\newblock \emph{\bibinfo{journal}{Advances in Neural Information Processing
  Systems}} \textbf{\bibinfo{volume}{34}}, \bibinfo{pages}{13757--13769}
  (\bibinfo{year}{2021}).

\bibitem{anderson2019cormorant}
\bibinfo{author}{Anderson, B.}, \bibinfo{author}{Hy, T.~S.} \&
  \bibinfo{author}{Kondor, R.}
\newblock \bibinfo{title}{Cormorant: Covariant molecular neural networks}.
\newblock \emph{\bibinfo{journal}{Advances in neural information processing
  systems}} \textbf{\bibinfo{volume}{32}} (\bibinfo{year}{2019}).

\bibitem{qiao2021unite}
\bibinfo{author}{Qiao, Z.} \emph{et~al.}
\newblock \bibinfo{title}{Unite: Unitary n-body tensor equivariant network with
  applications to quantum chemistry} (\bibinfo{year}{2021}).
\newblock \eprint{2105.14655}.

\bibitem{Schutt2019schnorb}
\bibinfo{author}{Sch{\"u}tt, K.~T.}, \bibinfo{author}{Gastegger, M.},
  \bibinfo{author}{Tkatchenko, A.}, \bibinfo{author}{M{\"u}ller, K.-R.} \&
  \bibinfo{author}{Maurer, R.~J.}
\newblock \bibinfo{title}{Unifying machine learning and quantum chemistry with
  a deep neural network for molecular wavefunctions}.
\newblock \emph{\bibinfo{journal}{Nature Communications}}
  \textbf{\bibinfo{volume}{10}}, \bibinfo{pages}{5024} (\bibinfo{year}{2019}).
\newblock \urlprefix\url{https://doi.org/10.1038/s41467-019-12875-2}.

\bibitem{Qiao.2020}
\bibinfo{author}{Qiao, Z.}, \bibinfo{author}{Welborn, M.},
  \bibinfo{author}{Anandkumar, A.}, \bibinfo{author}{Manby, F.~R.} \&
  \bibinfo{author}{Miller, T.~F.}
\newblock \bibinfo{title}{Orbnet: Deep learning for quantum chemistry using
  symmetry-adapted atomic-orbital features}.
\newblock \emph{\bibinfo{journal}{The Journal of Chemical Physics}}
  \textbf{\bibinfo{volume}{153}}, \bibinfo{pages}{124111}
  (\bibinfo{year}{2020}).
\newblock \urlprefix\url{https://aip.scitation.org/doi/full/10.1063/5.0021955}.

\bibitem{unke2021se3phisnet}
\bibinfo{author}{Unke, O.} \emph{et~al.}
\newblock \bibinfo{title}{Se (3)-equivariant prediction of molecular
  wavefunctions and electronic densities}.
\newblock \emph{\bibinfo{journal}{Advances in Neural Information Processing
  Systems}} \textbf{\bibinfo{volume}{34}}, \bibinfo{pages}{14434--14447}
  (\bibinfo{year}{2021}).

\bibitem{liu2019hyperbolic}
\bibinfo{author}{Liu, Q.}, \bibinfo{author}{Nickel, M.} \&
  \bibinfo{author}{Kiela, D.}
\newblock \bibinfo{title}{Hyperbolic graph neural networks}.
\newblock \emph{\bibinfo{journal}{Advances in Neural Information Processing
  Systems}} \textbf{\bibinfo{volume}{32}} (\bibinfo{year}{2019}).

\bibitem{wieder_compact_2020}
\bibinfo{author}{Wieder, O.} \emph{et~al.}
\newblock \bibinfo{title}{A compact review of molecular property prediction
  with graph neural networks}.
\newblock \emph{\bibinfo{journal}{Drug Discovery Today: Technologies}}
  \textbf{\bibinfo{volume}{37}} (\bibinfo{year}{2020}).
\newblock
  \urlprefix\url{https://www.sciencedirect.com/science/article/pii/S1740674920300305}.

\bibitem{ZHOU202057}
\bibinfo{author}{Zhou, J.} \emph{et~al.}
\newblock \bibinfo{title}{Graph neural networks: A review of methods and
  applications}.
\newblock \emph{\bibinfo{journal}{AI Open}} \textbf{\bibinfo{volume}{1}},
  \bibinfo{pages}{57--81} (\bibinfo{year}{2020}).
\newblock
  \urlprefix\url{https://www.sciencedirect.com/science/article/pii/S2666651021000012}.

\bibitem{SunReview2019}
\bibinfo{author}{Sun, M.} \emph{et~al.}
\newblock \bibinfo{title}{{Graph convolutional networks for computational drug
  development and discovery}}.
\newblock \emph{\bibinfo{journal}{Briefings in Bioinformatics}}
  \textbf{\bibinfo{volume}{21}}, \bibinfo{pages}{919--935}
  (\bibinfo{year}{2019}).
\newblock \urlprefix\url{https://doi.org/10.1093/bib/bbz042}.
\newblock
  \eprint{https://academic.oup.com/bib/article-pdf/21/3/919/33227266/bbz042.pdf}.

\bibitem{zhang_graph_2019}
\bibinfo{author}{Zhang, S.}, \bibinfo{author}{Tong, H.}, \bibinfo{author}{Xu,
  J.} \& \bibinfo{author}{Maciejewski, R.}
\newblock \bibinfo{title}{Graph convolutional networks: a comprehensive
  review}.
\newblock \emph{\bibinfo{journal}{Computational Social Networks}}
  \textbf{\bibinfo{volume}{6}}, \bibinfo{pages}{1--23} (\bibinfo{year}{2019}).

\bibitem{ComprehensiveSurvey}
\bibinfo{author}{Wu, Z.} \emph{et~al.}
\newblock \bibinfo{title}{A comprehensive survey on graph neural networks}.
\newblock \emph{\bibinfo{journal}{IEEE Transactions on Neural Networks and
  Learning Systems}} \textbf{\bibinfo{volume}{32}}, \bibinfo{pages}{4--24}
  (\bibinfo{year}{2021}).

\bibitem{Yuan2021GNNhyper}
\bibinfo{author}{Yuan, Y.}, \bibinfo{author}{Wang, W.} \&
  \bibinfo{author}{Pang, W.}
\newblock \bibinfo{title}{A systematic comparison study on hyperparameter
  optimisation of graph neural networks for molecular property prediction}.
\newblock \emph{\bibinfo{journal}{Proceedings of the Genetic and Evolutionary
  Computation Conference}}  (\bibinfo{year}{2021}).
\newblock \urlprefix\url{http://dx.doi.org/10.1145/3449639.3459370}.

\bibitem{jiang_could_2021}
\bibinfo{author}{Jiang, D.} \emph{et~al.}
\newblock \bibinfo{title}{Could graph neural networks learn better molecular
  representation for drug discovery? a comparison study of descriptor-based and
  graph-based models}.
\newblock \emph{\bibinfo{journal}{Journal of cheminformatics}}
  \textbf{\bibinfo{volume}{13}}, \bibinfo{pages}{1--23} (\bibinfo{year}{2021}).

\bibitem{Banitalebi2021GCNNAllyouNeed}
\bibinfo{author}{Banitalebi{-}Dehkordi, A.} \& \bibinfo{author}{Zhang, Y.}
\newblock \bibinfo{title}{{ML4CO:} is {GCNN} all you need? graph convolutional
  neural networks produce strong baselines for combinatorial optimization
  problems, if tuned and trained properly, on appropriate data}.
\newblock \emph{\bibinfo{journal}{CoRR}}
  \textbf{\bibinfo{volume}{abs/2112.12251}} (\bibinfo{year}{2021}).
\newblock \urlprefix\url{https://arxiv.org/abs/2112.12251}.
\newblock \eprint{2112.12251}.

\bibitem{liao2018lanczosnet}
\bibinfo{author}{Liao, R.}, \bibinfo{author}{Zhao, Z.},
  \bibinfo{author}{Urtasun, R.} \& \bibinfo{author}{Zemel, R.}
\newblock \bibinfo{title}{Lanczosnet: Multi-scale deep graph convolutional
  networks}.
\newblock In \emph{\bibinfo{booktitle}{International Conference on Learning
  Representations}} (\bibinfo{year}{2019}).
\newblock \urlprefix\url{https://openreview.net/forum?id=BkedznAqKQ}.

\bibitem{SpecConv1}
\bibinfo{author}{Bruna, J.}, \bibinfo{author}{Zaremba, W.},
  \bibinfo{author}{Szlam, A.} \& \bibinfo{author}{LeCun, Y.}
\newblock \bibinfo{title}{Spectral networks and locally connected networks on
  graphs} (\bibinfo{year}{2013}).
\newblock \urlprefix\url{https://arxiv.org/abs/1312.6203}.

\bibitem{SpecConv2}
\bibinfo{author}{Henaff, M.}, \bibinfo{author}{Bruna, J.} \&
  \bibinfo{author}{LeCun, Y.}
\newblock \bibinfo{title}{Deep convolutional networks on graph-structured
  data}.
\newblock \emph{\bibinfo{journal}{CoRR}}
  \textbf{\bibinfo{volume}{abs/1506.05163}} (\bibinfo{year}{2015}).
\newblock \urlprefix\url{http://arxiv.org/abs/1506.05163}.
\newblock \eprint{1506.05163}.

\bibitem{CayleyNets}
\bibinfo{author}{Levie, R.}, \bibinfo{author}{Monti, F.},
  \bibinfo{author}{Bresson, X.} \& \bibinfo{author}{Bronstein, M.~M.}
\newblock \bibinfo{title}{Cayleynets: Graph convolutional neural networks with
  complex rational spectral filters}.
\newblock \emph{\bibinfo{journal}{IEEE Transactions on Signal Processing}}
  \textbf{\bibinfo{volume}{67}}, \bibinfo{pages}{97--109}
  (\bibinfo{year}{2019}).

\bibitem{Schlichtkrull2018}
\bibinfo{author}{Schlichtkrull, M.} \emph{et~al.}
\newblock \bibinfo{title}{Modeling relational data with graph convolutional
  networks}.
\newblock In \bibinfo{editor}{Gangemi, A.} \emph{et~al.} (eds.)
  \emph{\bibinfo{booktitle}{The Semantic Web}}, \bibinfo{pages}{593--607}
  (\bibinfo{publisher}{Springer International Publishing},
  \bibinfo{address}{Cham}, \bibinfo{year}{2018}).

\bibitem{patchysan}
\bibinfo{author}{Niepert, M.}, \bibinfo{author}{Ahmed, M.} \&
  \bibinfo{author}{Kutzkov, K.}
\newblock \bibinfo{title}{Learning convolutional neural networks for graphs}.
\newblock In \bibinfo{editor}{Balcan, M.~F.} \& \bibinfo{editor}{Weinberger,
  K.~Q.} (eds.) \emph{\bibinfo{booktitle}{Proceedings of The 33rd International
  Conference on Machine Learning}}, vol.~\bibinfo{volume}{48} of
  \emph{\bibinfo{series}{Proceedings of Machine Learning Research}},
  \bibinfo{pages}{2014--2023} (\bibinfo{publisher}{PMLR}, \bibinfo{address}{New
  York, New York, USA}, \bibinfo{year}{2016}).
\newblock \urlprefix\url{https://proceedings.mlr.press/v48/niepert16.html}.

\bibitem{SpatialMolWang2019}
\bibinfo{author}{Wang, X.} \emph{et~al.}
\newblock \bibinfo{title}{Molecule property prediction based on spatial graph
  embedding}.
\newblock \emph{\bibinfo{journal}{Journal of Chemical Information and
  Modeling}} \textbf{\bibinfo{volume}{59}}, \bibinfo{pages}{3817--3828}
  (\bibinfo{year}{2019}).
\newblock \urlprefix\url{https://doi.org/10.1021/acs.jcim.9b00410}.
\newblock \bibinfo{note}{PMID: 31438677},
  \eprint{https://doi.org/10.1021/acs.jcim.9b00410}.

\bibitem{xieCrystalGraphConvolutional2018}
\bibinfo{author}{Xie, T.} \& \bibinfo{author}{Grossman, J.~C.}
\newblock \bibinfo{title}{Crystal graph convolutional neural networks for an
  accurate and interpretable prediction of material properties}.
\newblock \emph{\bibinfo{journal}{Physical review letters}}
  \textbf{\bibinfo{volume}{120}}, \bibinfo{pages}{145301}
  (\bibinfo{year}{2018}).

\bibitem{parkDevelopingImprovedCrystal2020}
\bibinfo{author}{Park, C.~W.} \& \bibinfo{author}{Wolverton, C.}
\newblock \bibinfo{title}{Developing an improved crystal graph convolutional
  neural network framework for accelerated materials discovery}.
\newblock \emph{\bibinfo{journal}{Physical Review Materials}}
  \textbf{\bibinfo{volume}{4}}, \bibinfo{pages}{063801} (\bibinfo{year}{2020}).

\bibitem{NEURIPS2020_217eedd1}
\bibinfo{author}{Yadati, N.}
\newblock \bibinfo{title}{Neural message passing for multi-relational ordered
  and recursive hypergraphs}.
\newblock In \bibinfo{editor}{Larochelle, H.}, \bibinfo{editor}{Ranzato, M.},
  \bibinfo{editor}{Hadsell, R.}, \bibinfo{editor}{Balcan, M.} \&
  \bibinfo{editor}{Lin, H.} (eds.) \emph{\bibinfo{booktitle}{Advances in Neural
  Information Processing Systems}}, vol.~\bibinfo{volume}{33},
  \bibinfo{pages}{3275--3289} (\bibinfo{publisher}{Curran Associates, Inc.},
  \bibinfo{year}{2020}).
\newblock
  \urlprefix\url{https://proceedings.neurips.cc/paper/2020/file/217eedd1ba8c592db97d0dbe54c7adfc-Paper.pdf}.

\bibitem{strathmann2021persistent}
\bibinfo{author}{Strathmann, H.}, \bibinfo{author}{Barekatain, M.},
  \bibinfo{author}{Blundell, C.} \& \bibinfo{author}{Veli{\v{c}}kovi{\'c}, P.}
\newblock \bibinfo{title}{Persistent message passing}.
\newblock In \emph{\bibinfo{booktitle}{ICLR 2021 Workshop on Geometrical and
  Topological Representation Learning}} (\bibinfo{year}{2021}).
\newblock \urlprefix\url{https://openreview.net/forum?id=HhOJZT--N23}.

\bibitem{molNetKIM2022}
\bibinfo{author}{Kim, Y.} \emph{et~al.}
\newblock \bibinfo{title}{Molnet: A chemically intuitive graph neural network
  for prediction of molecular properties}.
\newblock \emph{\bibinfo{journal}{Chemistry--An Asian Journal}}
  (\bibinfo{year}{2022}).

\bibitem{hu2021forcenet}
\bibinfo{author}{Hu, W.} \emph{et~al.}
\newblock \bibinfo{title}{Forcenet: A graph neural network for large-scale
  quantum calculations} (\bibinfo{year}{2021}).
\newblock \eprint{2103.01436}.

\bibitem{parkAccurateScalableGraph2021}
\bibinfo{author}{Park, C.} \emph{et~al.}
\newblock \bibinfo{title}{Accurate and scalable graph neural network force
  field and molecular dynamics with direct force architecture}.
\newblock \emph{\bibinfo{journal}{npj Computational Materials}}
  \textbf{\bibinfo{volume}{7}} (\bibinfo{year}{2021}).

\bibitem{SphereNet2018}
\bibinfo{author}{Coors, B.}, \bibinfo{author}{Condurache, A.~P.} \&
  \bibinfo{author}{Geiger, A.}
\newblock \bibinfo{title}{Spherenet: Learning spherical representations for
  detection and classification in omnidirectional images}.
\newblock In \bibinfo{editor}{Ferrari, V.}, \bibinfo{editor}{Hebert, M.},
  \bibinfo{editor}{Sminchisescu, C.} \& \bibinfo{editor}{Weiss, Y.} (eds.)
  \emph{\bibinfo{booktitle}{Computer Vision -- ECCV 2018}},
  \bibinfo{pages}{525--541} (\bibinfo{publisher}{Springer International
  Publishing}, \bibinfo{address}{Cham}, \bibinfo{year}{2018}).

\bibitem{GATv2_2021}
\bibinfo{author}{Brody, S.}, \bibinfo{author}{Alon, U.} \&
  \bibinfo{author}{Yahav, E.}
\newblock \bibinfo{title}{How attentive are graph attention networks?}
\newblock \emph{\bibinfo{journal}{CoRR}}
  \textbf{\bibinfo{volume}{abs/2105.14491}} (\bibinfo{year}{2021}).
\newblock \urlprefix\url{https://arxiv.org/abs/2105.14491}.
\newblock \eprint{2105.14491}.

\bibitem{2018attentionbased}
\bibinfo{author}{Thekumparampil, K.~K.}, \bibinfo{author}{Oh, S.},
  \bibinfo{author}{Wang, C.} \& \bibinfo{author}{Li, L.-J.}
\newblock \bibinfo{title}{Attention-based graph neural network for
  semi-supervised learning} (\bibinfo{year}{2018}).
\newblock \urlprefix\url{https://openreview.net/forum?id=rJg4YGWRb}.

\bibitem{withnall_building_2020}
\bibinfo{author}{Withnall, M.}, \bibinfo{author}{Lindel{\"o}f, E.},
  \bibinfo{author}{Engkvist, O.} \& \bibinfo{author}{Chen, H.}
\newblock \bibinfo{title}{Building attention and edge message passing neural
  networks for bioactivity and physical--chemical property prediction}.
\newblock \emph{\bibinfo{journal}{Journal of cheminformatics}}
  \textbf{\bibinfo{volume}{12}}, \bibinfo{pages}{1--18} (\bibinfo{year}{2020}).

\bibitem{xinyi2018capsule}
\bibinfo{author}{Xinyi, Z.} \& \bibinfo{author}{Chen, L.}
\newblock \bibinfo{title}{Capsule graph neural network}.
\newblock In \emph{\bibinfo{booktitle}{International Conference on Learning
  Representations}} (\bibinfo{year}{2019}).
\newblock \urlprefix\url{https://openreview.net/forum?id=Byl8BnRcYm}.

\bibitem{RGAT2019}
\bibinfo{author}{Busbridge, D.}, \bibinfo{author}{Sherburn, D.},
  \bibinfo{author}{Cavallo, P.} \& \bibinfo{author}{Hammerla, N.~Y.}
\newblock \bibinfo{title}{Relational graph attention networks}.
\newblock \emph{\bibinfo{journal}{CoRR}}
  \textbf{\bibinfo{volume}{abs/1904.05811}} (\bibinfo{year}{2019}).
\newblock \urlprefix\url{http://arxiv.org/abs/1904.05811}.
\newblock \eprint{1904.05811}.

\bibitem{wang2020bees}
\bibinfo{author}{Wang, F.} \emph{et~al.}
\newblock \bibinfo{title}{Graph attention convolutional neural network model
  for chemical poisoning of honey bees’ prediction}.
\newblock \emph{\bibinfo{journal}{Science Bulletin}}
  \textbf{\bibinfo{volume}{65}}, \bibinfo{pages}{1184--1191}
  (\bibinfo{year}{2020}).
\newblock
  \urlprefix\url{https://www.sciencedirect.com/science/article/pii/S2095927320302176}.

\bibitem{sacha2021molecule}
\bibinfo{author}{Sacha, M.} \emph{et~al.}
\newblock \bibinfo{title}{Molecule edit graph attention network: modeling
  chemical reactions as sequences of graph edits}.
\newblock \emph{\bibinfo{journal}{Journal of Chemical Information and
  Modeling}} \textbf{\bibinfo{volume}{61}}, \bibinfo{pages}{3273--3284}
  (\bibinfo{year}{2021}).

\bibitem{tang_self-attention_2020}
\bibinfo{author}{Tang, B.} \emph{et~al.}
\newblock \bibinfo{title}{A self-attention based message passing neural network
  for predicting molecular lipophilicity and aqueous solubility}.
\newblock \emph{\bibinfo{journal}{Journal of cheminformatics}}
  \textbf{\bibinfo{volume}{12}}, \bibinfo{pages}{1--9} (\bibinfo{year}{2020}).

\bibitem{li2021conformationguided}
\bibinfo{author}{Li, Z.}, \bibinfo{author}{Yang, S.}, \bibinfo{author}{Song,
  G.} \& \bibinfo{author}{Cai, L.}
\newblock \bibinfo{title}{Conformation-guided molecular representation with
  hamiltonian neural networks}.
\newblock In \emph{\bibinfo{booktitle}{International Conference on Learning
  Representations}} (\bibinfo{year}{2021}).
\newblock \urlprefix\url{https://openreview.net/forum?id=q-cnWaaoUTH}.

\bibitem{TFEquiv2018}
\bibinfo{author}{Thomas, N.} \emph{et~al.}
\newblock \bibinfo{title}{Tensor field networks: Rotation- and
  translation-equivariant neural networks for 3d point clouds}.
\newblock \emph{\bibinfo{journal}{CoRR}}
  \textbf{\bibinfo{volume}{abs/1802.08219}} (\bibinfo{year}{2018}).
\newblock \urlprefix\url{http://arxiv.org/abs/1802.08219}.
\newblock \eprint{1802.08219}.

\bibitem{ClebschGordanNets}
\bibinfo{author}{Kondor, R.}, \bibinfo{author}{Lin, Z.} \&
  \bibinfo{author}{Trivedi, S.}
\newblock \bibinfo{title}{Clebsch-gordan nets: a fully fourier space spherical
  convolutional neural network}.
\newblock In \bibinfo{editor}{Bengio, S.} \emph{et~al.} (eds.)
  \emph{\bibinfo{booktitle}{Advances in Neural Information Processing Systems
  31: Annual Conference on Neural Information Processing Systems 2018, NeurIPS
  2018, December 3-8, 2018, Montr{\'{e}}al, Canada}},
  \bibinfo{pages}{10138--10147} (\bibinfo{year}{2018}).
\newblock
  \urlprefix\url{https://proceedings.neurips.cc/paper/2018/hash/a3fc981af450752046be179185ebc8b5-Abstract.html}.

\bibitem{SEGNN2021}
\bibinfo{author}{Brandstetter, J.}, \bibinfo{author}{Hesselink, R.},
  \bibinfo{author}{van~der Pol, E.}, \bibinfo{author}{Bekkers, E.} \&
  \bibinfo{author}{Welling, M.}
\newblock \bibinfo{title}{Geometric and physical quantities improve {E(3)}
  equivariant message passing}.
\newblock \emph{\bibinfo{journal}{CoRR}}
  \textbf{\bibinfo{volume}{abs/2110.02905}} (\bibinfo{year}{2021}).
\newblock \urlprefix\url{https://arxiv.org/abs/2110.02905}.
\newblock \eprint{2110.02905}.

\bibitem{NEURIPS2020_15231a7c}
\bibinfo{author}{Fuchs, F.}, \bibinfo{author}{Worrall, D.},
  \bibinfo{author}{Fischer, V.} \& \bibinfo{author}{Welling, M.}
\newblock \bibinfo{title}{Se(3)-transformers: 3d roto-translation equivariant
  attention networks}.
\newblock In \bibinfo{editor}{Larochelle, H.}, \bibinfo{editor}{Ranzato, M.},
  \bibinfo{editor}{Hadsell, R.}, \bibinfo{editor}{Balcan, M.} \&
  \bibinfo{editor}{Lin, H.} (eds.) \emph{\bibinfo{booktitle}{Advances in Neural
  Information Processing Systems}}, vol.~\bibinfo{volume}{33},
  \bibinfo{pages}{1970--1981} (\bibinfo{publisher}{Curran Associates, Inc.},
  \bibinfo{year}{2020}).
\newblock
  \urlprefix\url{https://proceedings.neurips.cc/paper/2020/file/15231a7ce4ba789d13b722cc5c955834-Paper.pdf}.

\bibitem{GCNN}
\bibinfo{author}{Cohen, T.} \& \bibinfo{author}{Welling, M.}
\newblock \bibinfo{title}{Group equivariant convolutional networks}.
\newblock In \bibinfo{editor}{Balcan, M.~F.} \& \bibinfo{editor}{Weinberger,
  K.~Q.} (eds.) \emph{\bibinfo{booktitle}{Proceedings of The 33rd International
  Conference on Machine Learning}}, vol.~\bibinfo{volume}{48} of
  \emph{\bibinfo{series}{Proceedings of Machine Learning Research}},
  \bibinfo{pages}{2990--2999} (\bibinfo{publisher}{PMLR}, \bibinfo{address}{New
  York, New York, USA}, \bibinfo{year}{2016}).
\newblock \urlprefix\url{https://proceedings.mlr.press/v48/cohenc16.html}.

\bibitem{GCNN2}
\bibinfo{author}{Lu, Y.} \emph{et~al.}
\newblock \bibinfo{title}{Cnn-g: Convolutional neural network combined with
  graph for image segmentation with theoretical analysis}.
\newblock \emph{\bibinfo{journal}{IEEE Transactions on Cognitive and
  Developmental Systems}} \textbf{\bibinfo{volume}{13}},
  \bibinfo{pages}{631--644} (\bibinfo{year}{2021}).

\bibitem{Bekkers2020B-Spline}
\bibinfo{author}{Bekkers, E.~J.}
\newblock \bibinfo{title}{B-spline cnns on lie groups}.
\newblock In \emph{\bibinfo{booktitle}{International Conference on Learning
  Representations}} (\bibinfo{year}{2020}).
\newblock \urlprefix\url{https://openreview.net/forum?id=H1gBhkBFDH}.

\bibitem{DiffPool2018}
\bibinfo{author}{Ying, R.} \emph{et~al.}
\newblock \bibinfo{title}{Hierarchical graph representation learning with
  differentiable pooling}.
\newblock In \emph{\bibinfo{booktitle}{Proceedings of the 32nd International
  Conference on Neural Information Processing Systems}}, NIPS'18,
  \bibinfo{pages}{4805–4815} (\bibinfo{publisher}{Curran Associates Inc.},
  \bibinfo{address}{Red Hook, NY, USA}, \bibinfo{year}{2018}).

\bibitem{EdgePool2019}
\bibinfo{author}{Diehl, F.}
\newblock \bibinfo{title}{Edge contraction pooling for graph neural networks}
  (\bibinfo{year}{2019}).
\newblock \urlprefix\url{https://arxiv.org/abs/1905.10990}.

\bibitem{pmlr-v97-gao19a}
\bibinfo{author}{Gao, H.} \& \bibinfo{author}{Ji, S.}
\newblock \bibinfo{title}{Graph u-nets}.
\newblock In \bibinfo{editor}{Chaudhuri, K.} \& \bibinfo{editor}{Salakhutdinov,
  R.} (eds.) \emph{\bibinfo{booktitle}{Proceedings of the 36th International
  Conference on Machine Learning}}, vol.~\bibinfo{volume}{97} of
  \emph{\bibinfo{series}{Proceedings of Machine Learning Research}},
  \bibinfo{pages}{2083--2092} (\bibinfo{publisher}{PMLR},
  \bibinfo{year}{2019}).
\newblock \urlprefix\url{https://proceedings.mlr.press/v97/gao19a.html}.

\bibitem{Gao2021iPoolI}
\bibinfo{author}{Gao, X.}, \bibinfo{author}{Xiong, H.} \&
  \bibinfo{author}{Frossard, P.}
\newblock \bibinfo{title}{ipool - information-based pooling in hierarchical
  graph neural networks}.
\newblock \emph{\bibinfo{journal}{IEEE transactions on neural networks and
  learning systems}} \textbf{\bibinfo{volume}{PP}} (\bibinfo{year}{2021}).

\bibitem{EigenPooling2019}
\bibinfo{author}{Ma, Y.}, \bibinfo{author}{Wang, S.},
  \bibinfo{author}{Aggarwal, C.~C.} \& \bibinfo{author}{Tang, J.}
\newblock \bibinfo{title}{Graph convolutional networks with eigenpooling}.
\newblock In \emph{\bibinfo{booktitle}{Proceedings of the 25th ACM SIGKDD
  International Conference on Knowledge Discovery \& Data Mining}}, KDD '19,
  \bibinfo{pages}{723–731} (\bibinfo{publisher}{Association for Computing
  Machinery}, \bibinfo{address}{New York, NY, USA}, \bibinfo{year}{2019}).
\newblock \urlprefix\url{https://doi.org/10.1145/3292500.3330982}.

\bibitem{constrained-gvae}
\bibinfo{author}{Liu, Q.}, \bibinfo{author}{Allamanis, M.},
  \bibinfo{author}{Brockschmidt, M.} \& \bibinfo{author}{Gaunt, A.}
\newblock \bibinfo{title}{Constrained graph variational autoencoders for
  molecule design}.
\newblock \emph{\bibinfo{journal}{Advances in neural information processing
  systems}} \textbf{\bibinfo{volume}{31}} (\bibinfo{year}{2018}).

\bibitem{jt-vae}
\bibinfo{author}{Jin, W.}, \bibinfo{author}{Barzilay, R.} \&
  \bibinfo{author}{Jaakkola, T.}
\newblock \bibinfo{title}{Junction tree variational autoencoder for molecular
  graph generation}.
\newblock In \bibinfo{editor}{Dy, J.} \& \bibinfo{editor}{Krause, A.} (eds.)
  \emph{\bibinfo{booktitle}{Proceedings of the 35th International Conference on
  Machine Learning}}, vol.~\bibinfo{volume}{80} of
  \emph{\bibinfo{series}{Proceedings of Machine Learning Research}},
  \bibinfo{pages}{2323--2332} (\bibinfo{publisher}{PMLR},
  \bibinfo{year}{2018}).
\newblock \urlprefix\url{https://proceedings.mlr.press/v80/jin18a.html}.

\bibitem{gcpn}
\bibinfo{author}{You, J.}, \bibinfo{author}{Liu, B.}, \bibinfo{author}{Ying,
  Z.}, \bibinfo{author}{Pande, V.} \& \bibinfo{author}{Leskovec, J.}
\newblock \bibinfo{title}{Graph convolutional policy network for goal-directed
  molecular graph generation}.
\newblock \emph{\bibinfo{journal}{Advances in neural information processing
  systems}} \textbf{\bibinfo{volume}{31}} (\bibinfo{year}{2018}).

\bibitem{wang2017graphgan}
\bibinfo{author}{Wang, H.} \emph{et~al.}
\newblock \bibinfo{title}{Graphgan: Graph representation learning with
  generative adversarial nets}.
\newblock In \bibinfo{editor}{McIlraith, S.~A.} \& \bibinfo{editor}{Weinberger,
  K.~Q.} (eds.) \emph{\bibinfo{booktitle}{Proceedings of the Thirty-Second
  {AAAI} Conference on Artificial Intelligence, (AAAI-18), the 30th innovative
  Applications of Artificial Intelligence (IAAI-18), and the 8th {AAAI}
  Symposium on Educational Advances in Artificial Intelligence (EAAI-18), New
  Orleans, Louisiana, USA, February 2-7, 2018}}, \bibinfo{pages}{2508--2515}
  (\bibinfo{publisher}{{AAAI} Press}, \bibinfo{year}{2018}).
\newblock
  \urlprefix\url{https://www.aaai.org/ocs/index.php/AAAI/AAAI18/paper/view/16611}.

\bibitem{long2021constrained}
\bibinfo{author}{Long, T.} \emph{et~al.}
\newblock \bibinfo{title}{Constrained crystals deep convolutional generative
  adversarial network for the inverse design of crystal structures}.
\newblock \emph{\bibinfo{journal}{npj Computational Materials}}
  \textbf{\bibinfo{volume}{7}}, \bibinfo{pages}{1--7} (\bibinfo{year}{2021}).

\bibitem{design-generative-models}
\bibinfo{author}{Sanchez-Lengeling, B.} \& \bibinfo{author}{Aspuru-Guzik, A.}
\newblock \bibinfo{title}{Inverse molecular design using machine learning:
  Generative models for matter engineering}.
\newblock \emph{\bibinfo{journal}{Science}} \textbf{\bibinfo{volume}{361}},
  \bibinfo{pages}{360--365} (\bibinfo{year}{2018}).
\newblock
  \urlprefix\url{https://www.science.org/doi/abs/10.1126/science.aat2663}.
\newblock \eprint{https://www.science.org/doi/pdf/10.1126/science.aat2663}.

\bibitem{MoFlow2020}
\bibinfo{author}{Zang, C.} \& \bibinfo{author}{Wang, F.}
\newblock \bibinfo{title}{Moflow: An invertible flow model for generating
  molecular graphs}.
\newblock \emph{\bibinfo{journal}{Proceedings of the 26th ACM SIGKDD
  International Conference on Knowledge Discovery \& Data Mining}}
  (\bibinfo{year}{2020}).
\newblock \urlprefix\url{http://dx.doi.org/10.1145/3394486.3403104}.

\bibitem{gilbert1959random}
\bibinfo{author}{Gilbert, E.~N.}
\newblock \bibinfo{title}{Random graphs}.
\newblock \emph{\bibinfo{journal}{The Annals of Mathematical Statistics}}
  \textbf{\bibinfo{volume}{30}}, \bibinfo{pages}{1141--1144}
  (\bibinfo{year}{1959}).

\bibitem{watts_collective_1998}
\bibinfo{author}{Watts, D.~J.} \& \bibinfo{author}{Strogatz, S.~H.}
\newblock \bibinfo{title}{Collective dynamics of ‘small-world’ networks}.
\newblock \emph{\bibinfo{journal}{Nature}} \textbf{\bibinfo{volume}{393}},
  \bibinfo{pages}{440--442} (\bibinfo{year}{1998}).

\bibitem{leskovec2010}
\bibinfo{author}{Leskovec, J.}, \bibinfo{author}{Chakrabarti, D.},
  \bibinfo{author}{Kleinberg, J.}, \bibinfo{author}{Faloutsos, C.} \&
  \bibinfo{author}{Ghahramani, Z.}
\newblock \bibinfo{title}{Kronecker graphs: An approach to modeling networks}.
\newblock \emph{\bibinfo{journal}{Journal of Machine Learning Research}}
  \textbf{\bibinfo{volume}{11}}, \bibinfo{pages}{985--1042}
  (\bibinfo{year}{2010}).
\newblock \urlprefix\url{http://jmlr.org/papers/v11/leskovec10a.html}.

\bibitem{KingmaVAE}
\bibinfo{author}{Kingma, D.~P.} \& \bibinfo{author}{Welling, M.}
\newblock \bibinfo{title}{Auto-encoding variational bayes}.
\newblock In \bibinfo{editor}{Bengio, Y.} \& \bibinfo{editor}{LeCun, Y.} (eds.)
  \emph{\bibinfo{booktitle}{2nd International Conference on Learning
  Representations, {ICLR} 2014, Banff, AB, Canada, April 14-16, 2014,
  Conference Track Proceedings}} (\bibinfo{year}{2014}).
\newblock \urlprefix\url{http://arxiv.org/abs/1312.6114}.

\bibitem{NIPS2014_5ca3e9b1}
\bibinfo{author}{Goodfellow, I.} \emph{et~al.}
\newblock \bibinfo{title}{{Generative Adversarial Nets}}.
\newblock In \bibinfo{editor}{Ghahramani, Z.}, \bibinfo{editor}{Welling, M.},
  \bibinfo{editor}{Cortes, C.}, \bibinfo{editor}{Lawrence, N.} \&
  \bibinfo{editor}{Weinberger, K.~Q.} (eds.) \emph{\bibinfo{booktitle}{Advances
  in Neural Information Processing Systems}}, vol.~\bibinfo{volume}{27}
  (\bibinfo{publisher}{Curran Associates, Inc.}, \bibinfo{year}{2014}).
\newblock
  \urlprefix\url{https://proceedings.neurips.cc/paper/2014/file/5ca3e9b122f61f8f06494c97b1afccf3-Paper.pdf}.

\bibitem{sutton2018reinforcement}
\bibinfo{author}{Sutton, R.~S.} \& \bibinfo{author}{Barto, A.~G.}
\newblock \emph{\bibinfo{title}{Reinforcement learning: An introduction}}
  (\bibinfo{publisher}{MIT press}, \bibinfo{year}{2018}).

\bibitem{You2018}
\bibinfo{author}{You, J.}, \bibinfo{author}{Ying, R.}, \bibinfo{author}{Ren,
  X.}, \bibinfo{author}{Hamilton, W.} \& \bibinfo{author}{Leskovec, J.}
\newblock \bibinfo{title}{{G}raph{RNN}: Generating realistic graphs with deep
  auto-regressive models}.
\newblock In \bibinfo{editor}{Dy, J.} \& \bibinfo{editor}{Krause, A.} (eds.)
  \emph{\bibinfo{booktitle}{Proceedings of the 35th International Conference on
  Machine Learning}}, vol.~\bibinfo{volume}{80} of
  \emph{\bibinfo{series}{Proceedings of Machine Learning Research}},
  \bibinfo{pages}{5708--5717} (\bibinfo{publisher}{PMLR},
  \bibinfo{year}{2018}).
\newblock \urlprefix\url{https://proceedings.mlr.press/v80/you18a.html}.

\bibitem{dinh2014nice}
\bibinfo{author}{Dinh, L.}, \bibinfo{author}{Krueger, D.} \&
  \bibinfo{author}{Bengio, Y.}
\newblock \bibinfo{title}{{NICE:} non-linear independent components
  estimation}.
\newblock In \bibinfo{editor}{Bengio, Y.} \& \bibinfo{editor}{LeCun, Y.} (eds.)
  \emph{\bibinfo{booktitle}{3rd International Conference on Learning
  Representations, {ICLR} 2015, San Diego, CA, USA, May 7-9, 2015, Workshop
  Track Proceedings}} (\bibinfo{year}{2015}).
\newblock \urlprefix\url{http://arxiv.org/abs/1410.8516}.

\bibitem{dinh2016density}
\bibinfo{author}{Dinh, L.}, \bibinfo{author}{Sohl{-}Dickstein, J.} \&
  \bibinfo{author}{Bengio, S.}
\newblock \bibinfo{title}{Density estimation using real {NVP}}.
\newblock In \emph{\bibinfo{booktitle}{5th International Conference on Learning
  Representations, {ICLR} 2017, Toulon, France, April 24-26, 2017, Conference
  Track Proceedings}} (\bibinfo{publisher}{OpenReview.net},
  \bibinfo{year}{2017}).
\newblock \urlprefix\url{https://openreview.net/forum?id=HkpbnH9lx}.

\bibitem{kingma2018glow}
\bibinfo{author}{Kingma, D.~P.} \& \bibinfo{author}{Dhariwal, P.}
\newblock \bibinfo{title}{Glow: Generative flow with invertible 1x1
  convolutions}.
\newblock \emph{\bibinfo{journal}{Advances in neural information processing
  systems}} \textbf{\bibinfo{volume}{31}} (\bibinfo{year}{2018}).

\bibitem{vae-images}
\bibinfo{author}{Pu, Y.} \emph{et~al.}
\newblock \bibinfo{title}{Variational autoencoder for deep learning of images,
  labels and captions}.
\newblock In \bibinfo{editor}{Lee, D.}, \bibinfo{editor}{Sugiyama, M.},
  \bibinfo{editor}{Luxburg, U.}, \bibinfo{editor}{Guyon, I.} \&
  \bibinfo{editor}{Garnett, R.} (eds.) \emph{\bibinfo{booktitle}{Advances in
  Neural Information Processing Systems}}, vol.~\bibinfo{volume}{29}
  (\bibinfo{publisher}{Curran Associates, Inc.}, \bibinfo{year}{2016}).
\newblock
  \urlprefix\url{https://proceedings.neurips.cc/paper/2016/file/eb86d510361fc23b59f18c1bc9802cc6-Paper.pdf}.

\bibitem{kusner2017grammar}
\bibinfo{author}{Kusner, M.~J.}, \bibinfo{author}{Paige, B.} \&
  \bibinfo{author}{Hern{\'{a}}ndez{-}Lobato, J.~M.}
\newblock \bibinfo{title}{Grammar variational autoencoder}.
\newblock In \bibinfo{editor}{Precup, D.} \& \bibinfo{editor}{Teh, Y.~W.}
  (eds.) \emph{\bibinfo{booktitle}{Proceedings of the 34th International
  Conference on Machine Learning, {ICML} 2017, Sydney, NSW, Australia, 6-11
  August 2017}}, vol.~\bibinfo{volume}{70} of
  \emph{\bibinfo{series}{Proceedings of Machine Learning Research}},
  \bibinfo{pages}{1945--1954} (\bibinfo{publisher}{{PMLR}},
  \bibinfo{year}{2017}).
\newblock \urlprefix\url{http://proceedings.mlr.press/v70/kusner17a.html}.

\bibitem{dai2018syntaxdirected}
\bibinfo{author}{Dai, H.}, \bibinfo{author}{Tian, Y.}, \bibinfo{author}{Dai,
  B.}, \bibinfo{author}{Skiena, S.} \& \bibinfo{author}{Song, L.}
\newblock \bibinfo{title}{Syntax-directed variational autoencoder for
  structured data}.
\newblock In \emph{\bibinfo{booktitle}{6th International Conference on Learning
  Representations, {ICLR} 2018, Vancouver, BC, Canada, April 30 - May 3, 2018,
  Conference Track Proceedings}} (\bibinfo{publisher}{OpenReview.net},
  \bibinfo{year}{2018}).
\newblock \urlprefix\url{https://openreview.net/forum?id=SyqShMZRb}.

\bibitem{variational-graph-ae}
\bibinfo{author}{Kipf, T.~N.} \& \bibinfo{author}{Welling, M.}
\newblock \bibinfo{title}{Variational graph auto-encoders}
  (\bibinfo{year}{2016}).
\newblock \eprint{1611.07308}.

\bibitem{simonovsky2018graphvae}
\bibinfo{author}{Simonovsky, M.} \& \bibinfo{author}{Komodakis, N.}
\newblock \bibinfo{title}{Graphvae: Towards generation of small graphs using
  variational autoencoders}.
\newblock In \bibinfo{editor}{Kurkov{\'{a}}, V.},
  \bibinfo{editor}{Manolopoulos, Y.}, \bibinfo{editor}{Hammer, B.},
  \bibinfo{editor}{Iliadis, L.~S.} \& \bibinfo{editor}{Maglogiannis, I.} (eds.)
  \emph{\bibinfo{booktitle}{Artificial Neural Networks and Machine Learning -
  {ICANN} 2018 - 27th International Conference on Artificial Neural Networks,
  Rhodes, Greece, October 4-7, 2018, Proceedings, Part {I}}}, vol.
  \bibinfo{volume}{11139} of \emph{\bibinfo{series}{Lecture Notes in Computer
  Science}}, \bibinfo{pages}{412--422} (\bibinfo{publisher}{Springer},
  \bibinfo{year}{2018}).
\newblock \urlprefix\url{https://doi.org/10.1007/978-3-030-01418-6\_41}.

\bibitem{denton2015deep}
\bibinfo{author}{Denton, E.~L.}, \bibinfo{author}{Chintala, S.},
  \bibinfo{author}{Fergus, R.} \emph{et~al.}
\newblock \bibinfo{title}{Deep generative image models using a laplacian
  pyramid of adversarial networks}.
\newblock \emph{\bibinfo{journal}{Advances in neural information processing
  systems}} \textbf{\bibinfo{volume}{28}} (\bibinfo{year}{2015}).

\bibitem{yu2017seqgan}
\bibinfo{author}{Yu, L.}, \bibinfo{author}{Zhang, W.}, \bibinfo{author}{Wang,
  J.} \& \bibinfo{author}{Yu, Y.}
\newblock \bibinfo{title}{Seqgan: Sequence generative adversarial nets with
  policy gradient}.
\newblock \emph{\bibinfo{journal}{Proceedings of the AAAI Conference on
  Artificial Intelligence}} \textbf{\bibinfo{volume}{31}}
  (\bibinfo{year}{2017}).
\newblock
  \urlprefix\url{https://ojs.aaai.org/index.php/AAAI/article/view/10804}.

\bibitem{molgan}
\bibinfo{author}{Cao, N.~D.} \& \bibinfo{author}{Kipf, T.}
\newblock \bibinfo{title}{Molgan: An implicit generative model for small
  molecular graphs} (\bibinfo{year}{2018}).
\newblock \eprint{1805.11973}.

\bibitem{Li2018learningdeepgen}
\bibinfo{author}{Li, Y.}, \bibinfo{author}{Vinyals, O.}, \bibinfo{author}{Dyer,
  C.}, \bibinfo{author}{Pascanu, R.} \& \bibinfo{author}{Battaglia, P.}
\newblock \bibinfo{title}{Learning deep generative models of graphs}
  (\bibinfo{year}{2018}).
\newblock \urlprefix\url{https://arxiv.org/abs/1803.03324}.

\bibitem{atance2021}
\bibinfo{author}{Romeo~Atance, S.}, \bibinfo{author}{Viguera~Diez, J.},
  \bibinfo{author}{Engkvist, O.}, \bibinfo{author}{Olsson, S.} \&
  \bibinfo{author}{Mercado, R.}
\newblock \bibinfo{title}{De novo drug design using reinforcement learning with
  graph-based deep generative models}.
\newblock \emph{\bibinfo{journal}{ChemRxiv}}  (\bibinfo{year}{2021}).

\bibitem{Mercado2021}
\bibinfo{author}{Mercado, R.} \emph{et~al.}
\newblock \bibinfo{title}{Graph networks for molecular design}.
\newblock \emph{\bibinfo{journal}{Machine Learning: Science and Technology}}
  \textbf{\bibinfo{volume}{2}} (\bibinfo{year}{2021}).

\bibitem{Li2018MultiobjectiveDN}
\bibinfo{author}{Li, Y.}, \bibinfo{author}{Zhang, L.~R.} \&
  \bibinfo{author}{ming Liu, Z.}
\newblock \bibinfo{title}{Multi-objective de novo drug design with conditional
  graph generative model}.
\newblock \emph{\bibinfo{journal}{Journal of Cheminformatics}}
  \textbf{\bibinfo{volume}{10}} (\bibinfo{year}{2018}).

\bibitem{papamakarios2021normalizing}
\bibinfo{author}{Papamakarios, G.}, \bibinfo{author}{Nalisnick, E.~T.},
  \bibinfo{author}{Rezende, D.~J.}, \bibinfo{author}{Mohamed, S.} \&
  \bibinfo{author}{Lakshminarayanan, B.}
\newblock \bibinfo{title}{Normalizing flows for probabilistic modeling and
  inference.}
\newblock \emph{\bibinfo{journal}{J. Mach. Learn. Res.}}
  \textbf{\bibinfo{volume}{22}}, \bibinfo{pages}{1--64} (\bibinfo{year}{2021}).

\bibitem{madhawa2019graphnvp}
\bibinfo{author}{Madhawa, K.}, \bibinfo{author}{Ishiguro, K.},
  \bibinfo{author}{Nakago, K.} \& \bibinfo{author}{Abe, M.}
\newblock \bibinfo{title}{Graphnvp: An invertible flow model for generating
  molecular graphs} (\bibinfo{year}{2019}).
\newblock \eprint{1905.11600}.

\bibitem{bengio2021flow}
\bibinfo{author}{Bengio, E.}, \bibinfo{author}{Jain, M.},
  \bibinfo{author}{Korablyov, M.}, \bibinfo{author}{Precup, D.} \&
  \bibinfo{author}{Bengio, Y.}
\newblock \bibinfo{title}{Flow network based generative models for
  non-iterative diverse candidate generation}.
\newblock In \bibinfo{editor}{Ranzato, M.}, \bibinfo{editor}{Beygelzimer, A.},
  \bibinfo{editor}{Dauphin, Y.~N.}, \bibinfo{editor}{Liang, P.} \&
  \bibinfo{editor}{Vaughan, J.~W.} (eds.) \emph{\bibinfo{booktitle}{Advances in
  Neural Information Processing Systems 34: Annual Conference on Neural
  Information Processing Systems 2021, NeurIPS 2021, December 6-14, 2021,
  virtual}}, \bibinfo{pages}{27381--27394} (\bibinfo{year}{2021}).
\newblock
  \urlprefix\url{https://proceedings.neurips.cc/paper/2021/hash/e614f646836aaed9f89ce58e837e2310-Abstract.html}.

\bibitem{frey2022fastflows}
\bibinfo{author}{Frey, N.~C.}, \bibinfo{author}{Gadepally, V.} \&
  \bibinfo{author}{Ramsundar, B.}
\newblock \bibinfo{title}{Fastflows: Flow-based models for molecular graph
  generation} (\bibinfo{year}{2022}).
\newblock \eprint{2201.12419}.

\bibitem{wangDeepGenerativeModel2022}
\bibinfo{author}{Wang, S.}, \bibinfo{author}{Guo, X.} \& \bibinfo{author}{Zhao,
  L.}
\newblock \bibinfo{title}{Deep {{Generative Model}} for {{Periodic Graphs}}}.
\newblock \emph{\bibinfo{journal}{arXiv:2201.11932 [cs]}}
  (\bibinfo{year}{2022}).
\newblock \eprint{2201.11932}.

\bibitem{Feinberg2020drug}
\bibinfo{author}{Feinberg, E.~N.}, \bibinfo{author}{Joshi, E.},
  \bibinfo{author}{Pande, V.~S.} \& \bibinfo{author}{Cheng, A.~C.}
\newblock \bibinfo{title}{Improvement in admet prediction with multitask deep
  featurization}.
\newblock \emph{\bibinfo{journal}{Journal of Medicinal Chemistry}}
  \textbf{\bibinfo{volume}{63}}, \bibinfo{pages}{8835--8848}
  (\bibinfo{year}{2020}).
\newblock \urlprefix\url{https://doi.org/10.1021/acs.jmedchem.9b02187}.
\newblock \bibinfo{note}{PMID: 32286824},
  \eprint{https://doi.org/10.1021/acs.jmedchem.9b02187}.

\bibitem{gastegger2020fieldschnet}
\bibinfo{author}{Gastegger, M.}, \bibinfo{author}{Sch{\"u}tt, K.~T.} \&
  \bibinfo{author}{M{\"u}ller, K.-R.}
\newblock \bibinfo{title}{Machine learning of solvent effects on molecular
  spectra and reactions}.
\newblock \emph{\bibinfo{journal}{Chemical science}}
  \textbf{\bibinfo{volume}{12}}, \bibinfo{pages}{11473--11483}
  (\bibinfo{year}{2021}).

\bibitem{yan2020retroxpert}
\bibinfo{author}{Yan, C.} \emph{et~al.}
\newblock \bibinfo{title}{Retroxpert: Decompose retrosynthesis prediction like
  {A} chemist}.
\newblock In \bibinfo{editor}{Larochelle, H.}, \bibinfo{editor}{Ranzato, M.},
  \bibinfo{editor}{Hadsell, R.}, \bibinfo{editor}{Balcan, M.} \&
  \bibinfo{editor}{Lin, H.} (eds.) \emph{\bibinfo{booktitle}{Advances in Neural
  Information Processing Systems 33: Annual Conference on Neural Information
  Processing Systems 2020, NeurIPS 2020, December 6-12, 2020, virtual}}
  (\bibinfo{year}{2020}).
\newblock
  \urlprefix\url{https://proceedings.neurips.cc/paper/2020/hash/819f46e52c25763a55cc642422644317-Abstract.html}.

\bibitem{axelrod2021excited}
\bibinfo{author}{Axelrod, S.}, \bibinfo{author}{Shakhnovich, E.} \&
  \bibinfo{author}{G{\'o}mez-Bombarelli, R.}
\newblock \bibinfo{title}{Excited state non-adiabatic dynamics of large
  photoswitchable molecules using a chemically transferable machine learning
  potential}.
\newblock \emph{\bibinfo{journal}{Nature communications}}
  \textbf{\bibinfo{volume}{13}}, \bibinfo{pages}{1--11} (\bibinfo{year}{2022}).

\bibitem{li2020coarse}
\bibinfo{author}{Li, Z.} \emph{et~al.}
\newblock \bibinfo{title}{Graph neural network based coarse-grained mapping
  prediction}.
\newblock \emph{\bibinfo{journal}{Chem. Sci.}} \textbf{\bibinfo{volume}{11}},
  \bibinfo{pages}{9524--9531} (\bibinfo{year}{2020}).
\newblock \urlprefix\url{http://dx.doi.org/10.1039/D0SC02458A}.

\bibitem{gao2021gnes}
\bibinfo{author}{Gao, Y.} \emph{et~al.}
\newblock \bibinfo{title}{Gnes: Learning to explain graph neural networks}.
\newblock In \emph{\bibinfo{booktitle}{2021 IEEE International Conference on
  Data Mining (ICDM)}}, \bibinfo{pages}{131--140}
  (\bibinfo{organization}{IEEE}, \bibinfo{year}{2021}).

\bibitem{wangCombiningCrystalGraphs2022}
\bibinfo{author}{Wang, R.} \emph{et~al.}
\newblock \bibinfo{title}{Combining crystal graphs and domain knowledge in
  machine learning to predict metal-organic frameworks performance in methane
  adsorption}.
\newblock \emph{\bibinfo{journal}{Microporous and Mesoporous Materials}}
  \textbf{\bibinfo{volume}{331}} (\bibinfo{year}{2022}).

\bibitem{wangDeepTMCDeepLearning2022}
\bibinfo{author}{Wang, Z.}, \bibinfo{author}{Han, Y.}, \bibinfo{author}{Cai,
  J.}, \bibinfo{author}{Wu, S.} \& \bibinfo{author}{Li, J.}
\newblock \bibinfo{title}{{{DeepTMC}}: {{A}} deep learning platform to targeted
  design doped transition metal compounds}.
\newblock \emph{\bibinfo{journal}{Energy Storage Materials}}
  \textbf{\bibinfo{volume}{45}}, \bibinfo{pages}{1201--1211}
  (\bibinfo{year}{2022}).

\bibitem{freyMachineLearningEnabledDesign2020}
\bibinfo{author}{Frey, N.~C.}, \bibinfo{author}{Akinwande, D.},
  \bibinfo{author}{Jariwala, D.} \& \bibinfo{author}{Shenoy, V.~B.}
\newblock \bibinfo{title}{Machine {{Learning-Enabled Design}} of {{Point
  Defects}} in {{2D Materials}} for {{Quantum}} and {{Neuromorphic Information
  Processing}}}.
\newblock \emph{\bibinfo{journal}{ACS Nano}} \textbf{\bibinfo{volume}{14}},
  \bibinfo{pages}{13406--13417} (\bibinfo{year}{2020}).

\bibitem{daiGraphNeuralNetworks2021}
\bibinfo{author}{Dai, M.}, \bibinfo{author}{Demirel, M.},
  \bibinfo{author}{Liang, Y.} \& \bibinfo{author}{Hu, J.-M.}
\newblock \bibinfo{title}{Graph neural networks for an accurate and
  interpretable prediction of the properties of polycrystalline materials}.
\newblock \emph{\bibinfo{journal}{npj Computational Materials}}
  \textbf{\bibinfo{volume}{7}} (\bibinfo{year}{2021}).

\bibitem{swansonDeepLearningAutomated2020}
\bibinfo{author}{Swanson, K.}, \bibinfo{author}{Trivedi, S.},
  \bibinfo{author}{Lequieu, J.}, \bibinfo{author}{Swanson, K.} \&
  \bibinfo{author}{Kondor, R.}
\newblock \bibinfo{title}{Deep learning for automated classification and
  characterization of amorphous materials}.
\newblock \emph{\bibinfo{journal}{Soft Matter}} \textbf{\bibinfo{volume}{16}},
  \bibinfo{pages}{435--446} (\bibinfo{year}{2020}).

\bibitem{Pham2021}
\bibinfo{author}{Pham, T.-H.}, \bibinfo{author}{Qiu, Y.},
  \bibinfo{author}{Zeng, J.}, \bibinfo{author}{Xie, L.} \&
  \bibinfo{author}{Zhang, P.}
\newblock \bibinfo{title}{A deep learning framework for high-throughput
  mechanism-driven phenotype compound screening and its application to covid-19
  drug repurposing}.
\newblock \emph{\bibinfo{journal}{Nature Machine Intelligence}}
  \textbf{\bibinfo{volume}{3}}, \bibinfo{pages}{247--257}
  (\bibinfo{year}{2021}).
\newblock \urlprefix\url{https://doi.org/10.1038/s42256-020-00285-9}.

\bibitem{John2019HighThroughputPolymer}
\bibinfo{author}{St.~John, P.~C.} \emph{et~al.}
\newblock \bibinfo{title}{Message-passing neural networks for high-throughput
  polymer screening}.
\newblock \emph{\bibinfo{journal}{The Journal of Chemical Physics}}
  \textbf{\bibinfo{volume}{150}}, \bibinfo{pages}{234111}
  (\bibinfo{year}{2019}).
\newblock \urlprefix\url{http://dx.doi.org/10.1063/1.5099132}.

\bibitem{Peng2020drug}
\bibinfo{author}{Peng, Y.} \emph{et~al.}
\newblock \bibinfo{title}{Enhanced graph isomorphism network for molecular
  admet properties prediction}.
\newblock \emph{\bibinfo{journal}{IEEE Access}} \textbf{\bibinfo{volume}{8}},
  \bibinfo{pages}{168344--168360} (\bibinfo{year}{2020}).

\bibitem{Montanari2020drug}
\bibinfo{author}{Montanari, F.}, \bibinfo{author}{Kuhnke, L.},
  \bibinfo{author}{Ter~Laak, A.} \& \bibinfo{author}{Clevert, D.-A.}
\newblock \bibinfo{title}{Modeling physico-chemical admet endpoints with
  multitask graph convolutional networks}.
\newblock \emph{\bibinfo{journal}{Molecules}} \textbf{\bibinfo{volume}{25}}
  (\bibinfo{year}{2020}).
\newblock \urlprefix\url{https://www.mdpi.com/1420-3049/25/1/44}.

\bibitem{Yaowen2021drug}
\bibinfo{author}{Yaowen, G.}, \bibinfo{author}{Bowen, Z.}, \bibinfo{author}{Si,
  Z.}, \bibinfo{author}{Fengchun, Y.} \& \bibinfo{author}{Jiao, L.}
\newblock \bibinfo{title}{Predicting drug admet properties based on graph
  attention network}.
\newblock \emph{\bibinfo{journal}{Data Analysis and Knowledge Discovery}}
  \textbf{\bibinfo{volume}{5}}, \bibinfo{pages}{76--85} (\bibinfo{year}{2021}).

\bibitem{xiong2021graph}
\bibinfo{author}{Xiong, J.}, \bibinfo{author}{Xiong, Z.},
  \bibinfo{author}{Chen, K.}, \bibinfo{author}{Jiang, H.} \&
  \bibinfo{author}{Zheng, M.}
\newblock \bibinfo{title}{Graph neural networks for automated de novo drug
  design}.
\newblock \emph{\bibinfo{journal}{Drug Discovery Today}}
  \textbf{\bibinfo{volume}{26}}, \bibinfo{pages}{1382--1393}
  (\bibinfo{year}{2021}).
\newblock
  \urlprefix\url{https://www.sciencedirect.com/science/article/pii/S1359644621000787}.

\bibitem{cheung2020drugcovid}
\bibinfo{author}{Cheung, M.} \& \bibinfo{author}{Moura, J.~M.}
\newblock \bibinfo{title}{Graph neural networks for covid-19 drug discovery}.
\newblock In \emph{\bibinfo{booktitle}{2020 IEEE International Conference on
  Big Data (Big Data)}}, \bibinfo{pages}{5646--5648}
  (\bibinfo{organization}{IEEE}, \bibinfo{year}{2020}).

\bibitem{YU2021drugcovid}
\bibinfo{author}{Yu, X.}, \bibinfo{author}{Lu, S.}, \bibinfo{author}{Guo, L.},
  \bibinfo{author}{Wang, S.-H.} \& \bibinfo{author}{Zhang, Y.-D.}
\newblock \bibinfo{title}{Resgnet-c: A graph convolutional neural network for
  detection of covid-19}.
\newblock \emph{\bibinfo{journal}{Neurocomputing}}
  \textbf{\bibinfo{volume}{452}}, \bibinfo{pages}{592--605}
  (\bibinfo{year}{2021}).
\newblock
  \urlprefix\url{https://www.sciencedirect.com/science/article/pii/S0925231220319184}.

\bibitem{KUMAR2022drugcovid}
\bibinfo{author}{Kumar, A.}, \bibinfo{author}{Tripathi, A.~R.},
  \bibinfo{author}{Satapathy, S.~C.} \& \bibinfo{author}{Zhang, Y.-D.}
\newblock \bibinfo{title}{Sars-net: Covid-19 detection from chest x-rays by
  combining graph convolutional network and convolutional neural network}.
\newblock \emph{\bibinfo{journal}{Pattern Recognition}}
  \textbf{\bibinfo{volume}{122}}, \bibinfo{pages}{108255}
  (\bibinfo{year}{2022}).
\newblock
  \urlprefix\url{https://www.sciencedirect.com/science/article/pii/S0031320321004350}.

\bibitem{nakata_pubchemqc_2017}
\bibinfo{author}{Nakata, M.} \& \bibinfo{author}{Shimazaki, T.}
\newblock \bibinfo{title}{{PubChemQC} {Project}: {A} {Large}-{Scale}
  {First}-{Principles} {Electronic} {Structure} {Database} for {Data}-{Driven}
  {Chemistry}}.
\newblock \emph{\bibinfo{journal}{Journal of Chemical Information and
  Modeling}} \textbf{\bibinfo{volume}{57}}, \bibinfo{pages}{1300--1308}
  (\bibinfo{year}{2017}).
\newblock \urlprefix\url{https://doi.org/10.1021/acs.jcim.7b00083}.
\newblock \bibinfo{note}{Publisher: American Chemical Society}.

\bibitem{lee2021opt}
\bibinfo{author}{Lee, C.-K.} \emph{et~al.}
\newblock \bibinfo{title}{Transfer learning with graph neural networks for
  optoelectronic properties of conjugated oligomers}.
\newblock \emph{\bibinfo{journal}{The Journal of Chemical Physics}}
  \textbf{\bibinfo{volume}{154}}, \bibinfo{pages}{024906}
  (\bibinfo{year}{2021}).
\newblock \urlprefix\url{https://doi.org/10.1063/5.0037863}.

\bibitem{lu2020optoelec}
\bibinfo{author}{Lu, C.} \emph{et~al.}
\newblock \bibinfo{title}{Deep learning for optoelectronic properties of
  organic semiconductors}.
\newblock \emph{\bibinfo{journal}{The Journal of Physical Chemistry C}}
  \textbf{\bibinfo{volume}{124}}, \bibinfo{pages}{7048--7060}
  (\bibinfo{year}{2020}).
\newblock \urlprefix\url{https://doi.org/10.1021/acs.jpcc.0c00329}.
\newblock \eprint{https://doi.org/10.1021/acs.jpcc.0c00329}.

\bibitem{pronobis2018}
\bibinfo{author}{Pronobis, W.}, \bibinfo{author}{Sch{\"u}tt, K.~T.},
  \bibinfo{author}{Tkatchenko, A.} \& \bibinfo{author}{M{\"u}ller, K.-R.}
\newblock \bibinfo{title}{Capturing intensive and extensive dft/tddft molecular
  properties with machine learning}.
\newblock \emph{\bibinfo{journal}{The European Physical Journal B}}
  \textbf{\bibinfo{volume}{91}}, \bibinfo{pages}{178} (\bibinfo{year}{2018}).
\newblock \urlprefix\url{https://doi.org/10.1140/epjb/e2018-90148-y}.

\bibitem{atz2021}
\bibinfo{author}{Atz, K.}, \bibinfo{author}{Isert, C.},
  \bibinfo{author}{B{\"o}cker, M.~N.}, \bibinfo{author}{Jim{\'e}nez-Luna, J.}
  \& \bibinfo{author}{Schneider, G.}
\newblock \bibinfo{title}{$\delta$-quantum machine learning for medicinal
  chemistry}.
\newblock \emph{\bibinfo{journal}{ChemRxiv}}  (\bibinfo{year}{2021}).
\newblock
  \urlprefix\url{https://chemrxiv.org/engage/chemrxiv/article-details/61c02f7e7f367e306759a0fd}.

\bibitem{nakata_pubchemqc_2020}
\bibinfo{author}{Nakata, M.}, \bibinfo{author}{Shimazaki, T.},
  \bibinfo{author}{Hashimoto, M.} \& \bibinfo{author}{Maeda, T.}
\newblock \bibinfo{title}{{PubChemQC} {PM6}: {Data} {Sets} of 221 {Million}
  {Molecules} with {Optimized} {Molecular} {Geometries} and {Electronic}
  {Properties}}.
\newblock \emph{\bibinfo{journal}{Journal of Chemical Information and
  Modeling}} \textbf{\bibinfo{volume}{60}}, \bibinfo{pages}{5891--5899}
  (\bibinfo{year}{2020}).
\newblock \urlprefix\url{https://doi.org/10.1021/acs.jcim.0c00740}.
\newblock \bibinfo{note}{Publisher: American Chemical Society}.

\bibitem{gastegger2020basis}
\bibinfo{author}{Gastegger, M.}, \bibinfo{author}{McSloy, A.},
  \bibinfo{author}{Luya, M.}, \bibinfo{author}{Schütt, K.~T.} \&
  \bibinfo{author}{Maurer, R.~J.}
\newblock \bibinfo{title}{A deep neural network for molecular wave functions in
  quasi-atomic minimal basis representation}.
\newblock \emph{\bibinfo{journal}{The Journal of Chemical Physics}}
  \textbf{\bibinfo{volume}{153}}, \bibinfo{pages}{044123}
  (\bibinfo{year}{2020}).
\newblock \urlprefix\url{http://dx.doi.org/10.1063/5.0012911}.

\bibitem{burkholz2021}
\bibinfo{author}{Burkholz, R.}, \bibinfo{author}{Quackenbush, J.} \&
  \bibinfo{author}{Bojar, D.}
\newblock \bibinfo{title}{Using graph convolutional neural networks to learn a
  representation for glycans}.
\newblock \emph{\bibinfo{journal}{Cell Reports}} \textbf{\bibinfo{volume}{35}},
  \bibinfo{pages}{109251} (\bibinfo{year}{2021}).
\newblock \urlprefix\url{https://doi.org/10.1016/j.celrep.2021.109251}.

\bibitem{li2021fingerprint}
\bibinfo{author}{Li, S.} \emph{et~al.}
\newblock \bibinfo{title}{{MutagenPred}-{GCNNs}: A graph convolutional neural
  network-based classification model for mutagenicity prediction with
  data-driven molecular fingerprints}.
\newblock \emph{\bibinfo{journal}{Interdisciplinary Sciences: Computational
  Life Sciences}} \textbf{\bibinfo{volume}{13}}, \bibinfo{pages}{25--33}
  (\bibinfo{year}{2021}).
\newblock \urlprefix\url{https://doi.org/10.1007/s12539-020-00407-2}.

\bibitem{deng2021xgraphboost}
\bibinfo{author}{Deng, D.} \emph{et~al.}
\newblock \bibinfo{title}{Xgraphboost: Extracting graph neural network-based
  features for a better prediction of molecular properties}.
\newblock \emph{\bibinfo{journal}{Journal of Chemical Information and
  Modeling}} \textbf{\bibinfo{volume}{61}}, \bibinfo{pages}{2697--2705}
  (\bibinfo{year}{2021}).
\newblock \urlprefix\url{https://doi.org/10.1021/acs.jcim.0c01489}.
\newblock \bibinfo{note}{PMID: 34009965},
  \eprint{https://doi.org/10.1021/acs.jcim.0c01489}.

\bibitem{schweidtmann2020fuel}
\bibinfo{author}{Schweidtmann, A.~M.} \emph{et~al.}
\newblock \bibinfo{title}{Graph neural networks for prediction of fuel ignition
  quality}.
\newblock \emph{\bibinfo{journal}{Energy \& Fuels}}
  \textbf{\bibinfo{volume}{34}}, \bibinfo{pages}{11395--11407}
  (\bibinfo{year}{2020}).
\newblock \urlprefix\url{https://doi.org/10.1021/acs.energyfuels.0c01533}.
\newblock \eprint{https://doi.org/10.1021/acs.energyfuels.0c01533}.

\bibitem{kim2020gcicenet}
\bibinfo{author}{Kim, Q.}, \bibinfo{author}{Ko, J.-H.}, \bibinfo{author}{Kim,
  S.} \& \bibinfo{author}{Jhe, W.}
\newblock \bibinfo{title}{Gcicenet: a graph convolutional network for accurate
  classification of water phases}.
\newblock \emph{\bibinfo{journal}{Physical Chemistry Chemical Physics}}
  \textbf{\bibinfo{volume}{22}}, \bibinfo{pages}{26340–26350}
  (\bibinfo{year}{2020}).
\newblock \urlprefix\url{http://dx.doi.org/10.1039/D0CP03456H}.

\bibitem{ying2019gnnexplainer}
\bibinfo{author}{Ying, Z.}, \bibinfo{author}{Bourgeois, D.},
  \bibinfo{author}{You, J.}, \bibinfo{author}{Zitnik, M.} \&
  \bibinfo{author}{Leskovec, J.}
\newblock \bibinfo{title}{Gnnexplainer: Generating explanations for graph
  neural networks}.
\newblock \emph{\bibinfo{journal}{Advances in neural information processing
  systems}} \textbf{\bibinfo{volume}{32}} (\bibinfo{year}{2019}).

\bibitem{sanchez2019machine}
\bibinfo{author}{Sanchez-Lengeling, B.} \emph{et~al.}
\newblock \bibinfo{title}{Machine learning for scent: Learning generalizable
  perceptual representations of small molecules}.
\newblock \emph{\bibinfo{journal}{arXiv preprint arXiv:1910.10685}}
  (\bibinfo{year}{2019}).

\bibitem{yuan2022explainable}
\bibinfo{author}{Yuan, Q.}, \bibinfo{author}{Szczypi{\'n}ski, F.~T.} \&
  \bibinfo{author}{Jelfs, K.~E.}
\newblock \bibinfo{title}{Explainable graph neural networks for organic cages}.
\newblock \emph{\bibinfo{journal}{Digital discovery}}
  \textbf{\bibinfo{volume}{1}}, \bibinfo{pages}{127--138}
  (\bibinfo{year}{2022}).

\bibitem{haghighatlari2021newtonnet}
\bibinfo{author}{Haghighatlari, M.} \emph{et~al.}
\newblock \bibinfo{title}{Newtonnet: A newtonian message passing network for
  deep learning of interatomic potentials and forces}.
\newblock \emph{\bibinfo{journal}{Digital Discovery}}  (\bibinfo{year}{2022}).

\bibitem{unke2021spookynet}
\bibinfo{author}{Unke, O.~T.} \emph{et~al.}
\newblock \bibinfo{title}{Spookynet: Learning force fields with electronic
  degrees of freedom and nonlocal effects}.
\newblock \emph{\bibinfo{journal}{Nature Communications}}
  \textbf{\bibinfo{volume}{12}} (\bibinfo{year}{2021}).
\newblock \urlprefix\url{http://dx.doi.org/10.1038/s41467-021-27504-0}.

\bibitem{Wang.2019}
\bibinfo{author}{Wang, J.} \emph{et~al.}
\newblock \bibinfo{title}{Machine learning of coarse-grained molecular dynamics
  force fields}.
\newblock \emph{\bibinfo{journal}{ACS central science}}
  \textbf{\bibinfo{volume}{5}}, \bibinfo{pages}{755--767}
  (\bibinfo{year}{2019}).

\bibitem{husic2020coarse}
\bibinfo{author}{Husic, B.~E.} \emph{et~al.}
\newblock \bibinfo{title}{Coarse graining molecular dynamics with graph neural
  networks}.
\newblock \emph{\bibinfo{journal}{The Journal of chemical physics}}
  \textbf{\bibinfo{volume}{153}}, \bibinfo{pages}{194101}
  (\bibinfo{year}{2020}).

\bibitem{westermayr_deep_2020}
\bibinfo{author}{Westermayr, J.} \& \bibinfo{author}{Marquetand, P.}
\newblock \bibinfo{title}{Deep learning for uv absorption spectra with schnarc:
  First steps toward transferability in chemical compound space}.
\newblock \emph{\bibinfo{journal}{The Journal of Chemical Physics}}
  \textbf{\bibinfo{volume}{153}}, \bibinfo{pages}{154112}
  (\bibinfo{year}{2020}).

\bibitem{Westermayr.2020b}
\bibinfo{author}{Westermayr, J.}, \bibinfo{author}{Gastegger, M.} \&
  \bibinfo{author}{Marquetand, P.}
\newblock \bibinfo{title}{Combining schnet and sharc: The schnarc machine
  learning approach for excited-state dynamics}.
\newblock \emph{\bibinfo{journal}{The journal of physical chemistry letters}}
  \textbf{\bibinfo{volume}{11}}, \bibinfo{pages}{3828--3834}
  (\bibinfo{year}{2020}).

\bibitem{tabor2018accelerating}
\bibinfo{author}{Tabor, D.~P.} \emph{et~al.}
\newblock \bibinfo{title}{Accelerating the discovery of materials for clean
  energy in the era of smart automation}.
\newblock \emph{\bibinfo{journal}{Nature Reviews Materials}}
  \textbf{\bibinfo{volume}{3}}, \bibinfo{pages}{5--20} (\bibinfo{year}{2018}).

\bibitem{pande2017sm}
\bibinfo{author}{Liu, B.} \emph{et~al.}
\newblock \bibinfo{title}{Retrosynthetic reaction prediction using neural
  sequence-to-sequence models}.
\newblock \emph{\bibinfo{journal}{ACS Central Science}}
  \textbf{\bibinfo{volume}{3}}, \bibinfo{pages}{1103--1113}
  (\bibinfo{year}{2017}).
\newblock \urlprefix\url{https://doi.org/10.1021/acscentsci.7b00303}.
\newblock \bibinfo{note}{PMID: 29104927},
  \eprint{https://doi.org/10.1021/acscentsci.7b00303}.

\bibitem{yang2020sm}
\bibinfo{author}{Zheng, S.}, \bibinfo{author}{Rao, J.}, \bibinfo{author}{Zhang,
  Z.}, \bibinfo{author}{Xu, J.} \& \bibinfo{author}{Yang, Y.}
\newblock \bibinfo{title}{Predicting retrosynthetic reactions using
  self-corrected transformer neural networks}.
\newblock \emph{\bibinfo{journal}{J. Chem. Inf. Model.}}
  \textbf{\bibinfo{volume}{60}}, \bibinfo{pages}{47–55}
  (\bibinfo{year}{2020}).

\bibitem{lai2020sm}
\bibinfo{author}{Lin, K.}, \bibinfo{author}{Xu, Y.}, \bibinfo{author}{Pei, J.}
  \& \bibinfo{author}{Lai, L.}
\newblock \bibinfo{title}{Automatic retrosynthetic route planning using
  template-free models}.
\newblock \emph{\bibinfo{journal}{Chem. Sci.}} \textbf{\bibinfo{volume}{11}},
  \bibinfo{pages}{3355--3364} (\bibinfo{year}{2020}).
\newblock \urlprefix\url{http://dx.doi.org/10.1039/C9SC03666K}.

\bibitem{petraglia2020sm}
\bibinfo{author}{Schwaller, P.} \emph{et~al.}
\newblock \bibinfo{title}{Predicting retrosynthetic pathways using
  transformer-based models and a hyper-graph exploration strategy}.
\newblock \emph{\bibinfo{journal}{Chem. Sci.}} \textbf{\bibinfo{volume}{11}},
  \bibinfo{pages}{3316--3325} (\bibinfo{year}{2020}).
\newblock \urlprefix\url{http://dx.doi.org/10.1039/C9SC05704H}.

\bibitem{waller2017fp}
\bibinfo{author}{Segler, M. H.~S.} \& \bibinfo{author}{Waller, M.~P.}
\newblock \bibinfo{title}{Neural-symbolic machine learning for retrosynthesis
  and reaction prediction}.
\newblock \emph{\bibinfo{journal}{Chemistry – A European Journal}}
  \textbf{\bibinfo{volume}{23}}, \bibinfo{pages}{5966--5971}
  (\bibinfo{year}{2017}).
\newblock
  \urlprefix\url{https://chemistry-europe.onlinelibrary.wiley.com/doi/abs/10.1002/chem.201605499}.
\newblock
  \eprint{https://chemistry-europe.onlinelibrary.wiley.com/doi/pdf/10.1002/chem.201605499}.

\bibitem{waller2018fp}
\bibinfo{author}{Segler, M. H.~S.} \& \bibinfo{author}{Waller, M.~P.}
\newblock \bibinfo{title}{Planning chemical syntheses with deep neural networks
  and symbolic ai}.
\newblock \emph{\bibinfo{journal}{Nature}} \textbf{\bibinfo{volume}{555}},
  \bibinfo{pages}{604–610} (\bibinfo{year}{2018}).

\bibitem{jin2017predicting}
\bibinfo{author}{Jin, W.}, \bibinfo{author}{Coley, C.~W.},
  \bibinfo{author}{Barzilay, R.} \& \bibinfo{author}{Jaakkola, T.~S.}
\newblock \bibinfo{title}{Predicting organic reaction outcomes with
  weisfeiler-lehman network}.
\newblock In \bibinfo{editor}{Guyon, I.} \emph{et~al.} (eds.)
  \emph{\bibinfo{booktitle}{Advances in Neural Information Processing Systems
  30: Annual Conference on Neural Information Processing Systems 2017, December
  4-9, 2017, Long Beach, CA, {USA}}}, \bibinfo{pages}{2607--2616}
  (\bibinfo{year}{2017}).
\newblock
  \urlprefix\url{https://proceedings.neurips.cc/paper/2017/hash/ced556cd9f9c0c8315cfbe0744a3baf0-Abstract.html}.

\bibitem{jin2019predicting}
\bibinfo{author}{Coley, C.~W.} \emph{et~al.}
\newblock \bibinfo{title}{A graph-convolutional neural network model for the
  prediction of chemical reactivity}.
\newblock \emph{\bibinfo{journal}{Chem. Sci.}} \textbf{\bibinfo{volume}{10}},
  \bibinfo{pages}{370--377} (\bibinfo{year}{2019}).
\newblock \urlprefix\url{http://dx.doi.org/10.1039/C8SC04228D}.

\bibitem{struble2020wln}
\bibinfo{author}{Struble, T.~J.}, \bibinfo{author}{Coley, C.~W.} \&
  \bibinfo{author}{Jensen, K.~F.}
\newblock \bibinfo{title}{Multitask prediction of site selectivity in aromatic
  c–h functionalization reactions}.
\newblock \emph{\bibinfo{journal}{React. Chem. Eng.}}
  \textbf{\bibinfo{volume}{5}}, \bibinfo{pages}{896--902}
  (\bibinfo{year}{2020}).
\newblock \urlprefix\url{http://dx.doi.org/10.1039/D0RE00071J}.

\bibitem{guan2021wln}
\bibinfo{author}{Guan, Y.} \emph{et~al.}
\newblock \bibinfo{title}{Regio-selectivity prediction with a machine-learned
  reaction representation and on-the-fly quantum mechanical descriptors}.
\newblock \emph{\bibinfo{journal}{Chem. Sci.}} \textbf{\bibinfo{volume}{12}},
  \bibinfo{pages}{2198--2208} (\bibinfo{year}{2021}).
\newblock \urlprefix\url{http://dx.doi.org/10.1039/D0SC04823B}.

\bibitem{nikitin2020wln}
\bibinfo{author}{Nikitin, F.}, \bibinfo{author}{Isayev, O.} \&
  \bibinfo{author}{Strijov, V.}
\newblock \bibinfo{title}{Dracon: disconnected graph neural network for atom
  mapping in chemical reactions}.
\newblock \emph{\bibinfo{journal}{Phys. Chem. Chem. Phys.}}
  \textbf{\bibinfo{volume}{22}}, \bibinfo{pages}{26478--26486}
  (\bibinfo{year}{2020}).
\newblock \urlprefix\url{http://dx.doi.org/10.1039/D0CP04748A}.

\bibitem{do2018graph}
\bibinfo{author}{Do, K.}, \bibinfo{author}{Tran, T.} \&
  \bibinfo{author}{Venkatesh, S.}
\newblock \bibinfo{title}{Graph transformation policy network for chemical
  reaction prediction}.
\newblock In \bibinfo{editor}{Teredesai, A.} \emph{et~al.} (eds.)
  \emph{\bibinfo{booktitle}{Proceedings of the 25th {ACM} {SIGKDD}
  International Conference on Knowledge Discovery {\&} Data Mining, {KDD} 2019,
  Anchorage, AK, USA, August 4-8, 2019}}, \bibinfo{pages}{750--760}
  (\bibinfo{publisher}{{ACM}}, \bibinfo{year}{2019}).
\newblock \urlprefix\url{https://doi.org/10.1145/3292500.3330958}.

\bibitem{bradshaw2019generative}
\bibinfo{author}{Bradshaw, J.}, \bibinfo{author}{Kusner, M.~J.},
  \bibinfo{author}{Paige, B.}, \bibinfo{author}{Segler, M. H.~S.} \&
  \bibinfo{author}{Hern{\'{a}}ndez{-}Lobato, J.~M.}
\newblock \bibinfo{title}{A generative model for electron paths}.
\newblock In \emph{\bibinfo{booktitle}{7th International Conference on Learning
  Representations, {ICLR} 2019, New Orleans, LA, USA, May 6-9, 2019}}
  (\bibinfo{publisher}{OpenReview.net}, \bibinfo{year}{2019}).
\newblock \urlprefix\url{https://openreview.net/forum?id=r1x4BnCqKX}.

\bibitem{wen2021bond}
\bibinfo{author}{Wen, M.}, \bibinfo{author}{Blau, S.~M.},
  \bibinfo{author}{Spotte-Smith, E. W.~C.}, \bibinfo{author}{Dwaraknath, S.} \&
  \bibinfo{author}{Persson, K.~A.}
\newblock \bibinfo{title}{Bondnet: a graph neural network for the prediction of
  bond dissociation energies for charged molecules}.
\newblock \emph{\bibinfo{journal}{Chem. Sci.}} \textbf{\bibinfo{volume}{12}},
  \bibinfo{pages}{1858--1868} (\bibinfo{year}{2021}).
\newblock \urlprefix\url{http://dx.doi.org/10.1039/D0SC05251E}.

\bibitem{john2020bond}
\bibinfo{author}{St~John, P.~C.}, \bibinfo{author}{Guan, Y.},
  \bibinfo{author}{Kim, Y.}, \bibinfo{author}{Kim, S.} \&
  \bibinfo{author}{Paton, R.~S.}
\newblock \bibinfo{title}{Prediction of organic homolytic bond dissociation
  enthalpies at near chemical accuracy with sub-second computational cost}.
\newblock \emph{\bibinfo{journal}{Nature communications}}
  \textbf{\bibinfo{volume}{11}}, \bibinfo{pages}{1--12} (\bibinfo{year}{2020}).

\bibitem{pattanaik2020ts}
\bibinfo{author}{Pattanaik, L.}, \bibinfo{author}{Ingraham, J.~B.},
  \bibinfo{author}{Grambow, C.~A.} \& \bibinfo{author}{Green, W.~H.}
\newblock \bibinfo{title}{Generating transition states of isomerization
  reactions with deep learning}.
\newblock \emph{\bibinfo{journal}{Phys. Chem. Chem. Phys.}}
  \textbf{\bibinfo{volume}{22}}, \bibinfo{pages}{23618--23626}
  (\bibinfo{year}{2020}).
\newblock \urlprefix\url{http://dx.doi.org/10.1039/D0CP04670A}.

\bibitem{grambow2020ae}
\bibinfo{author}{Grambow, C.~A.}, \bibinfo{author}{Pattanaik, L.} \&
  \bibinfo{author}{Green, W.~H.}
\newblock \bibinfo{title}{Deep learning of activation energies}.
\newblock \emph{\bibinfo{journal}{The Journal of Physical Chemistry Letters}}
  \textbf{\bibinfo{volume}{11}}, \bibinfo{pages}{2992--2997}
  (\bibinfo{year}{2020}).
\newblock \urlprefix\url{https://doi.org/10.1021/acs.jpclett.0c00500}.
\newblock \bibinfo{note}{PMID: 32216310},
  \eprint{https://doi.org/10.1021/acs.jpclett.0c00500}.

\bibitem{dai2020retrosynthesis}
\bibinfo{author}{Dai, H.}, \bibinfo{author}{Li, C.}, \bibinfo{author}{Coley,
  C.~W.}, \bibinfo{author}{Dai, B.} \& \bibinfo{author}{Song, L.}
\newblock \bibinfo{title}{Retrosynthesis prediction with conditional graph
  logic network}.
\newblock In \bibinfo{editor}{Wallach, H.~M.} \emph{et~al.} (eds.)
  \emph{\bibinfo{booktitle}{Advances in Neural Information Processing Systems
  32: Annual Conference on Neural Information Processing Systems 2019, NeurIPS
  2019, December 8-14, 2019, Vancouver, BC, Canada}},
  \bibinfo{pages}{8870--8880} (\bibinfo{year}{2019}).
\newblock
  \urlprefix\url{https://proceedings.neurips.cc/paper/2019/hash/0d2b2061826a5df3221116a5085a6052-Abstract.html}.

\bibitem{ishida2019retro}
\bibinfo{author}{Ishida, S.}, \bibinfo{author}{Terayama, K.},
  \bibinfo{author}{Kojima, R.}, \bibinfo{author}{Takasu, K.} \&
  \bibinfo{author}{Okuno, Y.}
\newblock \bibinfo{title}{Prediction and interpretable visualization of
  retrosynthetic reactions using graph convolutional networks}.
\newblock \emph{\bibinfo{journal}{Journal of Chemical Information and
  Modeling}} \textbf{\bibinfo{volume}{59}}, \bibinfo{pages}{5026--5033}
  (\bibinfo{year}{2019}).
\newblock \urlprefix\url{https://doi.org/10.1021/acs.jcim.9b00538}.
\newblock \bibinfo{note}{PMID: 31769668},
  \eprint{https://doi.org/10.1021/acs.jcim.9b00538}.

\bibitem{chen2021retro}
\bibinfo{author}{Chen, S.} \& \bibinfo{author}{Jung, Y.}
\newblock \bibinfo{title}{Deep retrosynthetic reaction prediction using local
  reactivity and global attention}.
\newblock \emph{\bibinfo{journal}{JACS Au}} \textbf{\bibinfo{volume}{1}},
  \bibinfo{pages}{1612--1620} (\bibinfo{year}{2021}).
\newblock \urlprefix\url{https://doi.org/10.1021/jacsau.1c00246}.
\newblock \bibinfo{note}{PMID: 34723264},
  \eprint{https://doi.org/10.1021/jacsau.1c00246}.

\bibitem{somnath2021learning}
\bibinfo{author}{Somnath, V.~R.}, \bibinfo{author}{Bunne, C.},
  \bibinfo{author}{Coley, C.~W.}, \bibinfo{author}{Krause, A.} \&
  \bibinfo{author}{Barzilay, R.}
\newblock \bibinfo{title}{Learning graph models for retrosynthesis prediction}.
\newblock In \bibinfo{editor}{Ranzato, M.}, \bibinfo{editor}{Beygelzimer, A.},
  \bibinfo{editor}{Dauphin, Y.~N.}, \bibinfo{editor}{Liang, P.} \&
  \bibinfo{editor}{Vaughan, J.~W.} (eds.) \emph{\bibinfo{booktitle}{Advances in
  Neural Information Processing Systems 34: Annual Conference on Neural
  Information Processing Systems 2021, NeurIPS 2021, December 6-14, 2021,
  virtual}}, \bibinfo{pages}{9405--9415} (\bibinfo{year}{2021}).
\newblock
  \urlprefix\url{https://proceedings.neurips.cc/paper/2021/hash/4e2a6330465c8ffcaa696a5a16639176-Abstract.html}.

\bibitem{shi2021graph}
\bibinfo{author}{Shi, C.}, \bibinfo{author}{Xu, M.}, \bibinfo{author}{Guo, H.},
  \bibinfo{author}{Zhang, M.} \& \bibinfo{author}{Tang, J.}
\newblock \bibinfo{title}{A graph to graphs framework for retrosynthesis
  prediction}.
\newblock In \emph{\bibinfo{booktitle}{Proceedings of the 37th International
  Conference on Machine Learning, {ICML} 2020, 13-18 July 2020, Virtual
  Event}}, vol. \bibinfo{volume}{119} of \emph{\bibinfo{series}{Proceedings of
  Machine Learning Research}}, \bibinfo{pages}{8818--8827}
  (\bibinfo{publisher}{{PMLR}}, \bibinfo{year}{2020}).
\newblock \urlprefix\url{http://proceedings.mlr.press/v119/shi20d.html}.

\bibitem{bartelCriticalExaminationCompound2020}
\bibinfo{author}{Bartel, C.~J.} \emph{et~al.}
\newblock \bibinfo{title}{A critical examination of compound stability
  predictions from machine-learned formation energies}.
\newblock \emph{\bibinfo{journal}{npj Computational Materials}}
  \textbf{\bibinfo{volume}{6}}, \bibinfo{pages}{1--11} (\bibinfo{year}{2020}).

\bibitem{jorgensenMaterialsPropertyPrediction2019}
\bibinfo{author}{J{\o}rgensen, P.}, \bibinfo{author}{Garijo Del~R{\'i}o, E.},
  \bibinfo{author}{Schmidt, M.} \& \bibinfo{author}{Jacobsen, K.}
\newblock \bibinfo{title}{Materials property prediction using symmetry-labeled
  graphs as atomic position independent descriptors}.
\newblock \emph{\bibinfo{journal}{Physical Review B}}
  \textbf{\bibinfo{volume}{100}} (\bibinfo{year}{2019}).

\bibitem{nohUncertaintyQuantifiedHybridMachine2020}
\bibinfo{author}{Noh, J.}, \bibinfo{author}{Gu, G.}, \bibinfo{author}{Kim, S.}
  \& \bibinfo{author}{Jung, Y.}
\newblock \bibinfo{title}{Uncertainty-{{Quantified Hybrid Machine
  Learning}}/{{Density Functional Theory High Throughput Screening Method}} for
  {{Crystals}}}.
\newblock \emph{\bibinfo{journal}{Journal of Chemical Information and
  Modeling}} \textbf{\bibinfo{volume}{60}}, \bibinfo{pages}{1996--2003}
  (\bibinfo{year}{2020}).

\bibitem{pandeyPredictingEnergyStability2021}
\bibinfo{author}{Pandey, S.}, \bibinfo{author}{Qu, J.},
  \bibinfo{author}{Stevanovi{\'c}, V.}, \bibinfo{author}{St.~John, P.} \&
  \bibinfo{author}{Gorai, P.}
\newblock \bibinfo{title}{Predicting energy and stability of known and
  hypothetical crystals using graph neural network}.
\newblock \emph{\bibinfo{journal}{Patterns}} \textbf{\bibinfo{volume}{2}}
  (\bibinfo{year}{2021}).

\bibitem{jangStructureBasedSynthesizabilityPrediction2020}
\bibinfo{author}{Jang, J.}, \bibinfo{author}{Gu, G.~H.}, \bibinfo{author}{Noh,
  J.}, \bibinfo{author}{Kim, J.} \& \bibinfo{author}{Jung, Y.}
\newblock \bibinfo{title}{Structure-{{Based Synthesizability Prediction}} of
  {{Crystals Using Partially Supervised Learning}}}.
\newblock \emph{\bibinfo{journal}{Journal of the American Chemical Society}}
  \textbf{\bibinfo{volume}{142}}, \bibinfo{pages}{18836--18843}
  (\bibinfo{year}{2020}).

\bibitem{liGraphNetworkBased2021}
\bibinfo{author}{Li, X.-G.} \emph{et~al.}
\newblock \bibinfo{title}{Graph network based deep learning of bandgaps}.
\newblock \emph{\bibinfo{journal}{The Journal of Chemical Physics}}
  \textbf{\bibinfo{volume}{155}}, \bibinfo{pages}{154702}
  (\bibinfo{year}{2021}).

\bibitem{omprakashGraphRepresentationalLearning2021}
\bibinfo{author}{Omprakash, P.} \emph{et~al.}
\newblock \bibinfo{title}{Graph representational learning for bandgap
  prediction in varied perovskite crystals}.
\newblock \emph{\bibinfo{journal}{Computational Materials Science}}
  \textbf{\bibinfo{volume}{196}} (\bibinfo{year}{2021}).

\bibitem{naTuplewiseMaterialRepresentation2020}
\bibinfo{author}{Na, G.~S.}, \bibinfo{author}{Jang, S.}, \bibinfo{author}{Lee,
  Y.-L.} \& \bibinfo{author}{Chang, H.}
\newblock \bibinfo{title}{Tuplewise {{Material Representation Based Machine
  Learning}} for {{Accurate Band Gap Prediction}}}.
\newblock \emph{\bibinfo{journal}{The Journal of Physical Chemistry A}}
  \textbf{\bibinfo{volume}{124}}, \bibinfo{pages}{10616--10623}
  (\bibinfo{year}{2020}).

\bibitem{wangAcceleratingDiscoveryMetalOrganic2020}
\bibinfo{author}{Wang, R.}, \bibinfo{author}{Zhong, Y.}, \bibinfo{author}{Bi,
  L.}, \bibinfo{author}{Yang, M.} \& \bibinfo{author}{Xu, D.}
\newblock \bibinfo{title}{Accelerating {{Discovery}} of {{Metal-Organic
  Frameworks}} for {{Methane Adsorption}} with {{Hierarchical Screening}} and
  {{Deep Learning}}}.
\newblock \emph{\bibinfo{journal}{ACS Applied Materials and Interfaces}}
  \textbf{\bibinfo{volume}{12}}, \bibinfo{pages}{52797--52807}
  (\bibinfo{year}{2020}).

\bibitem{wilmerLargescaleScreeningHypothetical2012}
\bibinfo{author}{Wilmer, C.~E.} \emph{et~al.}
\newblock \bibinfo{title}{Large-scale screening of hypothetical
  metal\textendash organic frameworks}.
\newblock \emph{\bibinfo{journal}{Nature Chemistry}}
  \textbf{\bibinfo{volume}{4}}, \bibinfo{pages}{83--89} (\bibinfo{year}{2012}).

\bibitem{guPracticalDeepLearningRepresentation2020}
\bibinfo{author}{Gu, G.} \emph{et~al.}
\newblock \bibinfo{title}{Practical {{Deep-Learning Representation}} for {{Fast
  Heterogeneous Catalyst Screening}}}.
\newblock \emph{\bibinfo{journal}{Journal of Physical Chemistry Letters}}
  \textbf{\bibinfo{volume}{11}}, \bibinfo{pages}{3185--3191}
  (\bibinfo{year}{2020}).

\bibitem{goodallPredictingMaterialsProperties2020}
\bibinfo{author}{Goodall, R. E.~A.} \& \bibinfo{author}{Lee, A.~A.}
\newblock \bibinfo{title}{Predicting materials properties without crystal
  structure: Deep representation learning from stoichiometry}.
\newblock \emph{\bibinfo{journal}{Nature Communications}}
  \textbf{\bibinfo{volume}{11}}, \bibinfo{pages}{6280} (\bibinfo{year}{2020}).

\bibitem{curtaroloAFLOWAutomaticFramework2012}
\bibinfo{author}{Curtarolo, S.} \emph{et~al.}
\newblock \bibinfo{title}{{{AFLOW}}: {{An}} automatic framework for
  high-throughput materials discovery}.
\newblock \emph{\bibinfo{journal}{Computational Materials Science}}
  \textbf{\bibinfo{volume}{58}}, \bibinfo{pages}{218--226}
  (\bibinfo{year}{2012}).

\bibitem{luCouplingCrystalGraph2020}
\bibinfo{author}{Lu, S.} \emph{et~al.}
\newblock \bibinfo{title}{Coupling a {{Crystal Graph Multilayer Descriptor}} to
  {{Active Learning}} for {{Rapid Discovery}} of {{2D Ferromagnetic
  Semiconductors}}/{{Half-Metals}}/{{Metals}}}.
\newblock \emph{\bibinfo{journal}{Advanced Materials}}
  \textbf{\bibinfo{volume}{32}}, \bibinfo{pages}{2002658}
  (\bibinfo{year}{2020}).

\bibitem{catlow2012defects}
\bibinfo{author}{Catlow, R.}
\newblock \emph{\bibinfo{title}{Defects and disorder in crystalline and
  amorphous solids}}, vol. \bibinfo{volume}{418} (\bibinfo{publisher}{Springer
  Science \& Business Media}, \bibinfo{year}{2012}).

\bibitem{chenLearningPropertiesOrdered2021}
\bibinfo{author}{Chen, C.}, \bibinfo{author}{Zuo, Y.}, \bibinfo{author}{Ye,
  W.}, \bibinfo{author}{Li, X.} \& \bibinfo{author}{Ong, S.~P.}
\newblock \bibinfo{title}{Learning properties of ordered and disordered
  materials from multi-fidelity data}.
\newblock \emph{\bibinfo{journal}{Nature Computational Science}}
  \textbf{\bibinfo{volume}{1}}, \bibinfo{pages}{46--53} (\bibinfo{year}{2021}).

\bibitem{cianAtomisticGraphNeural2021}
\bibinfo{author}{Cian, L.} \emph{et~al.}
\newblock \bibinfo{title}{Atomistic {{Graph Neural Networks}} for metals:
  {{Application}} to bcc iron}.
\newblock \emph{\bibinfo{journal}{arXiv:2109.14012 [cond-mat]}}
  (\bibinfo{year}{2021}).
\newblock \eprint{2109.14012}.

\bibitem{bapstUnveilingPredictivePower2020}
\bibinfo{author}{Bapst, V.} \emph{et~al.}
\newblock \bibinfo{title}{Unveiling the predictive power of static structure in
  glassy systems}.
\newblock \emph{\bibinfo{journal}{Nature Physics}}
  \textbf{\bibinfo{volume}{16}}, \bibinfo{pages}{448--454}
  (\bibinfo{year}{2020}).

\bibitem{wangInverseDesignGlass2021}
\bibinfo{author}{Wang, Q.} \& \bibinfo{author}{Zhang, L.}
\newblock \bibinfo{title}{Inverse design of glass structure with deep graph
  neural networks}.
\newblock \emph{\bibinfo{journal}{Nature communications}}
  \textbf{\bibinfo{volume}{12}}, \bibinfo{pages}{1--11} (\bibinfo{year}{2021}).

\bibitem{park2022prediction}
\bibinfo{author}{Park, J.} \emph{et~al.}
\newblock \bibinfo{title}{Prediction and interpretation of polymer properties
  using the graph convolutional network}.
\newblock \emph{\bibinfo{journal}{ACS Polymers Au}}  (\bibinfo{year}{2022}).

\bibitem{zeng2018graph}
\bibinfo{author}{Zeng, M.} \emph{et~al.}
\newblock \bibinfo{title}{Graph convolutional neural networks for polymers
  property prediction}.
\newblock \emph{\bibinfo{journal}{arXiv preprint arXiv:1811.06231}}
  (\bibinfo{year}{2018}).

\bibitem{deringerGeneralpurposeMachinelearningForce2020}
\bibinfo{author}{Deringer, V.~L.}, \bibinfo{author}{Caro, M.~A.} \&
  \bibinfo{author}{Cs{\'a}nyi, G.}
\newblock \bibinfo{title}{A general-purpose machine-learning force field for
  bulk and nanostructured phosphorus}.
\newblock \emph{\bibinfo{journal}{Nature Communications}}
  \textbf{\bibinfo{volume}{11}}, \bibinfo{pages}{5461} (\bibinfo{year}{2020}).

\bibitem{wangSymmetryadaptedGraphNeural2021}
\bibinfo{author}{Wang, Z.} \emph{et~al.}
\newblock \bibinfo{title}{Symmetry-adapted graph neural networks for
  constructing molecular dynamics force fields}.
\newblock \emph{\bibinfo{journal}{Science China: Physics, Mechanics and
  Astronomy}} \textbf{\bibinfo{volume}{64}} (\bibinfo{year}{2021}).

\bibitem{razaMessagePassingNeural2020}
\bibinfo{author}{Raza, A.}, \bibinfo{author}{Sturluson, A.},
  \bibinfo{author}{Simon, C.~M.} \& \bibinfo{author}{Fern, X.}
\newblock \bibinfo{title}{Message {{Passing Neural Networks}} for {{Partial
  Charge Assignment}} to {{Metal}}\textendash{{Organic Frameworks}}}.
\newblock \emph{\bibinfo{journal}{The Journal of Physical Chemistry C}}
  \textbf{\bibinfo{volume}{124}}, \bibinfo{pages}{19070--19082}
  (\bibinfo{year}{2020}).

\bibitem{malikPredictingOutcomesMaterial2021}
\bibinfo{author}{Malik, S.}, \bibinfo{author}{Goodall, R.} \&
  \bibinfo{author}{Lee, A.}
\newblock \bibinfo{title}{Predicting the {{Outcomes}} of {{Material Syntheses}}
  with {{Deep Learning}}}.
\newblock \emph{\bibinfo{journal}{Chemistry of Materials}}
  \textbf{\bibinfo{volume}{33}}, \bibinfo{pages}{616--624}
  (\bibinfo{year}{2021}).

\bibitem{tremouilhac2017chemotion}
\bibinfo{author}{Tremouilhac, P.} \emph{et~al.}
\newblock \bibinfo{title}{Chemotion eln: an open source electronic lab notebook
  for chemists in academia}.
\newblock \emph{\bibinfo{journal}{Journal of cheminformatics}}
  \textbf{\bibinfo{volume}{9}}, \bibinfo{pages}{1--13} (\bibinfo{year}{2017}).

\bibitem{brandt2021kadi4mat}
\bibinfo{author}{Brandt, N.} \emph{et~al.}
\newblock \bibinfo{title}{Kadi4mat: A research data infrastructure for
  materials science}.
\newblock \emph{\bibinfo{journal}{Data Science Journal}}
  \textbf{\bibinfo{volume}{20}} (\bibinfo{year}{2021}).

\bibitem{xieCrystalDiffusionVariational2021}
\bibinfo{author}{Xie, T.}, \bibinfo{author}{Fu, X.}, \bibinfo{author}{Ganea,
  O.-E.}, \bibinfo{author}{Barzilay, R.} \& \bibinfo{author}{Jaakkola, T.}
\newblock \bibinfo{title}{Crystal {{Diffusion Variational Autoencoder}} for
  {{Periodic Material Generation}}}.
\newblock \emph{\bibinfo{journal}{arXiv:2110.06197 [cond-mat,
  physics:physics]}}  (\bibinfo{year}{2021}).
\newblock \eprint{2110.06197}.

\bibitem{bergerhoffInorganicCrystalStructure1983}
\bibinfo{author}{Bergerhoff, G.}, \bibinfo{author}{Hundt, R.},
  \bibinfo{author}{Sievers, R.} \& \bibinfo{author}{Brown, I.~D.}
\newblock \bibinfo{title}{The inorganic crystal structure data base}.
\newblock \emph{\bibinfo{journal}{Journal of Chemical Information and Computer
  Sciences}} \textbf{\bibinfo{volume}{23}}, \bibinfo{pages}{66--69}
  (\bibinfo{year}{1983}).

\bibitem{pickardAIRSSDataCarbon2020}
\bibinfo{author}{Pickard, C.~J.}
\newblock \bibinfo{title}{{{AIRSS}} data for carbon at {{10GPa}} and the
  {{C}}+{{N}}+{{H}}+{{O}} system at {{1GPa}}} (\bibinfo{year}{2020}).

\bibitem{NRELMaterialsDatabase}
\bibinfo{title}{{{NREL Materials Database}} ({{NRELMatDB}})}.
\newblock \urlprefix\url{https://materials.nrel.gov/}.

\bibitem{tranActiveLearningIntermetallics2018}
\bibinfo{author}{Tran, K.} \& \bibinfo{author}{Ulissi, Z.~W.}
\newblock \bibinfo{title}{Active learning across intermetallics to guide
  discovery of electrocatalysts for {{CO2}} reduction and {{H2}} evolution}.
\newblock \emph{\bibinfo{journal}{Nature Catalysis}}
  \textbf{\bibinfo{volume}{1}}, \bibinfo{pages}{696--703}
  (\bibinfo{year}{2018}).

\bibitem{kononovaTextminedDatasetInorganic2019}
\bibinfo{author}{Kononova, O.} \emph{et~al.}
\newblock \bibinfo{title}{Text-mined dataset of inorganic materials synthesis
  recipes}.
\newblock \emph{\bibinfo{journal}{Scientific Data}}
  \textbf{\bibinfo{volume}{6}}, \bibinfo{pages}{203} (\bibinfo{year}{2019}).

\bibitem{castelliNewCubicPerovskites2012}
\bibinfo{author}{Castelli, I.~E.} \emph{et~al.}
\newblock \bibinfo{title}{New cubic perovskites for one- and two-photon water
  splitting using the computational materials repository}.
\newblock \emph{\bibinfo{journal}{Energy \& Environmental Science}}
  \textbf{\bibinfo{volume}{5}}, \bibinfo{pages}{9034--9043}
  (\bibinfo{year}{2012}).

\bibitem{chungAdvancesUpdatesAnalytics2019}
\bibinfo{author}{Chung, Y.~G.} \emph{et~al.}
\newblock \bibinfo{title}{Advances, {{Updates}}, and {{Analytics}} for the
  {{Computation-Ready}}, {{Experimental Metal}}\textendash{{Organic Framework
  Database}}: {{CoRE MOF}} 2019}.
\newblock \emph{\bibinfo{journal}{Journal of Chemical \& Engineering Data}}
  \textbf{\bibinfo{volume}{64}}, \bibinfo{pages}{5985--5998}
  (\bibinfo{year}{2019}).

\bibitem{dragoniAchievingDFTAccuracy2018}
\bibinfo{author}{Dragoni, D.}, \bibinfo{author}{Daff, T.~D.},
  \bibinfo{author}{Cs{\'a}nyi, G.} \& \bibinfo{author}{Marzari, N.}
\newblock \bibinfo{title}{Achieving {{DFT}} accuracy with a machine-learning
  interatomic potential: {{Thermomechanics}} and defects in bcc ferromagnetic
  iron}.
\newblock \emph{\bibinfo{journal}{Physical Review Materials}}
  \textbf{\bibinfo{volume}{2}}, \bibinfo{pages}{013808} (\bibinfo{year}{2018}).

\bibitem{deringerDataDrivenLearningTotal2018}
\bibinfo{author}{Deringer, V.~L.}, \bibinfo{author}{Pickard, C.~J.} \&
  \bibinfo{author}{Cs{\'a}nyi, G.}
\newblock \bibinfo{title}{Data-{{Driven Learning}} of {{Total}} and {{Local
  Energies}} in {{Elemental Boron}}}.
\newblock \emph{\bibinfo{journal}{Physical Review Letters}}
  \textbf{\bibinfo{volume}{120}}, \bibinfo{pages}{156001}
  (\bibinfo{year}{2018}).

\bibitem{haastrupComputational2DMaterials2018}
\bibinfo{author}{Haastrup, S.} \emph{et~al.}
\newblock \bibinfo{title}{The {{Computational 2D Materials Database}}:
  High-throughput modeling and discovery of atomically thin crystals}.
\newblock \emph{\bibinfo{journal}{2D Materials}} \textbf{\bibinfo{volume}{5}},
  \bibinfo{pages}{042002} (\bibinfo{year}{2018}).

\bibitem{rasmussenComputational2DMaterials2015}
\bibinfo{author}{Rasmussen, F.~A.} \& \bibinfo{author}{Thygesen, K.~S.}
\newblock \bibinfo{title}{Computational {{2D Materials Database}}: {{Electronic
  Structure}} of {{Transition-Metal Dichalcogenides}} and {{Oxides}}}.
\newblock \emph{\bibinfo{journal}{The Journal of Physical Chemistry C}}
  \textbf{\bibinfo{volume}{119}}, \bibinfo{pages}{13169--13183}
  (\bibinfo{year}{2015}).

\bibitem{nazarianComprehensiveSetHighQuality2016}
\bibinfo{author}{Nazarian, D.}, \bibinfo{author}{Camp, J.~S.} \&
  \bibinfo{author}{Sholl, D.~S.}
\newblock \bibinfo{title}{A {{Comprehensive Set}} of {{High-Quality Point
  Charges}} for {{Simulations}} of {{Metal}}\textendash{{Organic Frameworks}}}.
\newblock \emph{\bibinfo{journal}{Chemistry of Materials}}
  \textbf{\bibinfo{volume}{28}}, \bibinfo{pages}{785--793}
  (\bibinfo{year}{2016}).

\bibitem{fey2019fast}
\bibinfo{author}{Fey, M.} \& \bibinfo{author}{Lenssen, J.~E.}
\newblock \bibinfo{title}{Fast graph representation learning with pytorch
  geometric}.
\newblock \emph{\bibinfo{journal}{arXiv preprint arXiv:1903.02428}}
  (\bibinfo{year}{2019}).

\bibitem{DGLLib}
\bibinfo{author}{Wang, M.} \emph{et~al.}
\newblock \bibinfo{title}{Deep graph library: A graph-centric,
  highly-performant package for graph neural networks} (\bibinfo{year}{2019}).
\newblock \urlprefix\url{https://arxiv.org/abs/1909.01315}.

\bibitem{reiser2021graph}
\bibinfo{author}{Reiser, P.}, \bibinfo{author}{Eberhard, A.} \&
  \bibinfo{author}{Friederich, P.}
\newblock \bibinfo{title}{Graph neural networks in tensorflow-keras with
  raggedtensor representation (kgcnn)}.
\newblock \emph{\bibinfo{journal}{Software Impacts}}
  \textbf{\bibinfo{volume}{9}}, \bibinfo{pages}{100095} (\bibinfo{year}{2021}).

\bibitem{gasteiger2022graph}
\bibinfo{author}{Gasteiger, J.} \emph{et~al.}
\newblock \bibinfo{title}{How do graph networks generalize to large and diverse
  molecular systems?}
\newblock \emph{\bibinfo{journal}{arXiv preprint arXiv:2204.02782}}
  (\bibinfo{year}{2022}).

\bibitem{krenn2022selfies}
\bibinfo{author}{Krenn, M.} \emph{et~al.}
\newblock \bibinfo{title}{Selfies and the future of molecular string
  representations}.
\newblock \emph{\bibinfo{journal}{arXiv preprint arXiv:2204.00056}}
  (\bibinfo{year}{2022}).

\bibitem{brammer2022tucan}
\bibinfo{author}{Brammer, J.~C.} \emph{et~al.}
\newblock \bibinfo{title}{Tucan: A molecular identifier and descriptor
  applicable to the whole periodic table from hydrogen to oganesson}.
\newblock \emph{\bibinfo{journal}{Research Square preprint}}
  (\bibinfo{year}{2022}).

\bibitem{yao2021inverse}
\bibinfo{author}{Yao, Z.} \emph{et~al.}
\newblock \bibinfo{title}{Inverse design of nanoporous crystalline reticular
  materials with deep generative models}.
\newblock \emph{\bibinfo{journal}{Nature Machine Intelligence}}
  \textbf{\bibinfo{volume}{3}}, \bibinfo{pages}{76--86} (\bibinfo{year}{2021}).

\bibitem{friederich2021scientific}
\bibinfo{author}{Friederich, P.}, \bibinfo{author}{Krenn, M.},
  \bibinfo{author}{Tamblyn, I.} \& \bibinfo{author}{Aspuru-Guzik, A.}
\newblock \bibinfo{title}{Scientific intuition inspired by machine
  learning-generated hypotheses}.
\newblock \emph{\bibinfo{journal}{Machine Learning: Science and Technology}}
  \textbf{\bibinfo{volume}{2}}, \bibinfo{pages}{025027} (\bibinfo{year}{2021}).

\bibitem{krenn2022scientific}
\bibinfo{author}{Krenn, M.} \emph{et~al.}
\newblock \bibinfo{title}{On scientific understanding with artificial
  intelligence}.
\newblock \emph{\bibinfo{journal}{arXiv preprint arXiv:2204.01467}}
  (\bibinfo{year}{2022}).

\bibitem{lavin2021simulation}
\bibinfo{author}{Lavin, A.} \emph{et~al.}
\newblock \bibinfo{title}{Simulation intelligence: Towards a new generation of
  scientific methods}.
\newblock \emph{\bibinfo{journal}{arXiv preprint arXiv:2112.03235}}
  (\bibinfo{year}{2021}).

\end{thebibliography}

\end{document}